\documentclass[aps, pra, twocolumn, superscriptaddress, amsmath, amssymb]{revtex4-2}

\usepackage{graphicx}
\usepackage{dcolumn}
\usepackage{bm}
\usepackage{mathptmx} 
\newcommand*{\hham}{\mathcal{H}}
\usepackage{color}
\usepackage[dvipsnames]{xcolor}
\usepackage[hidelinks,colorlinks=true, linkcolor=WildStrawberry, urlcolor=WildStrawberry, citecolor=Emerald, linktocpage=true, breaklinks=true, bookmarksopen=true]{hyperref}
\usepackage{multirow}
\usepackage[USenglish]{babel}

\begin{document}
	
	\title[Quantum plasmonics model of refractive index sensing using photon correlations]
	{Quantum plasmonics model of refractive index sensing using photon correlations} 
	
	\author{Luke C. Ugwuoke}
 \email{ lcugwuoke@gmail.com }
	\affiliation{
		Department of Physics, Stellenbosch University, 
		Private Bag X1, Matieland 7602, South Africa.}
	
	\author{Tjaart P. J. Kr\"{u}ger}
	\affiliation{ 
		Department of Physics, University of Pretoria, 
		Private Bag X20, Hatfield 0028, South Africa.}
        \affiliation{National Institute for Theoretical and Computational Sciences, (NITheCS), South Africa.}
	\author{Mark S. Tame}
	\affiliation{
		Department of Physics, Stellenbosch University, 
		Private Bag X1, Matieland 7602, South Africa.}
	\date{\today}
	\hyphenpenalty = 1000
	\begin{abstract}
		The interaction between the electric dipole moments of a quantum emitter and a metal nanoparticle gives rise to unique optical properties, such as interference-induced photon correlations, that could be useful for 
		enhanced intensity-based sensing.
		Using the quantum theory of photodetection, we propose a nanosensor system
		comprising a quantum emitter and a metal nanoparticle
		that explores the possibility of utilizing higher-order photon correlations for refractive index sensing.
		Both the refractive index sensitivity and resolution of the nanosensor, whose scattering spectrum lies within the visible region, are predicted.
		The sensor
		is supported by a substrate and driven weakly by a coherent field. 
		By calculating the mean photocount and its second factorial moment resulting from the scattered field of the system, the sensing performance of the intensity and intensity--intensity correlation, $g^{(2)}(0)$, are compared at optimal driving wavelengths. 
		The mean photocount was found to be inherently low, inhibiting the role of interference-induced photon antibunching in minimizing the sensor's intensity shot noise. 
		However, a regime in which the noise could be reduced below the shot noise limit is identified, leading to a quantum enhancement in the sensing performance.
	\end{abstract}
	
	\maketitle 

	\section{Introduction}\label{section1}
	Quantum plasmonic sensing has gained considerable attention recently within the plasmonics community 
	due to its potential to improve the resolution of plasmonic sensors~\cite{Lee11,Jiri08} beyond their classical limits~\cite{Mola19,Lee16}. 
	Of particular interest to researchers in the life sciences is plasmonic-based refractive index sensing, for use in sensing viruses and proteins, in addition to measuring drug kinetics~\cite{Cres12,Tay13,Chen07}. 
	The minimum change in the refractive index that leads to a detectable change in a sensor's output --- the refractive index resolution, also known as the limit of detection --- depends on the sensitivity of the sensor and to a large extent on the noise level in the sensor's output~\cite{Jiri08,Lee21}. The output noise arises from fluctuations in the intensity of the input driving light field used, as well as statistical features of the scattered light reaching a photodetector from the sensing region, both of which are generally Poissonian in nature. When combined, these sources of noise are referred to as the shot noise in classical setups~\cite{Mola19,Jiri08,Lee21}. 
	Additional noise stems from systematic elements, such as detector noise~\cite{Jiri08}, collectively known as technical noise~\cite{Lee21}. 
	Quantum plasmonic sensing strives to minimize the overall noise through several means, including the use of specialized quantum states of light to reduce intensity fluctuations below that of the shot noise~\cite{Lee16,Lee18}, designing highly-efficient photodetectors and measurement schemes~\cite{You20,Zhang20,Lubin22}, and utilizing 
	hybrid plasmonic systems that support scattered light with non-classical photon statistics~\cite{Lee21}.  

A particular class of quantum plasmonic sensor relies on the emitter--plasmon coupling effect in a hybrid system of a quantum dot (QD), or molecular dipole emitter, and a metal nanoparticle (MNP)~\cite{Hatef13,Lee21}.
	 These emerging plasmonic sensors exploit the sensitivity of different quantum effects exhibited by the hybrid system to minute changes in the refractive index~\cite{Hatef13,Mao13,Zheng23,Kong19,Lee21}. 
  
Amongst the quantum effects in a hybrid plasmonic system that have been investigated for plasmonic sensing is coherent exciton-plasmon dynamics~\cite{Hatef13,Mao13} and Rabi splitting \cite{Zheng23,Kong19}, where theoretical models show some promising results. 
However, an interesting quantum effect not yet explored for quantum plasmonic sensing is Fano interference~\cite{Luka10,Miro10} {--- }an interference phenomenon that occurs when two or more resonators interact. {Previous studies} have shown that Fano interference is exhibited by hybrid systems of single, or multiple QDs, and single plasmonic nanostructures~\cite{Ceng07,Waks10,Rous18,Dolf10,Gong15,Blaq18,Kam14,Sha13,Zang06,Ugler08,Kim14}. 
 Another intriguing property of these hybrid systems is their ability to support both Poissonian and sub-Poissonian photon statistics in the scattered light, induced by the Fano interference between the resonance of the MNP, for instance, and that of the QD~\cite{Waks10,Dolf10,Gong15,Blaq18}. The interaction between the broad linewidth of the localized surface plasmon resonance (LSPR) of the MNP supporting a continuum of energy and the narrow absorption, or emission, linewidth of the QD supporting a discrete energy gives rise to a Fano resonance and a modification of the quantum statistics of the scattered light~\cite{Wu10,Waks10}. The intensity--intensity correlation function, $g^{(2)}(\tau)$, is often used to characterize the photon statistics of the light scattered by the MNP component of the hybrid system~\cite{Waks10}, or by the entire hybrid system~\cite{Dolf10}. A rise in $g^{(2)}(\tau)$ with time ($\tau > 0$) indicates photon antibunching, i.e., the tendency of the scattered photons to arrive at a photodetector in a near-regular pattern, while the converse indicates photon bunching~\cite{Rod00,Milo95,Mik16,Mand90}. In Refs.~\cite{Gong15,Waks10}, it was shown that constructive Fano interference, which occurs within the Fano peak, is responsible for interference-induced photon antibunching and sub-Poissonian statistics. A question then naturally arises about whether the antibunching with sub-Poissonian statistics found within the Fano peak could possibly be exploited to reduce noise and gain an advantage in refractive index sensing.
	
	In this work, we {seek an answer to the above question} by proposing a theoretical model of an emitter-MNP quantum plasmonic sensor that makes use of the scattered intensity of the constructive Fano interference and the second-order correlation function at zero-delay time, $g^{(2)}(0)$, within the Fano spectrum, for refractive index sensing. The use of $g^{(2)}(0)$ and other higher-order intensity correlations has recently been considered for improving image resolution in confocal microscopy and quantum imaging~\cite{Tenn19,Lubin22,Chera19,Geno14}, thus providing additional motivation for its study in our context. By calculating the sensitivity and the resolution of the sensor based on the scattered intensity and $g^{(2)}(0)$, a comparison is made between their sensing performance at driving wavelengths within the plasmon and Fano spectra. Although the intensity has been thoroughly investigated for refractive index sensing~\cite{Celi21,Son14,Chen08,Jeon19,Lee11,Chen18} and applied to biosensing~\cite{Nath04,Chen20}, the antibunching effect could potentially minimize the shot noise component of the intensity noise~\cite{Treussart2002,Alleaume2004} and improve the resolution. Moreover, the comparisons between the intensity and $g^{(2)}(0)$ drawn here are necessary to investigate the possibility of an enhancement in each of their respective sensing parameters---refractive index sensitivity and resolution---predicted by our model. 

	This paper is organized as follows. In Sec.~\ref{section 2}, we introduce the model geometry of the quantum plasmonic sensor and the input parameters used in the model. In Sec.~\ref{section 3}, the equations of motion governing the expectation values of the system operators and their steady-state solutions are derived and discussed in the framework of cavity quantum electrodynamics (cQED), using an improved approach that takes substrate effects and the refractive index of the emitter into account, unlike in previous works~\cite{Waks10,Dolf10,Gong15}. In Sec.~\ref{section 4}, we introduce the photodetection theory following the approach in Ref.~\cite{Rod00}. The mean scattering rate of the system, the mean photocount, its second factorial moment, the second-order correlation function, and their standard deviations are discussed. In Sec.~\ref{section 5}, the sensing performance of the proposed sensor, using the scattered intensity and the intensity--intensity correlation as signals, 
	both within the plasmon and the Fano spectra, are discussed. We conclude the paper in Sec.~\ref{section 6} by mentioning the key results as well as some challenges and future directions. 

\section{Model geometry and Input parameters}\label{section 2}
\begin{figure*} [t!]
	\centering 
	\includegraphics[width = 0.40\textwidth]{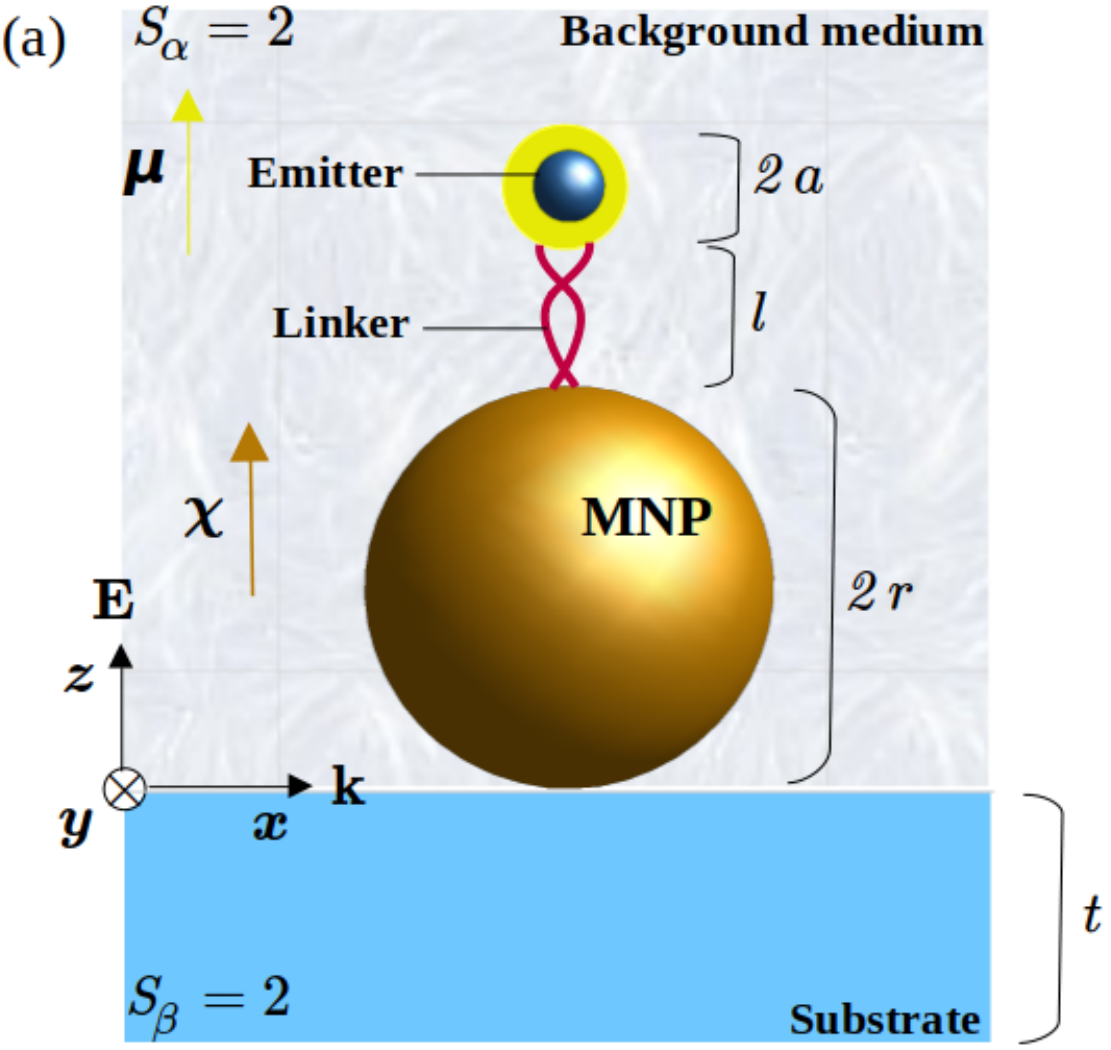}
 \qquad
    \includegraphics[width = 0.26\textwidth]{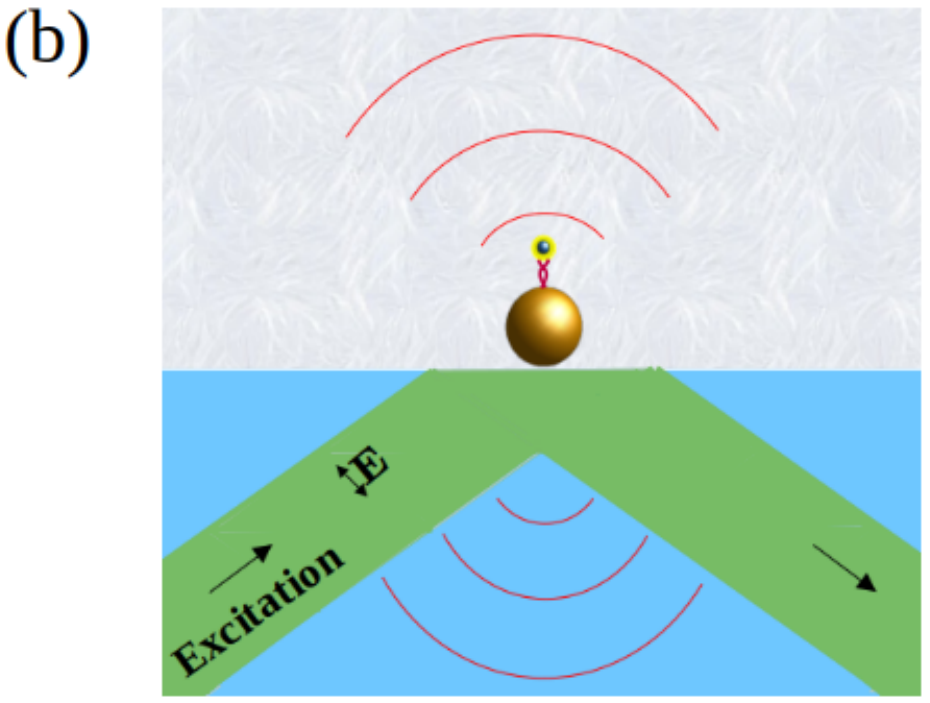}
	\caption{(Color online) (a) 2D schematic of the proposed quantum plasmonic nanosensor. It consists of a metal nanoparticle (MNP) and a  quantum emitter---shown here as a core-shell QD held by a linker attached to the MNP. The hybrid system is embedded in a medium whose refractive index changes lead to changes in the scattered intensity or $g^{(2)}(0)$, and supported by a substrate. (b) A possible experimental set-up for the realization of the theoretical model in (a), with the driving electric field as the transmitted evanescent field from total internal reflection.
	}\label{f1}
\end{figure*}

 The system we consider is shown in Fig.~\ref{f1}(a). It consists of an 
 emitter---which we model as a core-shell QD as an example---held by a linker attached to a spherical MNP. The system is embedded in a background medium and supported by a substrate.
 In the configuration shown, referred to as the longitudinal coupling (LC) configuration~\cite{Dolf10,Waks10,Ceng07}, the electric dipole moments of the MNP and QD are both parallel to the MNP-QD axis in the direction of polarization of the incident field, i.e., $\bm{\chi} = \chi \mathbf{\hat{e}}_{z}$ and $\bm{\mu} = \mu \mathbf{\hat{e}}_{z}$, respectively.  
 We will consider only the LC configuration, since it has been reported as the optimal configuration for atomic cooperativity between the MNP and the QD, as well as for  interference-induced effects in the hybrid system~\cite{Waks10,Dolf10,Gong15,Ceng07}, though our model can also be applied to the transverse coupling case.
 The parameter $S_{\beta}$ is the coupling factor of the MNP dipole moment, $\bm{\chi}$, to the image dipole formed in the substrate (perpendicular coupling: $\beta = z, S_{\beta} = 2$, 
parallel coupling: $\beta = x, y, S_{\beta} = 1$)~\cite{Zade17}, 
and $S_{\alpha}$ is the coupling factor of the QD to the MNP 
(longitudinal coupling: $\alpha = z, S_{\alpha} = 2$, 
transverse coupling: $\alpha = x, y, S_{\alpha} = -1$)~\cite{Waks10,Ceng07}. 
Here, we take $S_{\alpha} = S_{\beta} = 2$.

The QD is chosen to be a CdTe/CdS core-shell semiconductor nanocrystal whose emission spectrum is tunable within the visible region, as discussed in Ref.~\cite{Deng10}. 
The core-shell radius of the QD is denoted as $a$, where $a = r_{c} + t_{s}$, with $r_{c}$ being the core radius, and $t_{s}$ the shell thickness.
The QD is modeled as a two-level system with
ground and excited states denoted as $|0 \rangle$ and $|1 \rangle$, respectively. Here, the dipole dephasing rate of the QD is ignored and the non-radiative decay rate of the QD is assumed to be negligible. The latter is justified by their high intrinsic quantum yields~\cite{Deng10,Kawa20}, while the former is valid within the semi-classical, weak-field limit~\cite{Waks10}. 
The refractive index, free-space radiative decay rate, and emission frequency of the QD are denoted as
$n_{d}$, $\gamma_{ex}$, and  $\omega_{ex}$, respectively. 
The values of $r_{c}, t_{s}, \omega_{ex}$, and $\gamma_{ex}$ were taken from Ref.~\cite{Deng10}, while the value of $n_{d}$ is taken from Ref.~\cite{Hatef13}.
In Fig.~\ref{f1}(a), the surface-to-surface distance between the MNP and the QD, which we denote as $l$, is occupied by a linker, such that $l \ge 3$ nm, within which the dipole--dipole coupling is valid~\cite{Dolf10}, the atomic cooperativity is high~\cite{Waks10}, and charge transfer between the MNP and the QD does not occur~\cite{Lech16}. {As shown in Ref. \cite{Gong15}, both Fano interference and antibunching effects are supported by the hybrid system even beyond the dipole-dipole coupling limit.} 
 
A gold nanoparticle is used as the model MNP whose radius, denoted as $r$, is large enough such that the local response approximation is valid~\cite{Waks10,Dolf10,Kim14}, but small enough to ensure a non-vanishing Fano profile, and subsequently, a sizeable photon antibunching effect, as previously shown in Ref.~\cite{Dolf10}. 
The center-to-center distance between the MNP and the QD is denoted as $d$, where $d = r + l + a$. 
The nanosensor is driven weakly by a $z$-polarized electric field, 
$\mathbf{E}(t) = E_{0}\cos(\omega t)\mathbf{\hat{e}}_{z}$, with driving frequency $\omega$ and amplitude $E_{0} = (2I_{0}/cn\varepsilon_{0})^{\frac{1}{2}}$, 
where $I_{0}$ is the incident field intensity, $c$ is the speed of light in vacuum, $n$ is the refractive index of the background medium, and $\varepsilon_{0}$ is the permittivity of free space. The electric field is propagating in the $x$-direction with a wavevector of magnitude $k = \omega n/c$, as shown in Fig.~\ref{f1}(a). We have ignored the spatial dependence of $\mathbf{E}(t)$ since $kx \ll 1$ over the lateral extent of the sensor ($2r$ in the $x$-direction)  --- the quasi-static (long-wavelength) limit~\cite{Das07,Carm99,Wals08,Stef07}.
A weak-driving field ensures that the Fano peak is not suppressed relative to the plasmon resonance peak, since the Fano profile vanishes at strong-driving fields due to the saturation of the QD excited-state population~\cite{Dolf10,Waks10}.

The dielectric response of the MNP is modeled by a complex Drude permittivity with free-electron damping rate $\gamma_{p}$, bulk plasma frequency $\omega_{p}$, and high-frequency refractive index 
$n_{\infty}$. Here, $\omega_{p}$ was adjusted such that the LSPR of the MNP matches the experimental value for that of a MNP with the same radius~\cite{Chen08}. The background medium is water, and the substrate support is modeled as a weakly-reflecting glass slab with thickness $t \gg r$ and refractive index denoted as $n_{s}$.
In Table~\ref{t1}, the values of the input parameters used in our model are shown.

In Fig.~\ref{f1}(b), as a basic example, we provide a potentially more accessible experimental setting for the realization of the theoretical model shown in (a) and discussed above. It is based on a setup experimentally realized in Ref.~\cite{Eghlidi09}, thus making our scenario more realistic. Here, the driving electric field, $\mathbf{E}(t)$, is the evanescent field produced by total internal reflection. In this case, when the incident angle is greater than the critical angle, the dominant component of the transmitted field in the region above the substrate is in the $\mathbf{\hat{e}}_{z}$-direction. The amplitude of the electric field would then be modified as $E_0 = t_p (2I_{0}/cn_s\varepsilon_{0})^{\frac{1}{2}}(n_s/n)$, with $t_p$ obtained from the Fresnel equations at the boundary, as described in more detail in Ref.~\cite{BYU}.
\begin{table}[t!]
	\scalebox{0.9}{
		\begin{tabular}{ | c |c |c |c |c |c |l | }
			\cline{1-7}
			\multicolumn{1}{| c }{ \multirow{2}{*}{  } } &
			\multicolumn{1}{ c }{Sizes (nm)} &  & $r_{c}$ & $t_{s}$ & $a$ & $r$  \\ \cline{4-7}
			\multicolumn{1}{| c }{} &
			\multicolumn{1}{ c }{} &  & 0.8 & 0.7 & 1.5 & 25  \\ \cline{1-7}
			\multicolumn{1}{| c }{ \multirow{2}{*}{  } } &
			\multicolumn{1}{ c }{Refractive} &  & $n_{\infty}$ & $n_{d}$ & $n_{s}$ & $n$  \\ \cline{4-7}
			\multicolumn{1}{| c }{} &
			\multicolumn{1}{ c }{indices} &  & 3.16 & 2.45 & 1.5 & 1.3330 ---  \\ 

			\multicolumn{1}{| c }{} &
			\multicolumn{1}{ c }{} &  & & & & 1.3334  \\ 
   \cline{1-7}
   
			\multicolumn{1}{| c }{ \multirow{2}{*}{  } } &
			\multicolumn{1}{ c }{Frequencies/} &  & $\omega_{p}$ (meV) & $\gamma_{p}$ (meV) & $\omega_{ex}$ (meV)& $\gamma_{ex}$ (neV) \\ \cline{4-7}
			\multicolumn{1}{| c }{} &
			\multicolumn{1}{ c }{Damping and decay rates} &  & 8579 & 71 & 2149 & 118  \\ \cline{1-7}
			\multicolumn{1}{| c }{ \multirow{2}{*}{  } } &
			\multicolumn{1}{ c }{Others} &  & $I_{0}$ (W/cm$^{2}$) & $t$ (mm) & $l$ (nm) & $d$ (nm)  \\ \cline{4-7}
			\multicolumn{1}{| c }{} &
			\multicolumn{1}{ c }{} &  & 33.6 & 0.17 & 3.5 & 30  \\ \cline{1-7}
	\end{tabular} }
	\caption{List of input parameters used in the model as an example.
	}\label{t1}
\end{table}

\section{Equations of motion} \label{section 3}
In order to determine the mean photon scattering rate of the nanosensor as it is driven by the external field, we first write down the total Hamiltonian of the MNP-QD system. After applying the rotating-wave approximation, the total Hamiltonian of the coupled MNP-QD system in the dipole limit is obtained as~\cite{Gong15}
\begin{eqnarray}\label{e1}
		\hat{\hham} &=&  \hbar\Delta_{pl}\hat{a}^{\dagger}\hat{a} + 
		 \hbar\Delta_{ex}\hat{\sigma}^{\dagger}\hat{\sigma}
		 - \hbar g(\hat{\sigma} \hat{a}^{\dagger} + \hat{\sigma}^{\dagger}\hat{a}) \nonumber \\
		 & & - \hbar\Omega_{ex}(\hat{\sigma}  + \hat{\sigma}^{\dagger} )- 
		\hbar\Omega_{pl}(\hat{a}  + \hat{a}^{\dagger}).
\end{eqnarray}
Eq.~\eqref{e1} has been written in a frame rotating at the driving frequency, i.e., with respect to the Hamiltonian $\hbar \omega\left(\hat{a}^{\dagger}\hat{a} +
\hat{\sigma}^{\dagger}\hat{\sigma}\right)$, where the 
exciton (ex) raising and lowering operators of the QD,  $\hat{\sigma}^{\dagger} = |1\rangle\langle 0|$ and $\hat{\sigma} = |0 \rangle\langle 1|$, respectively, and the plasmon (pl) creation and annihilation operators of the MNP, $\hat{a}^{\dagger}$ and $\hat{a}$, respectively, are time-independent operators. The first two terms in Eq.~\eqref{e1} are the free Hamiltonians of the MNP and QD in the rotating frame, respectively. The third term is the Hamiltonian of the dipole--dipole interaction, while the fourth and fifth terms are the driving-field Hamiltonians of the QD and MNP, respectively.
In Eq.~\eqref{e1}, $\hbar$ is Dirac's constant, $\Delta_{pl} = \omega_{pl} - \omega$ is the plasmon-driving field detuning frequency, $\Delta_{ex} = \omega_{ex} - \omega$ is the exciton-driving field detuning frequency, $\Omega_{ex} = E_{0}\mu/2\hbar$ is the Rabi frequency, $\mu = er_{0}$ is the magnitude of the dipole moment of the QD, where $e$ is the electronic charge, and $\Omega_{pl} = E_{0}\chi/2\hbar$ is the driving frequency of the MNP. 
The LSPR of the MNP, denoted as $\omega_{pl}$, the magnitude of the MNP dipole moment, $\chi$, and the coupling rate of the QD to the MNP, denoted here as $g$, in the presence of a substrate, are obtained as (see Sections 1, 2, and 4 of the Supplemental Material \cite{Supp})
\begin{subequations}
	\begin{align}
		\omega_{pl} & = \sqrt{ \frac{\omega_{p}^{2}}{\epsilon_{\infty} + f\epsilon_{b} } - \gamma_{p}^{2} }, \label{e2a} \\
         \chi &  = \frac{1}{3}(f+1)\epsilon_{b}\sqrt{12\pi\varepsilon_{0}\hbar
         	\eta r^{3}  }, ~~~~  \text{and} \label{e2b} \\
         g & = \frac{1}{3}(f+1)\left(\frac{\mu S_{\alpha} }{d^{3} } \right)\left(\frac{\epsilon_{b} }{\epsilon^{\prime}_{b} } \right) \sqrt{ \frac{3\eta r^{3} }{4\pi\varepsilon_{0}\hbar }  }, \label{e2c}
	\end{align}
\end{subequations}
respectively. 
The LSPR of the MNP is obtained by accounting for the geometric factor of the MNP in the presence of a substrate in the evaluation of the Fr\"{o}hlich condition~\cite{Zade17}.
Both the MNP dipole moment, $\chi$, and the coupling rate of the QD to the MNP, $g$, due to the dipole--dipole interaction, are derived by comparing the classical model of the induced dipole field of the MNP in the presence of the QD to the quantum formulation --- an approach that is prevalent in the literature~\cite{Dolf10,Kim14,Ceng07,Gong15,Zang06}. 
In Eqs. \eqref{e2a}--\eqref{e2c}, $\varepsilon_{\infty} = 
n_{\infty}^{2}$ is the high-frequency dielectric constant of gold, 
$\varepsilon_{b} = n^{2}$ is the dielectric constant of the background medium, $\epsilon_{d} = n_{d}^{2}$ is the dielectic constant of the QD, $\epsilon^{\prime}_{b} = (2\epsilon_{b} + \epsilon_{d})/3$ is the effective dielectric constant of the medium, and $f = (1 - L)/L$, where
\begin{subequations}
	\begin{align}
		L & \approx \frac{1}{3}\left[1 - S_{\beta}\frac{\mathcal{R}}{8}\left( 1 - (1 - \mathcal{R}^{2})\Big(1 + \frac{t}{r}\Big)^{-3}\right)    \right], \label{e3a} ~~~~  \text{and} \\
		\eta &  = \frac{1}{2\omega_{pl}}\left(\frac{\omega_{p}}{\varepsilon_{\infty}+f\varepsilon_{b} } \right)^{2}. \label{e3b} 
	\end{align}
\end{subequations}
In Eq.~\eqref{e3a}, $L$ is the geometric factor of a spherical MNP supported by a substrate~\cite{Zade17}. This factor allows us to generalize previous work to the case where the system is placed on a substrate, which is more physically relevant. The reflectance of the substrate, denoted here as $\mathcal{R}$, is calculated from the quasi-static Fresnel reflection coefficient, 
 $\mathcal{R} = (\varepsilon_{s} - \varepsilon_{b})/(\varepsilon_{s} + \varepsilon_{b})$, as given in Ref.~\cite{Zade17}, where $\epsilon_{s} = n_{s}^{2}$ is the dielectric constant of the substrate. 
For a substrate whose refractive index is greater than that of the medium, $f > 2$, resulting in a redshift in the LSPR according to Eq.~\eqref{e2a}, while both the dipole moment of the MNP and the coupling rate of the QD to the MNP slightly increase, according to Eqs.~\eqref{e2b} and~\eqref{e2c} respectively. In our model, this is the case for a glass substrate.
However, a substrate whose refractive index is less than that of the medium will result in a converse effect.
In the absence of a substrate, or one closely matched to that of the medium, such as Teflon, we have $f = 2$, and Eqs.~\eqref{e2a}, \eqref{e2b}, \eqref{e2c}, and \eqref{e3b} reduce respectively to the previously reported expressions derived in Refs.~\cite{Dolf10,Gong15}. 
We have therefore extended the theory to the more practical case where the system is supported by a substrate, which is relevant to experimental setups.

In the presence of a driving field, the interaction of the MNP with the local environment leads to energy dissipation in both radiative and non-radiative forms. The latter is due to Ohmic heating in the MNP, while the former is caused by the scattering of light by the MNP dipole into the far field~\cite{Waks10,Kim14}. In addition, the decay rate of the QD due to an inevitable absorption--spontaneous emission event has to be accounted for. A complete description of the dynamics of the coupled MNP-QD system at optical frequencies, including the aforementioned dissipative pathways, is given by the following phenomenological master equation for the MNP-QD system in Lindblad form~\cite{Kim14,Gong15,Wals08,Carm99}:
\begin{equation}\label{e4}
\dot{\hat{ \rho }  } = \frac{i}{\hbar}[\hat{ \rho },\hat{ \hham }  ] 
 - \frac{\gamma_{pl}}{2}\mathcal{D}_{pl}[\hat{a},\hat{ \rho } ]
 - \frac{\gamma_{ex}}{2}\mathcal{D}_{ex}[\hat{\sigma},\hat{ \rho } ], 
\end{equation}
where 
\begin{equation}\label{e5}
\mathcal{D}_{j}[\hat{c}_{j},\hat{ \rho } ] =  \hat{ \rho }\hat{c}_{j}^{\dagger}\hat{c}_{j}  + \hat{c}_{j}^{\dagger}\hat{c}_{j}\hat{ \rho } - 2\hat{c}_{j}\hat{ \rho }\hat{c}_{j}^{\dagger}
\end{equation}
are the adjoint dissipators, with $j = pl, ex$, so that
$\hat{c}_{pl} = \hat{a}$, $\hat{c}_{ex} = \hat{\sigma}$, and $\hat{ \rho }$ is the density operator of the MNP-QD system in the rotating frame. 
In Eq.~\eqref{e4}, $\gamma_{pl} = \gamma_{nr} + \gamma_{r}$ is the decay rate of the MNP, where
\begin{subequations}
	\begin{align}
		\gamma_{nr} & = \gamma_{p}\left[ 1 +  \Big(\frac{\gamma_{p} }{\omega_{pl} } \Big)^{2} \right] \label{e6a} ~~~~  \text{and} \\
		\gamma_{r} &  = \frac{ 4}{9}(f+1)^{2}\eta n^{2}\Big(k\Big|_{\omega = \omega_{pl}}r\Big)^{3} \label{e6b} 
	\end{align}
\end{subequations}
are the non-radiative and radiative decay rates of the MNP, respectively, whose derivations are given in Sections 1 and 4 of the Supplemental Material \cite{Supp}.
The non-radiative decay rate of the MNP is obtained from a Lorentzian approximation of its quasi-static dipole polarizability in the presence of a substrate, and the radiative decay rate is treated phenomenologically by calculating it from the Wigner--Weisskopf formula~\cite{Lukas12}, which is consistent with a treatment using the radiation reaction field~\cite{Ford84}.
 
The coupled equations of motion describing the evolution of the system are obtained from the expectation values of the MNP and QD operators~\cite{Dolf10,Waks10,Kim14}: 
$\langle \hat{a} \rangle \equiv \mathrm{Tr}[\hat{a} \hat{ \rho }  ], 
\langle \hat{\sigma} \rangle \equiv \mathrm{Tr}[\hat{\sigma} \hat{ \rho }  ]$, and 
$\langle \hat{\sigma}_{z} \rangle \equiv \mathrm{Tr}[\hat{\sigma}_{z} \hat{ \rho }  ]$, where $\hat{\sigma}_{z} = 2\hat{\sigma}^{\dagger}\hat{\sigma} - \mathbb{I}$ is the Pauli Z operator, and 
$\langle\hat{\sigma}^{\dagger}\hat{\sigma}\rangle$ is the excited-state population of the QD. 
In the case of $\langle \hat{a} \rangle$ and $\langle \hat{\sigma} \rangle$, the solution in the laboratory frame is obtained by multiplying the expectation value obtained in the rotating frame by the phase factor $e^{-i\omega t}$. The other expectation values of interest are invariant. 
The equation of motion for the expectation value of the QD lowering operator, $\hat{\sigma}$, couples to higher-order expectation values leading to an infinite system of coupled equations. A semi-classical approach that assumes that the expectation value of the product of an MNP operator and a QD operator can be factored~\cite{Kim14,Waks10} allows us to approximate the equations and obtain the following Maxwell--Bloch equations using 
$\langle \overset{.}{\hat{a}} \rangle \equiv \partial_{t}\langle \hat{a} \rangle \equiv \mathrm{Tr}[\hat{a} \overset{.}{\hat{ \rho }}  ], $
and similarly
$
\langle \overset{.}{\hat{\sigma}} \rangle \equiv \partial_{t}\langle \hat{\sigma} \rangle \equiv \mathrm{Tr}[\hat{\sigma} \overset{.}{\hat{ \rho }}  ]$, 
$\langle \overset{.}{\hat{\sigma}}_{z} \rangle \equiv \partial_{t}\langle \hat{\sigma}_{z} \rangle \equiv \mathrm{Tr}[\hat{\sigma}_{z} \overset{.}{\hat{ \rho }}  ]$, and noting that the operators are time-independent, while their expectation values are not: 
\begin{subequations}
	\begin{align}
		\langle \overset{.}{\hat{a}} \rangle & = 
		-\Big(i\Delta_{pl}+\frac{1}{2}\gamma_{pl}\Big)\langle \hat{a} \rangle 
		+ i\Big(g\langle \hat{\sigma} \rangle + \Omega_{pl}\Big), \label{e7a}\\
		\langle \overset{.}{\hat{\sigma}} \rangle & = 
		-\Big(i[ \Delta_{ex} + \mathcal{F}\Delta_{pl}\langle \hat{\sigma}_{z} \rangle  ] 
		+\frac{1}{2}[\gamma_{ex} - \mathcal{F}\gamma_{pl}\langle \hat{\sigma}_{z} \rangle  ] \Big)\langle \hat{\sigma} \rangle \nonumber \\ 
  &\qquad +i\Omega\Big(1 -2\langle \hat{\sigma}^{\dagger}\hat{\sigma} \rangle \Big), \label{e7b}\\
		\langle \overset{.}{\hat{\sigma}_{z}} \rangle & = 
		-\Gamma\Big(\langle \hat{\sigma}_{z} \rangle + 1 \Big)
		+ 4 \Im[\Omega^{*}\langle\hat{\sigma}\rangle].  \label{e7c}
	\end{align}
\end{subequations}
Details of the derivation are given in Section 4 of the Supplemental Material \cite{Supp}.
The steady-state solution of Eq.~\eqref{e7a} is 
\begin{equation}
\langle \hat{a} \rangle = \frac{i[\Omega_{pl} + g\langle \hat{\sigma} \rangle ]}{i\Delta_{pl} +\frac{1}{2}\gamma_{pl} }. \label{e10c}
\end{equation}
Due to the short plasmon lifetime of the MNP dipole mode, i.e., $\gamma_{pl}^{-1}$ ($\sim 5.75$ fs)  $ \ll \gamma_{ex}^{-1}$ ($\sim 35$ ns),  $\langle \overset{.}{\hat{\sigma}} \rangle$ and 
$\langle \overset{.}{\hat{\sigma}_{z}} \rangle$ have been solved in the adiabatic limit. In the adiabatic approximation~\cite{Waks10,Ceng07}, Eq.~\eqref{e10c} is used to eliminate the dependence of $\langle \overset{.}{\hat{\sigma}} \rangle$ and 
$\langle \overset{.}{\hat{\sigma}_{z}} \rangle$, respectively, on $\langle \hat{a} \rangle$ to obtain Eqs.~\eqref{e7b} and \eqref{e7c}, as well as
the enhanced radiative decay rate of the QD, $\Gamma$, responsible for the Purcell effect, and the modified Rabi frequency, $\Omega$, respectively, as
\begin{subequations}
	\begin{align}
		\Gamma & = \gamma_{ex} + \mathcal{F}\gamma_{pl} \label{e8a} ~~~~  \text{and} \\
		\Omega &  = \Omega_{ex}\left[1 + \frac{ig\chi}{\mu\Big(i\Delta_{pl}+\frac{1}{2}\gamma_{pl}\Big)}\right], \label{e8b} 
	\end{align}
\end{subequations}
where 
\begin{equation}\label{e9}
\mathcal{F} = \frac{g^{2}}{ \Delta_{pl}^{2} + \frac{1}{4}\gamma_{pl}^{2}}
\end{equation}
is the plasmon-induced term.
In the weak-driving field limit, $\Gamma \gg \Omega$, we obtain $\langle \hat{\sigma}_{z} \rangle \approx -1$ from the steady-state solution of Eq.~\eqref{e7c}.
Setting $\langle \hat{\sigma}_{z} \rangle \approx  -1$ in Eq.~\eqref{e7b} leads to the following steady-state expectation value:
\begin{equation}\label{e11}
\langle \hat{\sigma} \rangle  \approx \frac{i\Omega[1 - 2\langle \hat{\sigma}^{\dagger}\hat{\sigma} \rangle ] }{i\Delta + \frac{1}{2}\Gamma }.  
\end{equation}
Finally, in order to obtain an analytical expression for the QD excited-state population, 
$\langle \hat{\sigma}^{\dagger}\hat{\sigma} \rangle$, 
that depends on the driving field intensity, we substitute Eq.~\eqref{e11} for $\langle \hat{\sigma} \rangle$ in the steady-state solution of Eq.~\eqref{e7c} to obtain
\begin{equation}
\langle \hat{\sigma}^{\dagger}\hat{\sigma} \rangle  \approx  \frac{\mathcal{P}}{1 + 2\mathcal{P}} , \label{e10a}\\
\end{equation}
where $\mathcal{P}$ is the saturation parameter of the QD excited-state population, similar to that of a driven two-level atom~\cite{Das07}, but with plasmon-induced terms, and $\Delta$ is the plasmon-induced exciton shift, obtained respectively as
\begin{subequations}
\begin{align}
	\mathcal{P} & = \frac{2|\Omega/\Gamma|^{2} }{1 + 2(\Delta/\Gamma)^{2}  } , \label{e11a}~~~~  \text{and}\\
	\Delta & = \Delta_{ex} - \mathcal{F}\Delta_{pl}.  \label{e11b}
\end{align}
\end{subequations}

The detailed derivation of the above analytical solutions is
given in Section 4 of the Supplemental Material \cite{Supp}. They were validated by performing simulations using the open-source quantum simulation toolbox QuTiP~\cite{Nori13,Nati11}. 
The simulations were performed within a cQED framework where
the MNP is treated as a cavity and the QD as a two-level atom~\cite{Das07,Wals08,Nati11}. Using number states with up to $10$ photons to truncate the infinite Hilbert space of the MNP field operator, the optical response of the driven atom-cavity system was found to agree with the analytical solution, as shown via the scattered flux observable in Fig.~\ref{f2}(a) (see next section for details). 
Besides saving computational time, 
the analytical solutions are more insightful than simulations since they reveal how the QD excited-state population,  $\langle \hat{\sigma}^{\dagger}\hat{\sigma} \rangle$, and the coherence, 
$\langle \hat{\sigma} \rangle$, are modified by the presence of the MNP. As we show in the next section, these modified terms are important in the description of the photon statistics of the light scattered by the system when operating as a nanosensor. 

\section{Photodetection} \label{section 4}
Following the approach in Refs.~\cite{Waks10,Rous18}, 
the scattering operator
for the far-field radiating spatial mode of the MNP-QD system is given by 
\begin{equation}\label{e14-4}
\hat{b}  \approx \sqrt{\gamma_{r} } \hat{a},
\end{equation}
where we have ignored the contribution from the QD since 
$\gamma_{ex}~(\sim 2.86\times10^{7}$ s$^{-1})$ $\ll  \gamma_{r}~(\sim 2.33\times10^{11}$ s$^{-1})$. 
We note that the QD has its {\it decay rate} enhanced by the Purcell effect due to the presence of the MNP; however, its {\it emission rate} into free space, $\gamma_{ex}$, is an intrinsic property and this is modified only slightly by the presence of the MNP, as described in more detail in Ref.~\cite{Pelton2015}.
We also note that although Eq.~\eqref{e14-4} has no direct contribution from the QD, the dynamics of the QD are imprinted indirectly onto the MNP plasmon dynamics and picked up in 
$\hat{ \rho }$, which gives a non-trivial time dependence to the expectation value of the plasmon operator $\hat{a}$, or a combination of it. 
The mean rate (flux) at which the incident photons are scattered into free-space is then~\cite{Waks10,Rous18}
\begin{equation}\label{e12}
\langle \hat{b}^{\dagger}\hat{b} \rangle 
\approx \gamma_{r}\langle \hat{a}^{\dagger}\hat{a} \rangle, 
\end{equation}
with the steady-state mean photon number for any given time in the near field of the MNP 
as (see Supplemental Material, Section 5 \cite{Supp})
\begin{equation}\label{e13}
	\langle \hat{a}^{\dagger}\hat{a} \rangle  = 
	\frac{\Omega_{pl}^{2} + g^{2}\langle \hat{\sigma}^{\dagger}\hat{\sigma} \rangle + 2\Omega_{pl}g\Re\langle \hat{\sigma} \rangle   }{\Delta_{pl}^{2} + \frac{1}{4}\gamma_{pl}^{2}  }.
\end{equation}

Alternatively, the polarization operator, given in Refs.~\cite{Dolf10,Gong15} as $\hat{P}^{+} = \chi \hat{a} + \mu\hat{\sigma} \approx \chi \hat{a}$ (since $\chi \gg\mu$, for example, in our model, $\chi \approx 64 \mu$ and $\mu = 72$ D),  can be used to calculate the scattered intensity, which is proportional to the expectation value of $\hat{P}^{-}\hat{P}^{+}$, 
where $\hat{P}^{+}$ is the operator that corresponds to the radiative 
plasmon mode (since $\chi \propto \sqrt{\gamma_{r}}$ via the Wigner--Weisskopf formula~\cite{Waks10,Lukas12}---see Supplemental Material, Section 5 \cite{Supp}).
However, the scattered flux in Eq.~\eqref{e12} is more useful, since in addition to representing the radiative part, it can be converted to the mean photocount --- a dimensionless parameter that allows us to calculate fluctuations in the scattered light via photodetection theory~\cite{Rod00}. 

It should also be mentioned that $\langle \hat{P}^{-}\hat{P}^{+} \rangle    (\text{cf. }  \langle\hat{b}^{\dagger}\hat{b} \rangle)$ is proportional to the total scattering spectrum, comprising both the coherent, $|\langle \hat{P}^{+} \rangle|^{2}  
(\text{cf. }  |\langle\hat{b}\rangle|^{2})$, and the incoherent, $\langle \hat{P}^{-}\hat{P}^{+} \rangle - |\langle \hat{P}^{+} \rangle|^{2}   (\text{cf.}  \langle\hat{b}^{\dagger}\hat{b} \rangle - |\langle\hat{b}\rangle|^{2})$, scattering components~\cite{Dolf10,Das07,Wals08}. 
For low light intensities, coherent scattering (scattered light with the same frequency, $\omega_{s}$, as the driving frequency, $\omega$) dominates~\cite{Dolf10}, although the light is only coherent to first order as it is monochromatic. At second order, this monochromatic light can be bunched or antibunched~\cite{Das07}, as we show shortly. 
Thus, in the steady state, we can treat the radiating mode associated with the operator $\hat{b}$ as a single frequency mode 
($\omega_{s} = \omega$) for each $\omega$ of the driving field considered, and
convert Eq.~\eqref{e12} to the mean photocount, $\langle \hat{m} \rangle$, recorded by a photodetector with a photon collection efficiency $\xi $ (representing the solid angle capture of the radiation and detection efficiency), within some integration time $\mathrm{T}_{int}$ (see Supplemental Material, Section 5 \cite{Supp}), to obtain
\begin{equation}\label{e14}
\langle \hat{m} \rangle = \xi \mathrm{T}_{int} \langle \hat{b}^{\dagger}\hat{b} \rangle ,
\end{equation}
where $\hat{m}$ is the photocount operator. 

\begin{figure}[t!]
	\centering 
	\includegraphics[width = 0.485\textwidth]{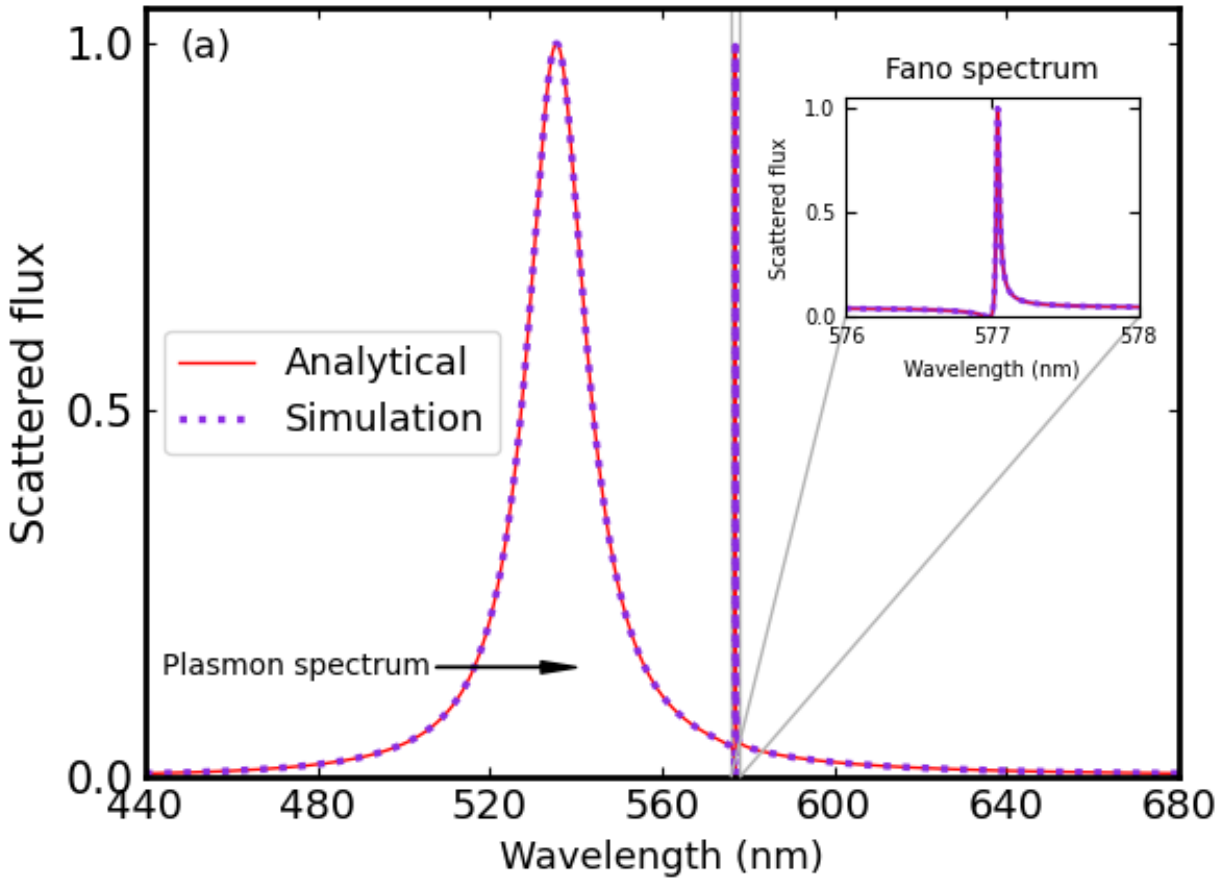}
	\includegraphics[width = 0.485\textwidth]{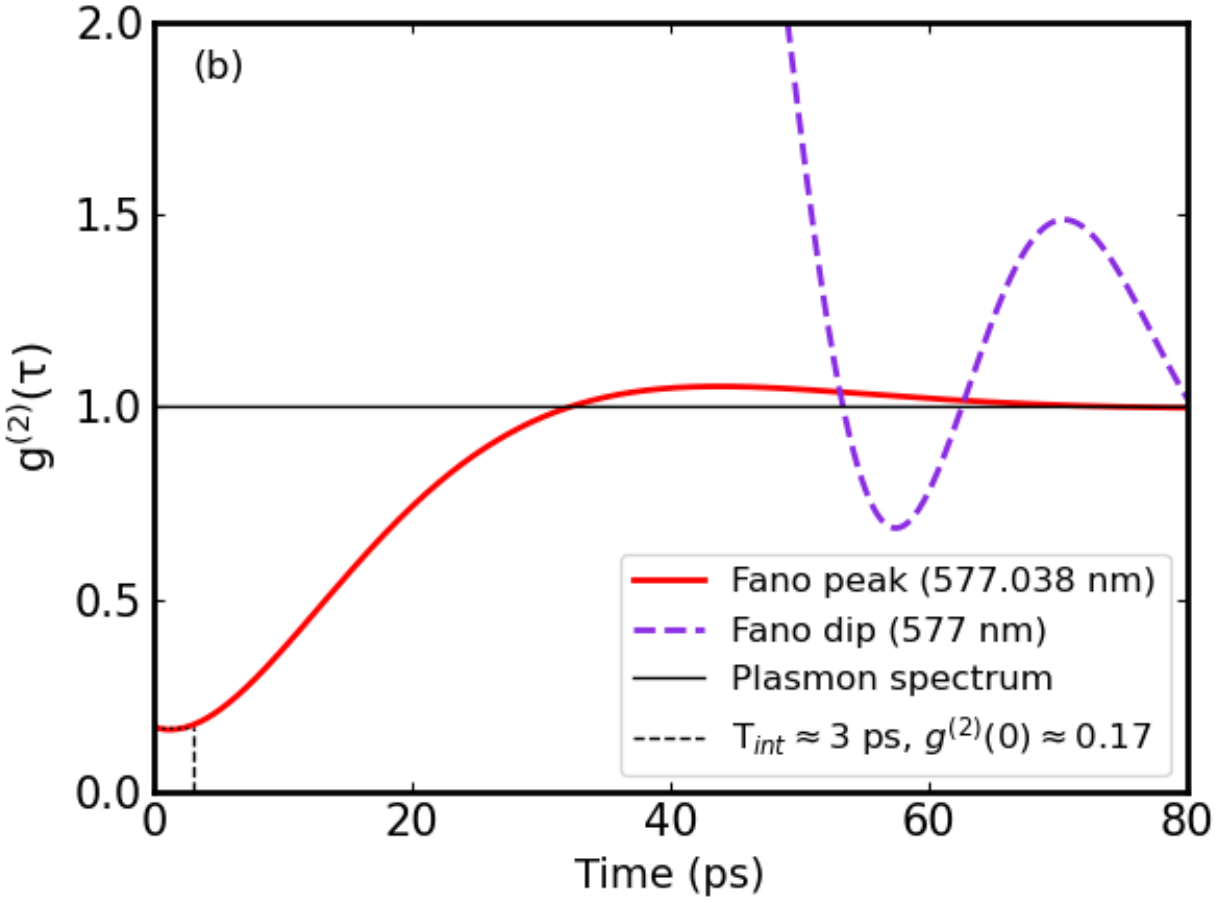}
	\caption{(Color online) (a) Normalized scattered flux of the MNP-QD system, $\langle\hat{b}^{\dagger}\hat{b} \rangle$. Inset shows the Fano profile. 
		The agreement between the analytically-calculated spectrum and the simulated spectrum is shown.
		(b) The second-order correlation function, $g^{(2)}(\tau)$, at different regions of the scattering spectrum, obtained from a QuTiP simulation.
	}\label{f2}
\end{figure}

In the presence of the QD, a transparency region, where the 
scattered flux approaches zero, appears in the scattering 
spectrum at $\sim 577$ nm, as shown in Fig.~\ref{f2}(a). 
This is referred to as the Fano dip~\cite{Dolf10,Wu10} and it is due to destructive interference between the broad plasmon scattering spectrum of the MNP and the narrow QD emission lineshape. The inset of Fig.~\ref{f2}(a) shows the asymmetric Fano spectrum, which begins at the Fano dip and reaches a peak at $\sim 577.038$ nm, referred to as the Fano peak~\cite{Dolf10,Wu10}, where the plasmon spectrum interferes constructively with the QD emission. 
The driving field intensity, $I_{0} \sim 33.6$ W/cm$^{2}$, is chosen in our analysis, as it leads to a scattering spectrum in which the ratio of the scattered flux at the LSPR, $\sim 535.5$ nm, to the scattered flux at the Fano resonance, $\sim 577.038$ nm, tends to unity, as shown in Fig.~\ref{f2}(a). 
This is useful in our plasmonic sensing model since differences in intensity can lead to differences in the relative noise (and thus the signal-to-noise ratio which influences the sensor's resolution) and we would like to draw an unbiased comparison of the intensity-based sensing at driving wavelengths within the plasmon and Fano spectra. The scattered flux of the Fano resonance is large and comparable to the plasmon resonance here mainly due to the size of the dipole moment taken for the QD ($\mu=72$ D) and the driving intensity $I_0$ used, which keeps us in the weak-excitation limit.

The standard deviation in the mean photocount for a single frequency mode is~\cite{Rod00} 
\begin{equation}\label{e15}
\Delta m = \sqrt{\langle \hat{m}(\hat{m}-1) \rangle + \langle \hat{m} \rangle -  \langle \hat{m} \rangle^{2} },
\end{equation}
where
\begin{equation}\label{e16}
\langle \hat{m}(\hat{m}-1) \rangle \approx (\xi \mathrm{T}_{int})^{2}\langle(\hat{b}^{\dagger})^{2}\hat{b}^{2} \rangle
\end{equation}
is the second factorial moment of the photocount operator in the steady state for small ${\rm T}_{int}$ and the two-photon scattering probability density, $\langle(\hat{b}^{\dagger})^{2}\hat{b}^{2} \rangle$, is obtained as 
(see Supplemental Material, Section 5 \cite{Supp})
\begin{equation}\label{e16x}
\langle(\hat{b}^{\dagger})^{2}\hat{b}^{2} \rangle = \gamma_{r}^{2}
\frac{\Omega_{pl}^{4} + 4g^{2}\Omega_{pl}^{2}\langle \hat{\sigma}^{\dagger}\hat{\sigma} \rangle + 4\Omega_{pl}^{3}g\Re\langle \hat{\sigma} \rangle   }{(\Delta_{pl}^{2} + \frac{1}{4}\gamma_{pl}^{2})^{2}  }.
\end{equation}

As mentioned in Sec.~\ref{section1}, in addition to the mean photocount (intensity), we are also interested in utilizing the intensity--intensity correlation as a sensing signal. 
The steady-state, second-order correlation function with a delay time $\tau>0$ is independent of the detector efficiency and is given by~\cite{Rod00,Waks10,Nori13}
\begin{equation}\label{e17}
g^{(2)}(\tau) = \lim_{t\rightarrow \infty} \frac{ \langle \hat{b}^{\dagger}(t) \hat{b}^{\dagger}(t+\tau)\hat{b}(t+\tau) \hat{b}(t) \rangle }{ \langle \hat{b}^{\dagger}(t)\hat{b}(t) \rangle\langle \hat{b}^{\dagger}(t+\tau)\hat{b}(t+\tau) \rangle }.
\end{equation}
Note that Eq.~\eqref{e17} is written in the Heisenberg picture for the operator $\hat{b}$. However, we are using the Schrödinger picture where $\lim_{t\rightarrow \infty}\langle \hat{b}^{\dagger}(t+\tau)\hat{b}(t+\tau) \rangle = \lim_{t\rightarrow \infty}\langle \hat{b}^{\dagger}(t)\hat{b}(t) \rangle = \text{Tr}[\hat{b}^{\dagger}\hat{V}(t,t)\hat{b}\hat{\rho} ]$ and 
$\lim_{t\rightarrow \infty} \langle \hat{b}^{\dagger}(t) \hat{b}^{\dagger}(t+\tau)\hat{b}(t+\tau) \hat{b}(t) \rangle =  
 \text{Tr}[B\hat{V}(t+\tau,t)C\hat{\rho}A ]$, with $A = \hat{b}^{\dagger}, B = \hat{b}^{\dagger}\hat{b}, C = \hat{b}$, and $\hat{V}(t_2,t_1)$ being the evolution operator between time $t_1$ and $t_2$ \cite{Gard04,Nati11,Nori13}.
Then, for zero-delay time, $\tau=0$, we have $\hat{V}=I$ and Eq.~\eqref{e17} reduces to $g^{(2)}(0) = \langle (\hat{b}^{\dagger})^{2}\hat{b}^{2} \rangle/\langle \hat{b}^{\dagger}\hat{b} \rangle^{2}$, which, when simplified using Eqs. 
\eqref{e12}, \eqref{e13}, and \eqref{e16x}, leads to 
\begin{equation}\label{e20}
	g^{(2)}(0) = \frac{\Omega_{pl}^{4} + 4\Omega_{pl}^{3}g\Re\langle \hat{\sigma} \rangle + 4\Omega_{pl}^{2}g^{2}\langle \hat{\sigma}^{\dagger}\hat{\sigma} \rangle    }
	{\Big( \Omega_{pl}^{2} + 2\Omega_{pl}g\Re\langle \hat{\sigma} \rangle + g^{2}\langle \hat{\sigma}^{\dagger}\hat{\sigma} \rangle     \Big)^{2}  }.
\end{equation}

We now have expressions for $\langle \hat{m} \rangle$, $\Delta m$, 
and $g^{(2)}(0)$ using Eqs.~\eqref{e14}, \eqref{e15} and \eqref{e20}. The final parameter needed is the noise in $g^{(2)}(0)$, i.e., $\Delta g^{(2)}(0)$, which we discuss soon. First, we study the behaviour of $g^{(2)}(0)$. The photon statistics of the scattered light in terms of $g^{(2)}(\tau)$ is shown in Fig.~\ref{f2}(b), where we have used QuTiP as we only have an analytical formula for  $g^{(2)}(0)$, as given in Eq.~(\ref{e20}). Studying $g^{(2)}(\tau)$ provides additional information about the temporal behavior of the second-order statistics of the scattered light from the system, although ultimately we are interested in using $g^{(2)}(0)$ only as a sensor signal. Note that due to the time delay, $\tau$, $g^{(2)}(\tau)$ depends on the 
intermediate dynamics of the MNP-QD system, in which changes in $\hat{\rho}$ can occur on a timescale as small as $\gamma_{pl}^{-1}$ ($\sim 5.75$ fs), even though the radiative flux is determined by 
the lifetime $\gamma_{r}^{-1}$ ($\sim 4.29$ ps).

Within the plasmon spectrum, and especially at the LSPR in Fig.~\ref{f2}(a), the scattered light has the same second-order statistics as that of the incident light, i.e., coherent light, and $g^{(2)}(\tau) = 1$ at all times, associated with a random arrival of photons at the detector according to Poissonian statistics. At the Fano peak and dip it is necessary to check the full time dependence of $g^{(2)}(\tau)$ for the purposes of using $g^{(2)}(0)$ as a sensor signal, as we must consider how the numerator and the denominator in 
Eq.~\eqref{e17} are measured in an experiment. Both are measured using photon counting within a finite time period, or integration time, $\mathrm{T}_{int}$, 
which must be fixed at the same value for the correct normalization.  
At the Fano peak, $g^{(2)}(\tau)$ varies on the $5-10$ ps timescale ($\sim \gamma_r^{-1}$), as seen in Fig.~\ref{f2}(b), and as we require a stable signal for sensing, the integration time, $\mathrm{T}_{int}=3$ ps, is chosen such that $\mathrm{T}_{int}<\gamma_r^{-1}$, within which
$g^{(2)}(\tau = 0) = g^{(2)}(\tau = \mathrm{T}_{int}) \approx 0.17$ (see Supplemental Material, Section 6 \cite{Supp}).
From $\tau = 0$ to $\tau = \mathrm{T}_{int}$, the scattered light is heavily bunched at the Fano dip, $g^{(2)}(\tau) > 1$, but approximately constant, and heavily antibunched at the Fano peak, $g^{(2)}(\tau) \sim 0.17 $. 
At longer delay times, $\tau > \mathrm{T}_{int}$, the second-order statistics of the scattered light at the Fano peak ceases to be uniform as it approaches that of the incident light, while the scattered light at the Fano dip oscillates around $g^{(2)}(\tau) = 1$ before second-order coherent scattering finally dominates, in agreement with the behaviour found in Ref.~\cite{Dolf10}.

For studying the noise in the mean photocount (intensity) signal, $\langle \hat{m} \rangle$, it is useful to rewrite Eq.~\eqref{e15} in terms of the detected second-order correlation function for zero-delay time~\cite{Rod00}, $g_D^{(2)}(0)$, as 
\begin{equation}\label{e19}
\Delta m = \sqrt{ \langle \hat{m} \rangle } \sqrt{1 + (g_D^{(2)}(0)-1 )\langle \hat{m} \rangle }.
\end{equation}
Here, $g_D^{(2)}(0)=g^{(2)}(0)$ if ${\rm T}_{int}$ is small enough such that $g^{(2)}(0)=g^{(2)}(\tau={\rm T}_{int})$, which is satisfied for ${\rm T}_{int}=3$ ps, as discussed above. From Eq.~\eqref{e19}, it can be seen that a value of $g^{(2)}(0) < 1$ reduces the value of $\Delta m$ below that of the shot noise, $\Delta m = \sqrt{ \langle \hat{m} \rangle }$, if the photocount, $\langle \hat{m} \rangle$, is not too small. Using Eq.~\eqref{e19}, and ensuring ${\rm T}_{int}$ is the same for all quantities, the photon statistics of the scattered light, as captured by the different values of $g^{(2)}(0)$, can be used to investigate the wavelength dependence of the intensity noise at different regions of the scattering spectrum (from the LSPR to the Fano dip and Fano peak), allowing us to deduce the following: at driving wavelengths within the plasmon spectrum, $\Delta m = \sqrt{ \langle \hat{m} \rangle }$, since $g^{(2)}(0) = 1$, so that the noise is Poissonian and the same as the shot noise in the driving intensity, while within the Fano spectrum, $\Delta m < \sqrt{ \langle \hat{m} \rangle }$, since $g^{(2)}(0) < 1$, characterized by a near-regular photocount pattern on the detector, so that the noise is sub-Poissonian and reduced below the shot noise. At the Fano dip, $\Delta m > \sqrt{ \langle \hat{m} \rangle }$, since $g^{(2)}(0) > 1$, so that the noise is super-Poissonian and thus the refractive index resolution cannot be improved at driving wavelengths within the Fano dip using the intensity as the signal.  

\begin{figure} [t!]
	\centering 
	\includegraphics[width = 0.48\textwidth]{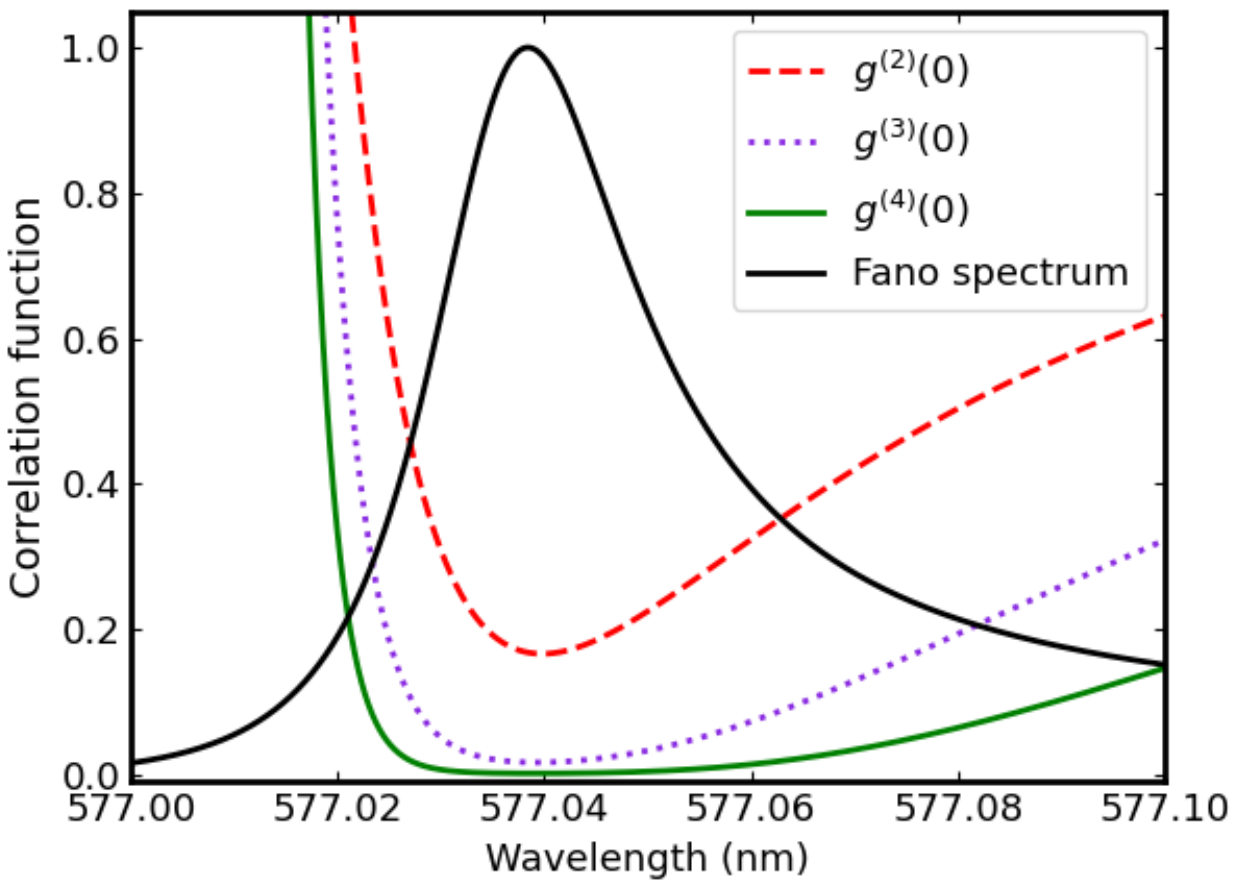}
	\caption{(Color online) The zero-delay time steady-state correlation functions, $g^{(2)}(0)$, $g^{(3)}(0)$, and $g^{(4)}(0)$, at wavelengths within the Fano spectrum. 
	{Here, the Fano spectrum is an enlarged version of the inset in Fig.~\ref{f2}(a) showing only the Fano peak.}
	}\label{f3}
\end{figure}

Before showing the results of the intensity as a signal and its noise, we complete the derivation of the noise in $g^{(2)}(0)$. When it is used as a sensing signal we have that fluctuations in $g^{(2)}(0)$ will affect the resolution of the sensor. Though $g^{(2)}(0)=\langle \hat{m}(\hat{m}-1) \rangle/\langle \hat{m} \rangle^{2}$ is independent of the detector efficiency, as both 
$\langle \hat{m}(\hat{m}-1) \rangle$ and $\langle \hat{m} \rangle^{2}$ 
are proportional to $\xi^{2}$, the standard deviation in 
$g^{(2)}(0)$ is obtained via the propagation of uncertainties in both 
$\langle \hat{m}(\hat{m}-1) \rangle$ and $\langle \hat{m} \rangle^{2}$ using the standard error propagation rule in Ref.~\cite{Forn08}, as (see Supplemental Material, Section 6 \cite{Supp})
\begin{equation}\label{e21}
\Delta g^{(2)}(0) = \left(2g^{(2)}(0)\frac{\Delta m }{ \langle \hat{m} \rangle }\right)\sqrt{1 + \left(\frac{\Delta m_{2} }{2g^{(2)}(0) \langle \hat{m} \rangle\Delta m }\right)^{2} },
\end{equation}
where 
\begin{eqnarray}
\Delta m_{2} &=& \langle \hat{m} \rangle^{2}
\bigg[g^{(4)}(0) - [g^{(2)}(0)]^{2}   \nonumber \\
& & + 4g^{(3)}(0)\frac{\xi}{ \langle \hat{m} \rangle}  + 2g^{(2)}(0)\left(\frac{\xi}{ \langle \hat{m} \rangle} \right)^{2} \bigg]^{1/2} \label{e22}
\end{eqnarray}
is the standard deviation in the second factorial moment of the photocount, $\langle \hat{m}(\hat{m}-1) \rangle$, and
$g^{(3)}(0)$ and $g^{(4)}(0)$ are the third- and fourth-order steady-state correlation functions at zero-delay time, obtained respectively as (see Supplemental Material, 
Section 5 \cite{Supp})
\begin{subequations}
\begin{align}
g^{(3)}(0) &  = \frac{\Omega_{pl}^{6} +  6g\Omega_{pl}^{5}\Re\langle \hat{\sigma} \rangle + 9g^{2}\Omega_{pl}^{4}\langle \hat{\sigma}^{\dagger}\hat{\sigma} \rangle   }
{\Big( \Omega_{pl}^{2} + 2\Omega_{pl}g\Re\langle \hat{\sigma} \rangle + g^{2}\langle \hat{\sigma}^{\dagger}\hat{\sigma} \rangle     \Big)^{3}  } ~~~~ \text{and} \label{e24a}\\
g^{(4)}(0) &  = \frac{\Omega_{pl}^{8} + 8g\Omega_{pl}^{7}\Re\langle \hat{\sigma} \rangle + 16g^{2}\Omega_{pl}^{6}\langle \hat{\sigma}^{\dagger}\hat{\sigma} \rangle    }
{\Big( \Omega_{pl}^{2} + 2\Omega_{pl}g\Re\langle \hat{\sigma} \rangle + g^{2}\langle \hat{\sigma}^{\dagger}\hat{\sigma} \rangle    \Big)^{4}  }.
\label{e24b}
\end{align}
\end{subequations}
The second-, third-, and fourth-order steady-state correlation functions within the Fano spectrum at zero-delay time are shown in Fig.~\ref{f3}. Both $g^{(3)}(0)$ and $g^{(4)}(0)$ show that the scattered light is antibunched at the Fano resonance and at driving wavelengths near the Fano resonance, i.e., $g^{(n)}(0) < 1, n = 3, 4$,  in accordance with $g^{(2)}(0)$, where $g^{(4)}(0) < g^{(3)}(0) < g^{(2)}(0) <1$, in agreement with previous works~\cite{Miri14}. {Note that the narrow wavelength range in Fig.~\ref{f3} is achievable with the current state of comb laser technology \cite{Buyalo24}.}

Both $\Delta m$ and $\Delta g^{(2)}(0)$ depend on the detector efficiency, $\xi$, and in what follows, 
we set $\xi = 70\%$, a value that is achievable by contemporary single photon detection and collection optics~\cite{You20,Zhang20}.
The standard deviations in Eqs. \eqref{e19} and \eqref{e21}, i.e., the signal noises, obtained within the integration time, $\mathrm{T}_{int}$, correspond to a single shot measurement and can be large compared to the mean value of their respective signal. The noise can be reduced by averaging a time series of single measurements, 
referred to as temporal averaging~\cite{Mola19,Jiri08}. 
The time-averaging of $\mathcal{N}$ measurements reduces the noises as
\begin{subequations}
\begin{align}
\sigma_{m} & = \frac{\Delta m }{\sqrt{ \mathcal{N} } },~~~\text{and} \label{e27xa}\\ 
\sigma_{g^{(2)}(0) } & = \frac{\Delta g^{(2)}(0) }{\sqrt{ \mathcal{N} }\label{e27xb} },
\end{align}
\end{subequations}
respectively, where $\mathcal{N} = 1/\mathrm{T}_{int}$ is the number of measurements possible in one second, and
for simplicity the role of the detector recovery time in these measurements has been ignored. 
In an experiment, $\mathrm{T}_{int}$ would be the collection time for 
$\langle \hat{m} \rangle$ using one single-photon detector and the coincidence window for $g^{(2)}(0)$ using a pair of single-photon detectors. 
For example, $\langle \hat{m} \rangle \approx 12.42\times10^{-5}$ in $\mathrm{T}_{int}$ at the LSPR (equivalent to $4.14\times10^{7}$ photon counts in one second of measurements), as we show in Sec.~\ref{section 5}.
This corresponds to $\sigma_{m} \approx \langle \hat{m} \rangle^{1/2}(1/{\rm T}_{int})^{-1/2} = (12.42\times10^{-5})^{1/2}(3\times10^{-12})^{1/2} \sim 1.93\times 10^{-8}$ in ${\rm T}_{int}$. Thus, the use of repeated measurements reduces the noise $\sigma_{m}$ to below that of the signal $\langle \hat{m} \rangle$.

We note that setting $\mathrm{T}_{int} \sim 3$ ps is challenging but achievable with state-of-the-art detector technology where $\sim $ps accuracy is possible~\cite{You20,Zhang20}. Such a short measurement time is necessary for improving both $\langle \hat{m} \rangle$- and $g^{(2)}(0)$-based refractive index sensing resolutions, since both $\Delta m$ and 
$\Delta m_{2}$ depend on $g^{(2)}(\tau)$, which is neither uniform nor close to zero beyond $\tau = \mathrm{T}_{int}$ (as the probability of two-photon scattering increases), as shown in Fig.~\ref{f2}(b). The time-delayed higher-order steady-state correlations, $g^{(3)}(\tau)$ and 
$g^{(4)}(\tau)$, which $\Delta m_{2}$ also depends on, 
vary more slowly than $g^{(2)}(\tau)$ beyond $\tau = \mathrm{T}_{int}$, as shown in Fig.~S2 of the Supplemental Material \cite{Supp}. We therefore use the values of $g^{(3)}(0)$ and $g^{(4)}(0)$ over the period $ \mathrm{T}_{int}$ in the calculation of $\Delta g^{(2)}(0)$. 
Finally, we note that $\mathrm{T}_{int}$ can, in principle, be increased arbitrarily for $\langle \hat{m} \rangle$; however, the value of $\Delta m$ is determined by $g^{(2)}(0)$ within $\mathrm{T}_{int}$ according to Eq.~(\ref{e19}) and $g^{(2)}(0) \rightarrow 1$ as $\mathrm{T}_{int}$ increases. Thus by increasing $\mathrm{T}_{int}$ any potential reduction in noise due to antibunching is lost. This is the main reason we set $\mathrm{T}_{int}$ to be $3$ ps for both $\langle \hat{m} \rangle$ and $g^{(2)}(0)$, and use $\mathcal{N}$ to obtain the signals from these variables and their noise over a total measurement period longer than a single measurement time.

\section{Sensing performance} \label{section 5}
We now use the method of intensity interrogation~\cite{Mola19,Lee21,Jiri08} to determine the ratio of the change in the scattered intensity, $I$ (since $I \propto \langle \hat{m} \rangle$), or the change in the second-order correlation function for zero-delay time, $g^{(2)}(0)$, to the change in the refractive index, $n$, of the local environment.
\begin{figure*} [t!] 
	\centering 
	\includegraphics[width = 0.458\textwidth]{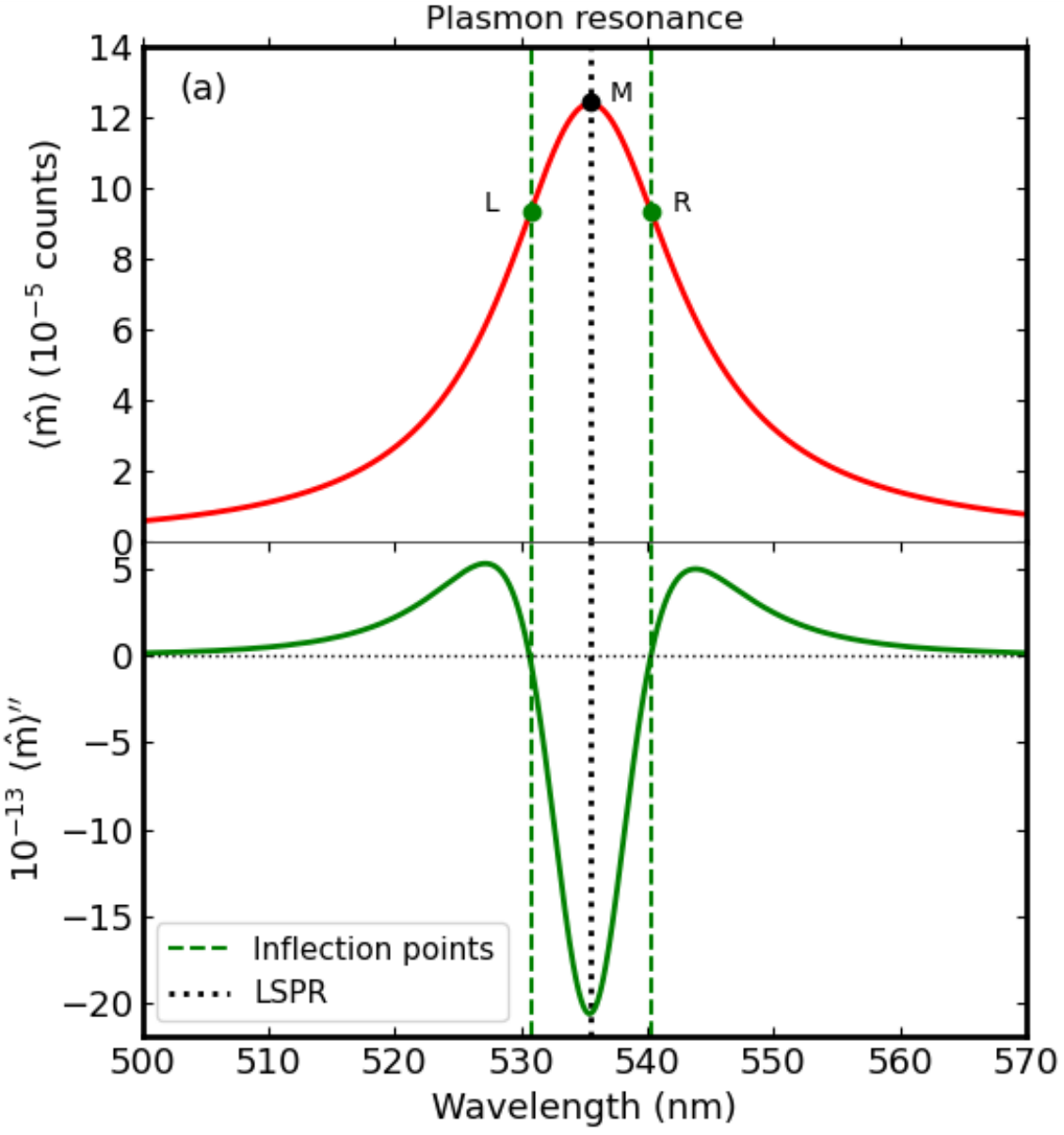}
	\includegraphics[width = 0.465\textwidth]{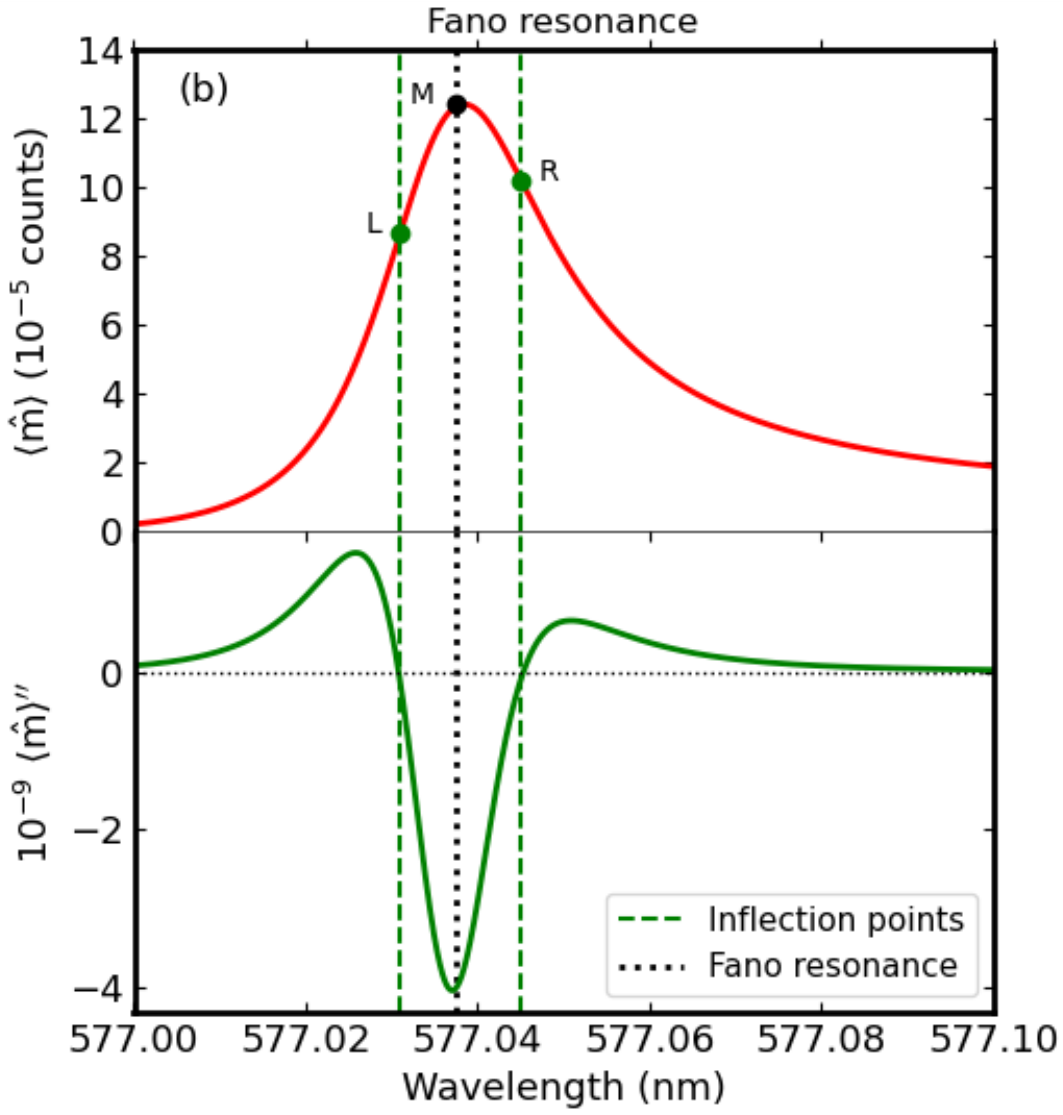}
	\caption{(Color online)  Optimal driving wavelengths for intensity sensing at $n = 1.3330$ within the (a) plasmon spectrum and (b) Fano spectrum {showing the Fano peak only}. Upper panels: photocounts, $\langle \hat{m} \rangle$, obtained in $\mathrm{T}_{int}$. Lower panels: second derivative (with respect to $\lambda$) of the photocounts, $\langle \hat{m} \rangle^{\prime\prime}$. 
	}\label{f4}
\end{figure*}
With $I$ replaced by the mean photocount, $\langle \hat{m} \rangle$, 
we follow the approach in Ref.~\cite{Lee21} and define the intensity sensitivity and the intensity--intensity sensitivity, respectively, as 
\begin{subequations}
\begin{align}
\mathrm{S}_{I} &  = \left|\left(  \dfrac{ d\langle \hat{m} \rangle }{ dn }\Big|_{\lambda_{D}}  \right)\right| ~~~\text{and}  \label{e25a} \\
\mathrm{S}_{I-I} &  = \left|\left(  \dfrac{ dg^{(2)}(0) }{ dn }\Big|_{\lambda_{D}}  \right)\right|, \label{e25b}
\end{align}
\end{subequations}
each of which is in units of RIU$^{-1}$, where RIU stands for refractive index unit. In Eq.~\eqref{e25a}, the driving wavelength (equivalent to the scattered wavelength), $\lambda_{D}$, is a wavelength either within the plasmon spectrum or the {Fano spectrum in Fig.~\ref{f3}}, while the driving wavelength in Eq.~\eqref{e25b} is a wavelength within the {Fano spectrum in Fig.~\ref{f3}}, 
since the $g^{(2)}(0)$ values within the plasmon spectrum are independent of the refractive index of the medium, i.e., $g^{(2)}(0) = 1$ across the plasmon spectrum, as shown in Fig.~\ref{f2}(b).  
It was reported in Refs.~\cite{Jeon19,Jiri08,Mola19} that driving wavelengths corresponding to the inflection points on the plasmon spectrum are the optimal wavelengths for plasmonic sensing. Mathematically, they are points where the second derivative of the intensity with respect to the wavelength is zero. In other words, they correspond to the steepest points on the intensity curve. At these fixed wavelength points, as the plasmon peak shifts left or right with the refractive index, the intensity changes maximally. We therefore define the enhancement in the refractive index sensitivity as
\begin{equation} \label{e26}
	\mathcal{E}_{\mathrm{Si}} = \frac{ \mathrm{S}_{I}( \lambda_{D = \mathrm{Fi}}  )   }{\mathrm{S}_{I} ( \lambda_{D = \mathrm{Pi}} )   }, 
\end{equation}
where F (P) stands for Fano resonance (plasmon resonance), and i = L, M, and R, corresponding to the inflection point on the left-hand side, the middle point, and the right-hand side on the Fano resonance (plasmon resonance), respectively. 
The LSPR (the middle point, PM), and the plasmon inflection points (PL and PR) are shown in Fig.~\ref{f4}(a), while the middle Fano point (FM) and the Fano inflection points (FL and FR) are shown in Fig.~\ref{f4}(b). 
Since $g^{(2)}(0)$ does not vary across the plasmon spectrum, we cannot define a similar enhancement factor for 
$\mathrm{S}_{I-I}$. 

In either the plasmon or Fano spectrum, the inflection points correspond to wavelengths where the second derivative (with respect to $\lambda$) of the mean photocount is zero, while the resonance corresponds to the wavelength where the second derivative of the mean photocount reaches a minimum, as indicated 
in the lower panels of Fig.~\ref{f4}. Due to the symmetric (plasmon spectrum, Fig.~\ref{f4}(a)) versus asymmetric (Fano spectrum, Fig.~\ref{f4}(b)) lineshapes of the scattering spectrum, the mean photocounts vary at the inflection points.
However, the driving intensity has been tuned so that
the mean photocount obtained at the LSPR is approximately the same as the value obtained at the middle point of the Fano resonance, i.e., $\langle \hat{m} \rangle \sim 12.42\times 10^{-5}$ counts in $\mathrm{T}_{int}$, in both Figs. \ref{f4}(a) and \ref{f4}(b), which allows us to fairly compare the sensing performance at the two regions. 
On the other hand, the system can be driven such that the plasmon peak is enhanced relative to the Fano peak (increase in $I_{0}$), or such that the plasmon peak is suppressed relative to the Fano peak 
(decrease in $I_{0}$). Either of these two approaches will lead to an unfair comparison of the sensing performance at the plasmon and Fano resonances since any definition of an enhancement factor will be flawed. More information regarding the dependence of the Fano resonance on the driving field intensity can be found in Refs.~\cite{Dolf10,Sha13}. In the top part of Table~\ref{t2}, the wavelengths, $\lambda_D$, corresponding to the inflection points have been listed, along with the sensitivities, $S_I$, and enhancement factors, $\mathcal{E}$, at these points.

\begin{figure} [t!]
	\centering 
	\includegraphics[width = 0.46\textwidth]{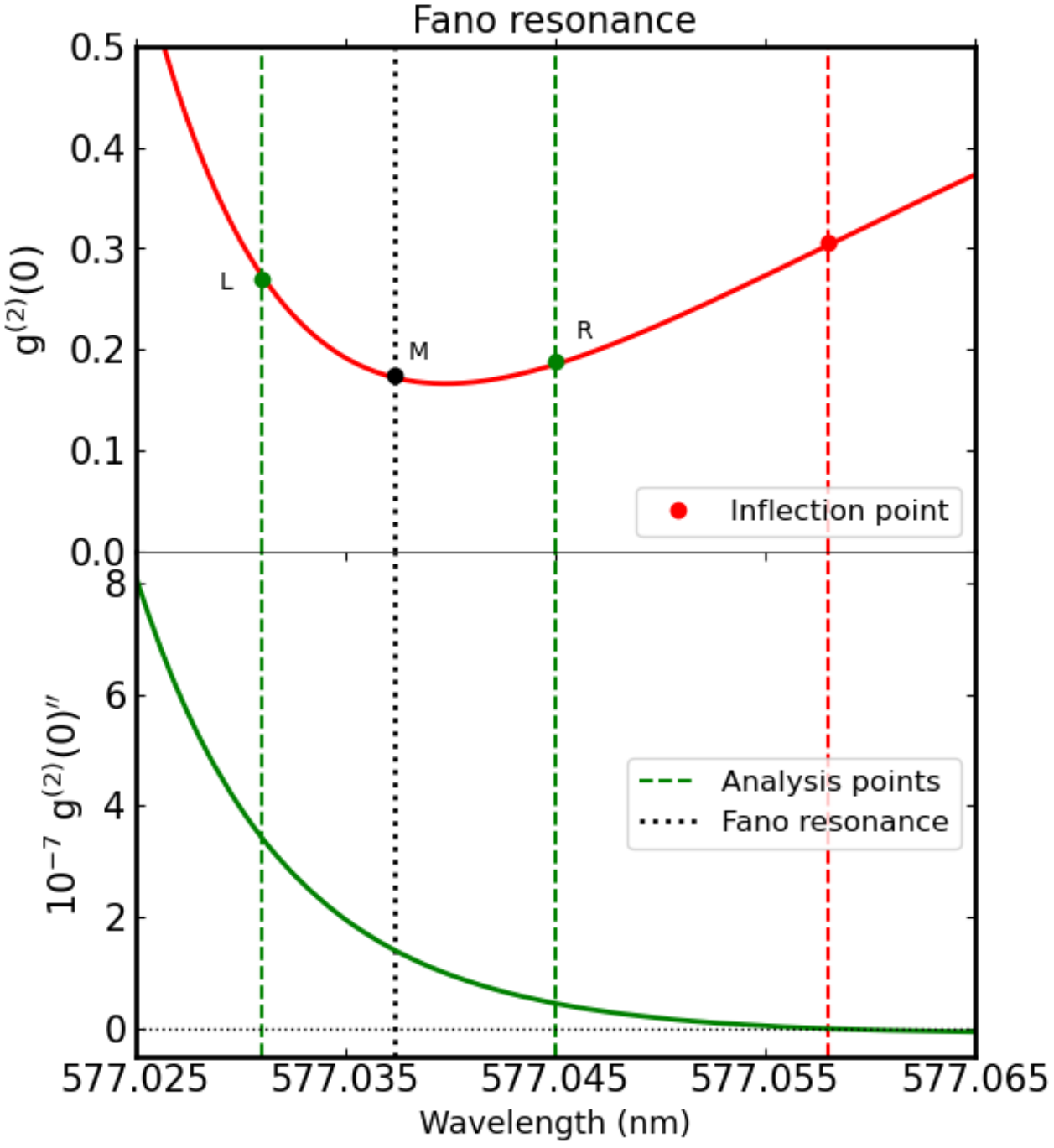}
	\caption{(Color online)  Driving wavelengths for $g^{(2)}(0)$-based refractive index sensing within the Fano spectrum ({wavelengths around the Fano peak only}) at $n = 1.3330$.
		Upper panel: $g^{(2)}(0)$. Lower panel: second derivative of $g^{(2)}(0)$ with respect to $\lambda$, $g^{(2)}(0)^{\prime\prime}$.
	}\label{f5}
\end{figure}

In Fig.~\ref{f5}, the inflection point on the $g^{(2)}(0)$ curve is similarly obtained by finding the wavelength corresponding to the zero-value of the second derivative of $g^{(2)}(0)$, 
where we have restricted the driving wavelengths to those where the scattered light shows antibunching photon statistics only. 
Although the huge bunching effect near the Fano dip (see Fig.~\ref{f2}(b)) will lead to high values of $\mathrm{S}_{I-I}$, plasmonic sensing within the Fano dip is not
 realistic since $\langle \hat{m} \rangle \rightarrow 0$ in this region, and as we mentioned earlier, the noise also increases at the Fano dip. 
Compared to Fig.~\ref{f4}, there is only one inflection point on the 
second-order correlation function at the right-hand side of FR, 
\begin{table}[t!]
	\scalebox{0.85}{
		\begin{tabular}{ | c |c |c |c |c |l | }
			\cline{1-6}
			\multicolumn{1}{| c }{ \multirow{2}{*}{  } } &
			\multicolumn{1}{ c }{$\lambda_{D}$ (nm) } &  & FL $ = 577.031$ & FM $ = 577.038$ & FR $ = 577.045$ \\ \cline{4-6}
			\multicolumn{1}{| c }{} &
			\multicolumn{1}{ c }{} &  & PL $ = 530.770$ & PM $ = 535.500$  & PR $ = 540.270$   \\ \cline{1-6}
			\multicolumn{1}{| c }{ \multirow{2}{*}{  } } &
			\multicolumn{1}{ c }{ $\mathrm{S}_{I} \times10^{-4}$  } &  & @FL $: 4.04 $ & @FM $: 4.15 $ & @FR $: 12.17$ \\ \cline{4-6}
			\multicolumn{1}{| c }{} &
			\multicolumn{1}{ c }{[RIU $^{-1}$] } &  & @PL $: 5.87 $ & @PM $: 3.93$  & @PR $: 11.37 $   \\ \cline{1-6}
			\multicolumn{1}{| c }{ \multirow{2}{*}{  } } &
			\multicolumn{1}{ c }{ $\mathcal{E}_{S}$ } &  & $ \sim 0.69 $ & $ \sim 1.06 $ & $ ~~~\sim 1.07 $ \\ \cline{1-6}	
			\multicolumn{1}{| c }{ \multirow{2}{*}{  } } &
			\multicolumn{1}{ c }{$\mathrm{S}_{I-I}$ [RIU $^{-1}$] } &  & @FL $: 3.85 $ & @FM $: 1.05 $ & @FR $: 0.50$ \\ \cline{1-6}
			\multicolumn{1}{| c }{ \multirow{2}{*}{  } } &
			\multicolumn{1}{ c }{ $\Delta n_{I}\times10^{-5}$  } &  & @FL $: 4.60$ & @FM $: 5.50$ & @FR $:  2.10$ \\ \cline{4-6}
			\multicolumn{1}{| c }{} &
			\multicolumn{1}{ c }{ [RIU] } &  & @PL $: 3.90$ & @PM $: 6.90$  & @PR $:2.10 $   \\ \cline{1-6}
			\multicolumn{1}{| c }{ \multirow{2}{*}{  } } &
			\multicolumn{1}{ c }{ $\mathcal{E}_{\Delta n}$ } &  & $ \sim 0.84 $ & $ \sim 1.25 $ & $ ~~~\sim 1.00 $ \\ \cline{1-6}		
			\multicolumn{1}{| c }{ \multirow{2}{*}{  } } &
			\multicolumn{1}{ c }{$\Delta n_{I-I} \times10^{-3}$ [RIU] } &  & @FL $: 1.3 $ & @FM $: 2.8 $ & @FR $: 7.6 $ \\ \cline{1-6}			
	\end{tabular} }
	\caption{Refractive index sensitivities, resolution, and enhancement factors obtained at the optimal driving wavelengths. 
	}\label{t2}
\end{table}
as indicated in Fig.~\ref{f5}.
\begin{figure*} [t!]
	\centering 
	\includegraphics[width = 0.45\textwidth]{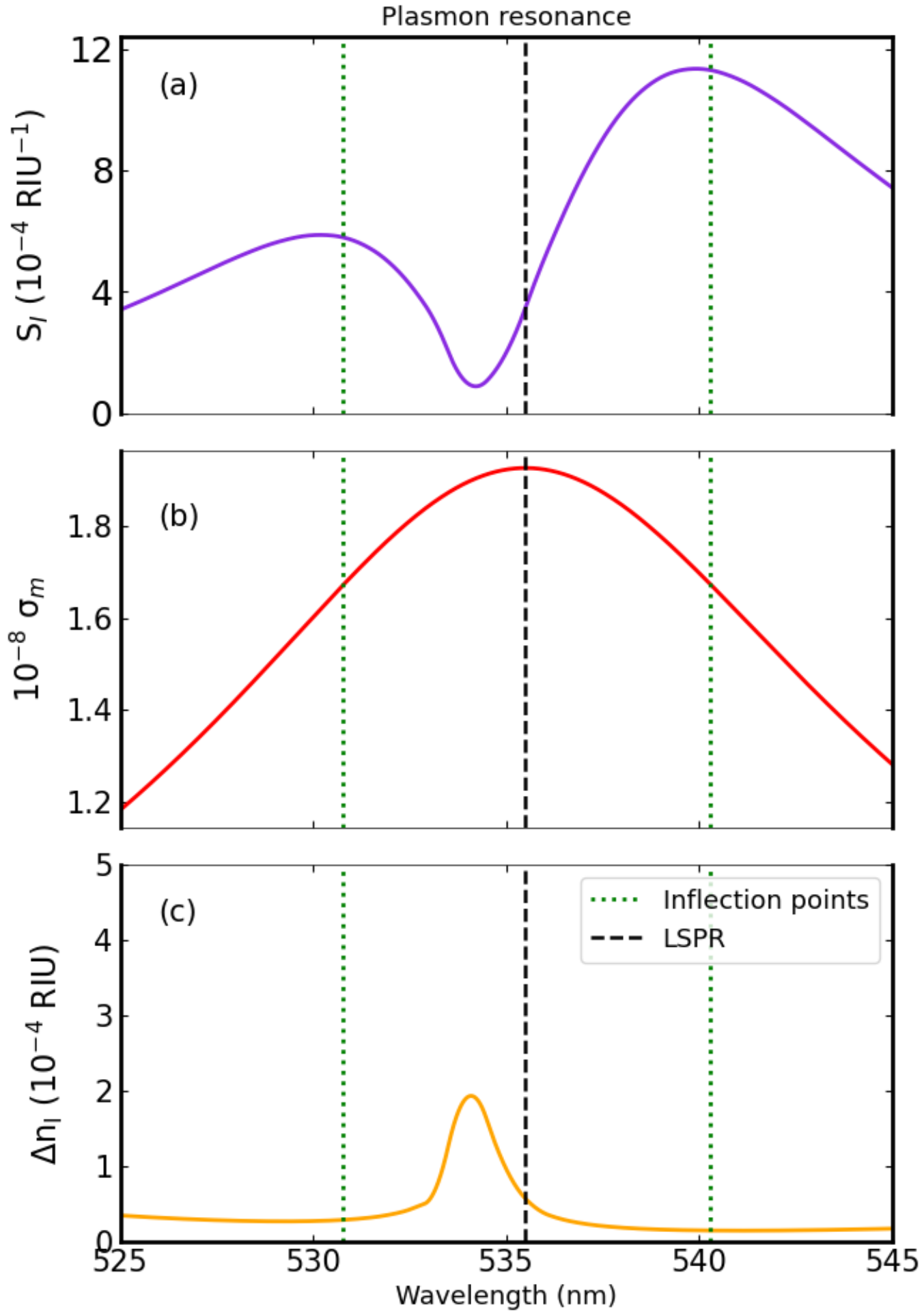}~~
	\includegraphics[width = 0.44\textwidth]{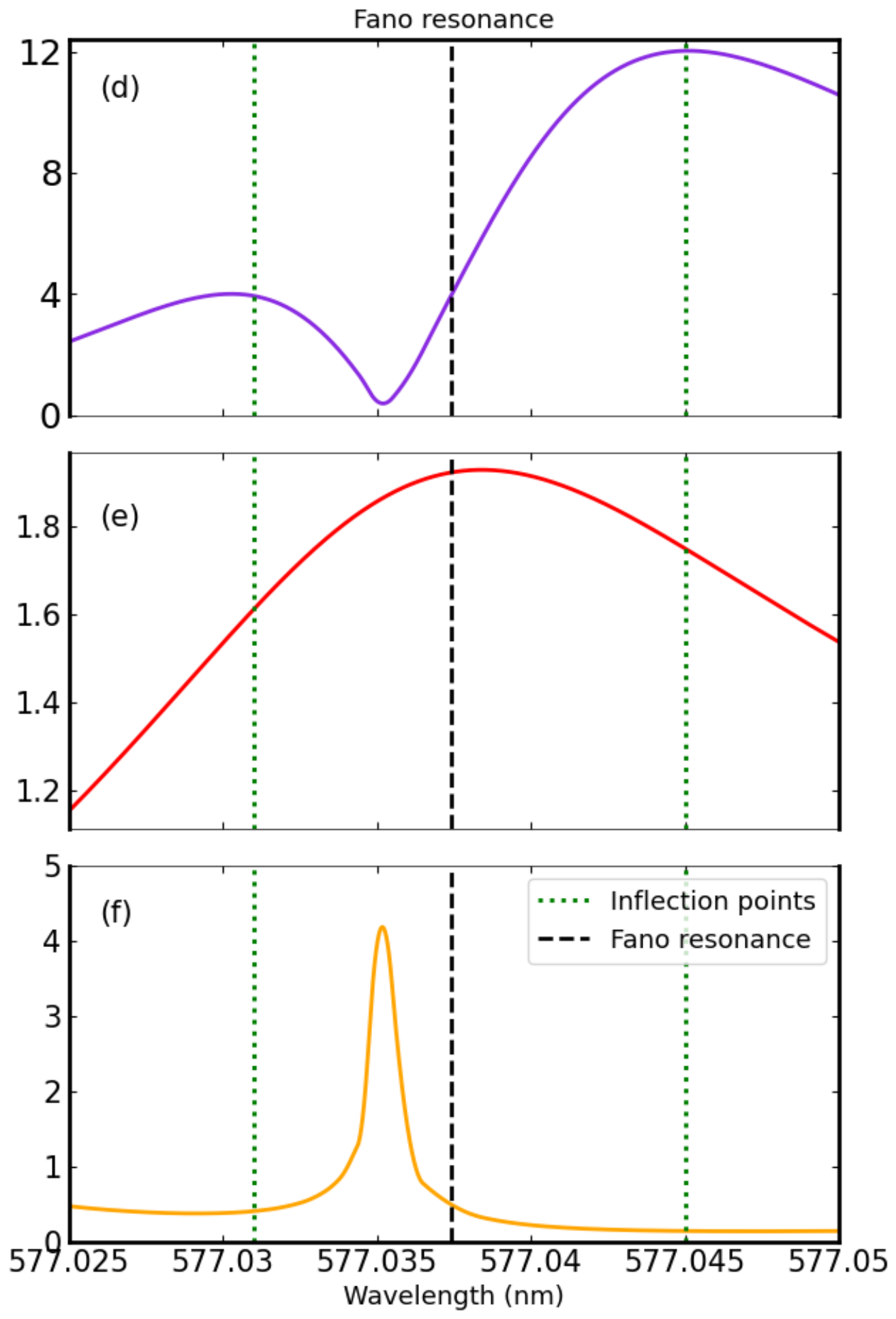}
	\caption{(Color online)  
		Intensity sensitivity, standard deviation (noise) for one second of measurements, and sensing resolution (i.e., limit of detection) at driving wavelengths within the plasmon spectrum [(a), (b), (c)], and Fano spectrum ({wavelengths around the Fano peak only}) [(d), (e), (f)], obtained within the refractive index range, $n = 1.3330 - 1.3334$. The sensing resolution can be improved by taking measurements for more than one second, with the results for the plasmon and Fano resonances scaling by the same factor.
	}\label{f6}
\end{figure*}
Unlike the intensity sensitivity, the inflection point on the $g^{(2)}(0)$ curve does not give the highest sensitivity, $S_{I-I}$, due to a lower gradient than on the left-hand side of FM, and is therefore not the optimal driving wavelength for refractive index sensing with a $g^{(2)}(0)$ signal. The sensitivities at the different points FL, FM, and FR are given in Table~\ref{t2}.

Following the approach in Ref.~\cite{Mola19}, the resolution (or limit of detection) of the sensor for a repeated series of measurements lasting one second is calculated using the time-averaged standard deviations and sensitivities, as 
\begin{subequations}
	\begin{align}
		\Delta n_{I} &  = \frac{\sigma_{m} }{ \mathrm{S}_{I} }~~~~\text{and}  \label{e27a} \\
		\Delta n_{I-I} &  = \frac{\sigma_{g^{(2)}(0)} }{ \mathrm{S}_{I-I} }. \label{e27b}
	\end{align}
\end{subequations}
These resolutions are obtained by using a simple linear error propagation formula~\cite{Mola19}. However, a more rigorous approach is to calculate the Fisher information and obtain the Cram\'er-Rao bound, which is a lower bound on the estimation error, or resolution using our terminology. The Fisher information for an unknown parameter $\theta$ measured using observable $X$ with a Gaussian probability distribution is $I(\theta)=\Delta^{-2}$, where $\Delta$ is the standard deviation of the probability distribution for $X$~\cite{FisherTut,Paris09}. For ${\cal N}$ independent and identically distributed measurements we then have $I_{\cal N}(\theta)={\cal N}I(\theta)$. The Cram\'er-Rao bound is then $[I_{\cal N}(\theta)]^{-1/2}=\Delta/\sqrt{\cal N}$. Taking the sensitivities $S_I$ and $S_{I-I}$ to be linear for small changes in the refractive index (see Supplemental Material, Section 5 \cite{Supp}) and the probability distribution for $\langle \hat{m} \rangle$ and $g^{(2)}(0)$ to be Gaussian with standard deviations $\Delta m$ and $\Delta g^{(2)}(0)$, we have that the Cram\'er-Rao bounds for the resulting refractive index measurements are equal to Eqs.~\eqref{e27a} and \eqref{e27b}, respectively. The sensitivities $S_I$ and $S_{I-I}$ therefore allow the propagation of the bounds on the resolution of the photocount and second-order correlation to bounds on the resolution of the refractive index.

The enhancement in the refractive index resolution is defined as
\begin{equation} \label{e26b}
	\mathcal{E}_{\mathrm{\Delta ni}} = \frac{\Delta n_{I}( \lambda_{D = \mathrm{Pi}} )   }{ \Delta n_{I}( \lambda_{D = \mathrm{Fi}}  )   },
\end{equation}
where, as before, we are unable to define a similar enhancement for $\Delta n_{I-I}$ due to a constant $g^{(2)}(0)$ within the plasmon spectrum.

In Fig.~\ref{f6}, the sensing performance of the scattered intensity at driving wavelengths within the plasmon spectrum (Figs. \ref{f6}(a), (b), and (c)) is compared to the performance at wavelengths within the Fano spectrum (Figs. \ref{f6}(d), (e), and (f)). 
The sensing performance was investigated using the refractive index range $n = 1.3330 - 1.3334$ (corresponding to a typical range in a biosensing context~\cite{Jiri08}), where the mean photocount has a linear dependence on $n$, for each $\lambda_{D}$, as shown in Fig.~S1 of the Supplemental Material \cite{Supp}.

The intensity sensitivities, Fig.~\ref{f6}(a) and Fig.~\ref{f6}(d), have a similar dependence on the driving wavelengths --- on the left-hand side of the plasmon resonance (Fano resonance), PL (FL) is the optimal driving wavelength, while PR (FR) is the optimal driving wavelength on the right-hand side of the plasmon resonance (Fano resonance). The dips show that intensity sensing should not be carried out around the wavelength corresponding to the plasmon peak (and Fano peak), where the intensity has a smaller variation with the refractive index, {\it i.e.}~the derivative of the intensity with respect to the refractive index is small at a fixed wavelength near the initial peak position. This is a common feature of intensity sensing at a resonance peak in an intensity spectrum --- the intensity at a fixed wavelength near the initial peak does not vary significantly with refractive index, while the intensity on either side of the fixed wavelength varies more significantly (as the peak red or blue shifts).

In each spectrum, the sensitivity is on the order of $\sim 10^{-4}$ RIU$^{-1}$ and reaches a maximum value at PR (FR). As summarized in Table~\ref{t2}, there is no significant enhancement in the sensitivity since $\mathcal{E}_{S} \approx 1$ at the optimal driving wavelengths. Hence, the scattered intensity at the Fano spectrum is not necessarily more sensitive to changes in the refractive index than the scattered 
intensity at the plasmon spectrum, given that the scattering peaks at the two regions lead to mean photocounts with the same order of magnitude.
\begin{figure} [t!]
	\centering 
	\includegraphics[width = 0.45\textwidth]{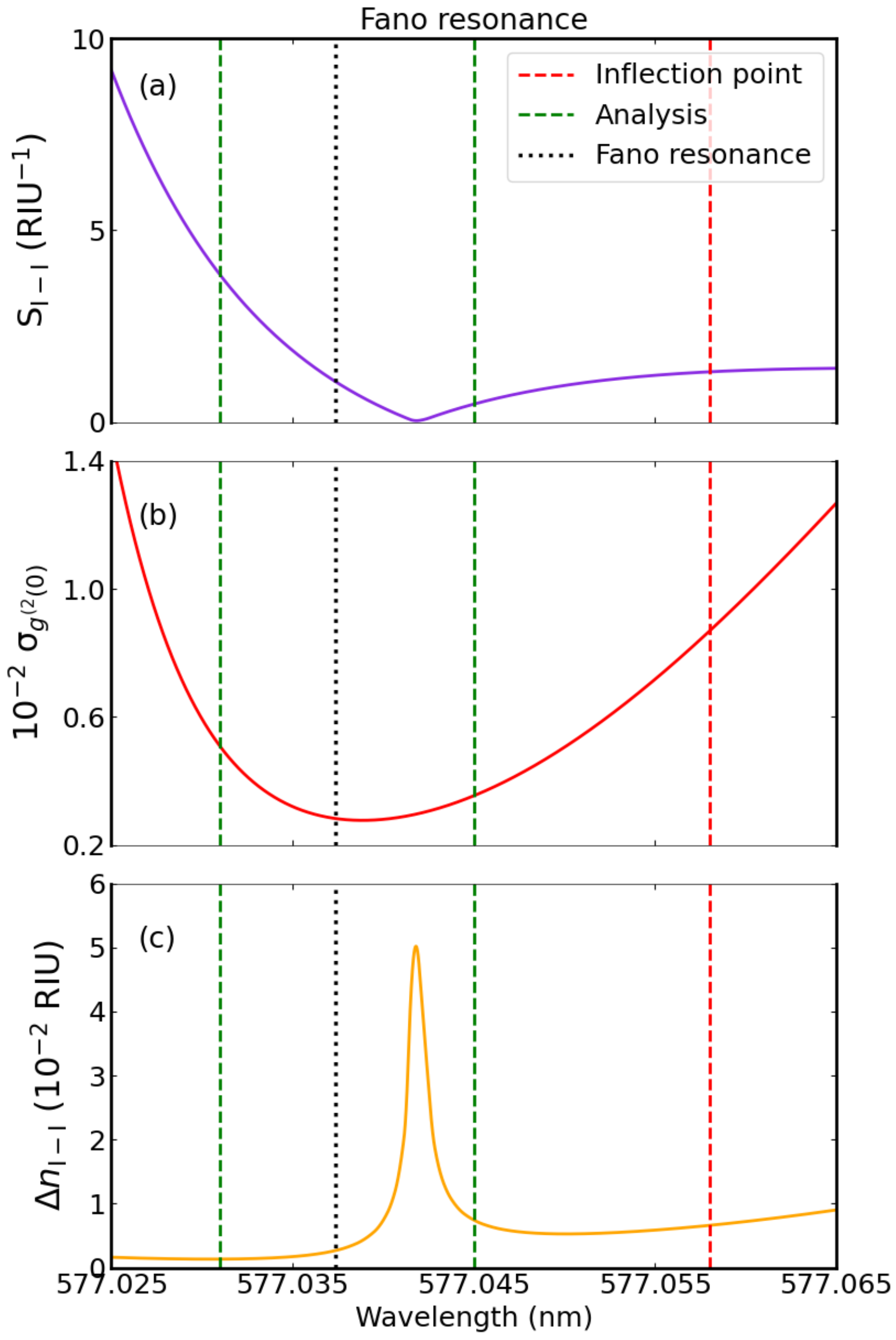}
	\caption{(Color online) 
		(a) Intensity--intensity refractive index sensitivity, (b) standard deviation (noise) in the $g^{(2)}(0)$ signal for one second of measurements, and (c) sensing resolution (i.e., limit of detection) at driving wavelengths within the Fano spectrum ({wavelengths around the Fano peak only}), obtained within the refractive index range, $n = 1.3330 - 1.3334$. The sensing resolution can be improved by taking measurements for more than one second.
	}\label{f7}
\end{figure}
\begin{figure} [t!]
	\centering 
	\includegraphics[width = 0.45\textwidth]{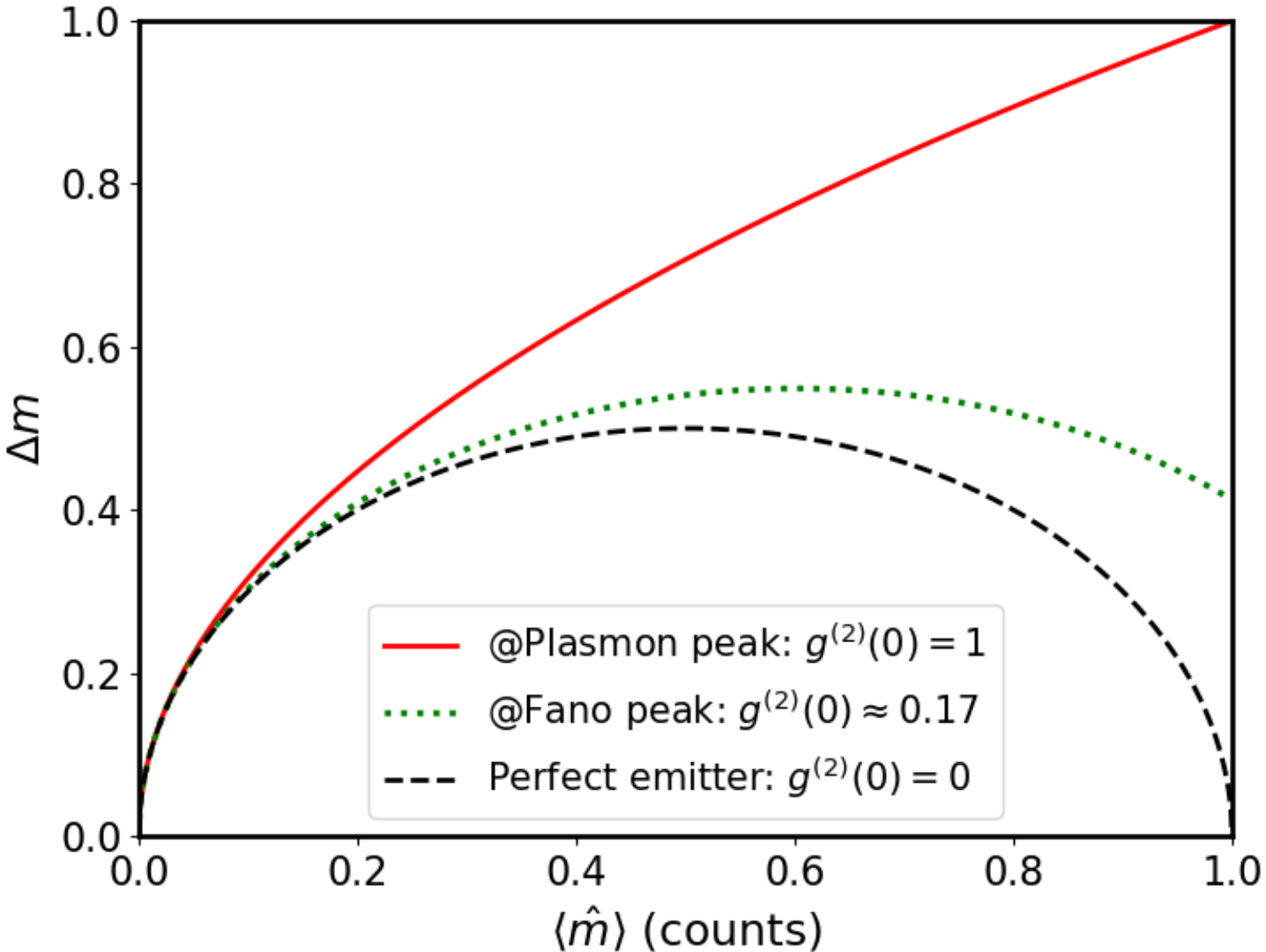}
	\caption{Effects of the statistical property of the scattered light on the noise level assuming that photocounts in the range $0.0 - 1.0$ and $g^{(2)}(0) \approx 0.17$ ($g^{(2)}(0) = 1$) at the Fano resonance (plasmon resonance) are obtainable within the integration time, $\mathrm{T}_{int}$.
	}\label{f8}
\end{figure}

The dependence of the noise on driving wavelengths within the plasmon and Fano spectrum are shown in Fig.~\ref{f6}(b) and Fig.~\ref{f6}(e), respectively. In each spectrum, the standard deviation in the mean photocount, $\sigma_m$, is on the order of $\sim 10^{-8}$, and reaches a maximum value at PM (FM). Hence, in order to minimize the noise,
the wavelengths, PM (FM), should not be used as driving wavelengths. 
As shown in Fig.~\ref{f6}(e), the scattered light within the Fano spectrum has the same first-order photon statistics as the scattered light within the plasmon spectrum, i.e., coherent light. This is because the 
low mean photocount within $\mathrm{T}_{int}$, $\langle \hat{m} \rangle \sim 10^{-4}$, renders the photon antibunching effect on $\Delta m$, and thus $\sigma_{m}$, insignificant within the Fano spectrum, i.e., $(g^{(2)}(0)-1 )\langle \hat{m} \rangle \approx 0$ in Eq.~\eqref{e19} and Eq.~\eqref{e27xa}, so that the noise within the two spectra remains comparable.  As shown in Fig.~\ref{f6}(c) and Fig.~\ref{f6}(f), the sensing resolution is not optimal near PM (FM) due to the combined effect of the low sensitivity and high noise.
However, the resolutions are on the same order of magnitude,
$\sim 10^{-5}$ RIU at the inflection points {(which is within the range of those of start-of-the-art plasmonic sensors \cite{Lee21})}, resulting in no significant enhancement, since $\mathcal{E}_{\Delta n} \approx 1$, as shown in Table~\ref{t2}. Finally, we note that the sensing resolutions shown in Fig.~\ref{f6}(c) and \ref{f6}(f) correspond to one second of measurements, and that both can be improved in principle by simply performing temporal averaging over a larger set of measurements. This, however, may lead to an increase in other sources of noise in the signal, such as technical noise in the setup (e.g. vibrations, equipment response etc.), which would impact the sensing resolution.

We now focus on using $g^{(2)}(0)$ as a signal for refractive index sensing, with the results shown in Fig.~\ref{f7}. The observed wavelength dependence of the sensing parameters is quite different from that of the intensity-based sensing. As shown in Fig.~\ref{f7}(a) and Fig.~\ref{f7}(c), FL rather than the inflection point on the $g^{(2)}(0)$ curve is the optimal driving wavelength for $g^{(2)}(0)$-based plasmonic sensing. In terms of performance, this is followed by the driving wavelength at FM, which now minimizes the noise in contrast to intensity sensing. However, this result is due to the noise being proportional to the sensing signal (low for $g^{(2)}(0)$ or high for $\langle \hat{m} \rangle$ at FM). Note that the sensitivity of $g^{(2)}(0)$, given by $S_{I-I}$, is several orders of magnitude higher than that of the intensity, $S_{I}$. This is because the latter is defined with respect to the time interval ${\mathrm T}_{int}$, where the signal $\langle \hat{m} \rangle \sim 10^{-4}$. The sensing resolution, given by the formulas in Eq.~(\ref{e27a}) and (\ref{e27b}), and shown in Figs. \ref{f6}(c) and \ref{f7}(c), respectively, takes into account the time averaging of the signal $\langle \hat{m} \rangle$ and provides a more fair comparison between the sensing performance of intensity- and $g^{(2)}(0)$-based sensing. Apart from the scale difference of the sensitivities, it can be seen from Fig.~\ref{f7} that similar to intensity-based sensing, the sensing resolution of $g^{(2)}(0)$ is not optimal at the driving wavelength where the sensitivity goes to zero. 

Due to a low mean photocount $\langle \hat{m} \rangle$ within ${\mathrm T}_{int}$, and thus the minor role of the antibunching effect in reducing the noise, the uncertainty in the
mean photocount, $\sigma_{m}$, is almost linearly propagated in $g^{(2)}(0)$, 
i.e., $\sigma_{g^{(2)}(0)} \approx 2g^{(2)}(0) \sigma_{m}/\langle \hat{m} \rangle \sim 10^{-3} - 10^{-2}$ across the Fano spectrum (see Eq.~\eqref{e21} and Fig.~\ref{f7}(b)), resulting in a relatively more noisy $g^{(2)}(0)$ signal than the intensity.
Hence, compared to intensity sensing, when including noise, the resolution of $g^{(2)}(0)$-based sensing is surpassed by that of intensity-based sensing for the system. See Table~\ref{t2} for specific values for comparison.

In Fig.~\ref{f8}, we show, using Eq.~\eqref{e19}, that the scattered light at both the Fano and plasmon resonances has the same noise level up to $\langle \hat{m} \rangle \approx 10^{-1}$, beyond which a significant reduction in the noise level begins to occur at the Fano resonance. 
The perfect emitter trend in Fig.~\ref{f8} is similar to the behaviour of the intensity of transmitted light through a medium and its noise for a fixed photon number of one probing the medium~\cite{Lee16}.
A 1000-fold increase in the mean photocount is needed in our hybrid system to achieve a decrease in the signal noise at the Fano resonance that is similar to a perfect emitter, regardless of the system being an efficient single photon source with $g^{(2)}(0) \approx 0.17$. We note that although a perfect emitter can provide the needed 1000-fold increase in the mean photocount, on its own it is not appreciably affected by a change in the refractive index~\cite{Lee21}, as in our hybrid system, and is therefore not a suitable candidate to use as a sensor. 

Finally, while fine-tuning of the model parameters in our hybrid system can lead to an enhanced Fano profile relative to the plasmon spectrum, resulting in an enhancement in the sensing parameters at the Fano resonance, this cannot be regarded as a quantum advantage in sensing since the comparative performance is then biased. On the other hand, replacing the MNP-QD system with other kinds of hybrid plasmonic systems, are possible options to explore, however these approaches are beyond the scope of this work, which is a starting point for such studies.

\section{Conclusion} \label{section 6}
We developed a quantum plasmonics 
model of a refractive index sensor operating at visible wavelengths, within the framework of cQED. The model is formulated in the dipole limit using the semi-classical, weak-field approximation. 
The proposed model geometry is experimentally realistic, with model parameters clearly defined.
The semi-classical model enabled us to obtain the sensing features of 
the steady-state scattered intensity and intensity--intensity correlation of the MNP-QD system.
Using photodetection theory, our results show that although the light scattered by the system could exhibit photon antibunching statistics within the Fano spectrum, the low mean photocount due to the small scattering efficiency of the MNP renders the quantum properties of the scattered light too small for improving the sensing resolution, as the noise is not reduced below that of the shot noise. Though we could increase the MNP size in order to improve its scattering efficiency and enhance the photocount, this approach suppresses the Fano profile, which subsequently leads to a disappearance of the antibunching effect. 
Therefore, in order to utilize the sub-shot-noise behaviour due to the constructive Fano interference in an MNP-QD hybrid system to improve the refractive index resolution of a quantum plasmonic sensor, the scattering efficiency or the mean photocount has to be enhanced without significantly altering the antibunching properties of the system. Future work in this direction is needed to understand how a quantum advantage may be obtained with such a system.

\section*{Acknowledgements}
This research was supported financially by the Department of Science and Innovation (DSI) through the South African Quantum Technology Initiative (SA QuTI), Stellenbosch University (SU), the National Research Foundation (NRF), and the Council for Scientific and Industrial Research (CSIR).

\bibliographystyle{unsrt}
\bibliography{manuscript.bib}
\end{document}


\maketitle 
\newpage 
Only new symbols will be defined. Symbols used in the manuscript retain their previously defined meanings.

\section{Dipole polarizability of the MNP}
The driving frequency-dependent and complex quasi-static dipole polarizability of a spherical MNP is given by \cite{Zade17,Ford84,Stef07}: 
\begin{equation} \label{e1}
\alpha(\omega) = \frac{4\pi r^{3}}{3} \frac{\epsilon(\omega)-\epsilon_{b}}{L\epsilon(\omega)+(1-L)\epsilon_{b}}, 
\end{equation}
with $\epsilon(\omega)$ as the dielectric function of the metal, modeled by the Drude permittivity \cite{Celi21,Kim14}:
\begin{equation}\label{e2}
	\epsilon(\omega) = \epsilon_{r}(\omega)+i\epsilon_{i}(\omega) = 
	\left(\epsilon_{\infty}-\frac{\omega_{p}^{2}}{\omega^{2}+\gamma_{p}^{2}}\right)
	+ i \left(\frac{\omega_{p}^{2}\gamma_{p}}{\omega(\omega^{2}+\gamma_{p}^{2})}\right),
\end{equation}
where $\epsilon_{r}(\omega)$ and $\epsilon_{i}(\omega)$ are respectively the real and imaginary parts of the dielectric function. The effects of a substrate on the dipole plasmon of the MNP are characterized by an effective geometric factor, $L$, given in Ref. \cite{Zade17}, for a weakly-reflecting substrate, $t\gg r$, as: 
\begin{equation}\label{e2a}
L \approx \frac{1}{3}\left[1 - \frac{S_{\beta}}{8}\left(\frac{\epsilon_{s} - \epsilon_{b} }{\epsilon_{s} + \epsilon_{b} } \right)\left(1 - \frac{4\epsilon_{s}\epsilon_{b} }{(\epsilon_{s} + \epsilon_{b})^{2} }\Big(1 + \frac{t}{r} \Big)^{-3}\right)    \right].
\end{equation}
We can re-write Eq. \eqref{e2a} in terms of the quasi-static (wavenumber-independent) reflection coefficient of the substrate: 
$\mathcal{R} = (\epsilon_{s} - \epsilon_{b})/(\epsilon_{s} + \epsilon_{b})$, with $1 - \mathcal{R}^{2} = 4\epsilon_{s} \epsilon_{b}/(\epsilon_{s} + \epsilon_{b})^{2}$, to obtain
\begin{equation}
L \approx \frac{1}{3}\left[1 - S_{\beta}\frac{\mathcal{R}}{8}\left( 1 - (1 - \mathcal{R}^{2})\Big(1 + \frac{t}{r}\Big)^{-3}\right)    \right].
\end{equation}
We can re-write Eq. \eqref{e1} in terms of a refined geometric factor, $f = (1-L)/L$, to obtain 
\begin{equation} \label{e3}
	\alpha(\omega) = \frac{4\pi(f+1)r^{3}}{3} \frac{\epsilon(\omega)-\epsilon_{b}}{\epsilon(\omega)+f\epsilon_{b}}.
\end{equation}
In the simple case of no substrate ($\mathcal{R} = 0$), $L$ reduces to $1/3$, leading to $f = 2$. 
According to the Fr\"{o}hlich condition, at the LSPR, $\omega = \omega_{pl}$, $\epsilon_{r}(\omega_{pl}) = -f\epsilon_{b}$ is satisfied for the MNP. 
Solving for $\omega_{pl}$ 
in the following equation
\[\epsilon_{r}(\omega_{pl}) =  \epsilon_{\infty}-\frac{\omega_{p}^{2}}{\omega_{pl}^{2}+\gamma_{p}^{2}} = -f\epsilon_{b}
\]
 leads to 
\begin{equation}\label{e4}
\omega_{pl} = \sqrt{\frac{\omega_{p}^{2}}{\epsilon_{\infty}+f\epsilon_{b}}-\gamma_{p}^{2}}.
\end{equation}
In quantum plasmonics, the dipole polarizability of the MNP has to be expressed in Lorentzian form \cite{Waks10,Kim14,Dolf10}. This form ensures that the polarizability has an explicit dependence on both  
$\omega_{pl}$ and $\gamma_{p}$, which is suitable for a cQED treatment.
First, we Taylor-expand $\epsilon_{r}(\omega)$ around $\omega_{pl}$ and truncate at the second term to obtain
\begin{equation}\label{e5}
	\epsilon_{r}(\omega) \approx 
	\epsilon_{r}(\omega_{pl})+\epsilon'_{r}(\omega_{pl})(\omega-\omega_{pl}), 
\end{equation}
where $\epsilon'_{r}(\omega_{pl}) = \dfrac{\partial \epsilon_{r}(\omega)}
{\partial \omega}\Big|_{\omega = \omega_{pl}} = \frac{2\omega_{pl} \omega_{p}^{2}}{(\omega_{pl}^{2}+\gamma_{p}^{2})^{2}}$. 
Then we write $\epsilon(\omega)$ in terms of $\epsilon_{r}(\omega)$ and $\epsilon_{i}(\omega)$ in Eq. \eqref{e3}, substitute Eq. \eqref{e5} for $\epsilon_{r}(\omega)$, and retain only the $(f+1)\epsilon_{b}$ term in the numerator of the resulting equation to obtain 
\begin{equation}\label{e6}
	\alpha(\omega) \approx \frac{i\frac{4}{3}\pi\epsilon_{b}(f+1)^{2}r^{3}
		\eta }
	{i\Delta_{pl} + \frac{1}{2}\gamma
	}, 
\end{equation}
where $\eta = 1/\epsilon'_{r}(\omega_{pl})$ and $\gamma = 
2\eta\epsilon_{i}(\omega_{pl}) = 2\eta\frac{\omega_{p}^{2}\gamma_{p}}{\omega_{pl}(\omega_{pl}^{2}+\gamma_{p}^{2})}$. Here, $\gamma$ is identified as the non-radiative plasmon decay rate, i.e. $\gamma = 
\gamma_{nr}$, since it originates from the intrinsic plasmon decay rate, $\gamma_{p}$, and it is independent of the speed of light in the medium.
We simplify $\eta$ and $\gamma_{nr}$ respectively to obtain
\begin{subequations}
\begin{align}
\eta & = \frac{1}{2\omega_{pl}}\left(\frac{\omega_{p}}{\varepsilon_{\infty}+f\varepsilon_{b} } \right)^{2}, \label{e7a}\\
\gamma_{nr} & = \gamma_{p}\left[ 1 +  \Big(\frac{\gamma_{p} }{\omega_{pl} } \Big)^{2} \right]. \label{e7b}
\end{align}
\end{subequations}
In the absence of a substrate, $f = 2$, and Eq. \eqref{e6} simplifies to the previously reported expression in Ref. \cite{Dolf10}. The radiative decay rate of the MNP is treated phenomenologically by obtaining it from the Wigner-Weisskopf formula given in Ref. \cite{Lukas12} with the background medium included. It is
\begin{equation}\label{e8}
\gamma_{r} = \frac{ \chi^{2}\sqrt{\epsilon_{b}} }{ 3\pi\epsilon_{0} \hslash   }\left(\frac{\omega_{pl}}{c}\right)^{3}, 
\end{equation}
where the dipole moment, $\chi$, of the MNP is derived in Sec. \ref{sec4}. The plasmon damping rate, $\gamma_{pl}$, is obtained by adding $\gamma_{r}$ to $\gamma_{nr}$ $(\gamma_{pl} = \gamma_{r} + \gamma_{nr})$, allowing us to re-write the dipole polarizability of the MNP in Lorentzian form as
\begin{equation}\label{e9}
	\alpha(\omega) \approx \frac{i\frac{4}{3}\pi\epsilon_{b}(f+1)^{2}r^{3}
		\eta }
	{i\Delta_{pl} + \frac{1}{2}\gamma_{pl} 
	}.
\end{equation}
Physically, the polarizability should have contributions from both internal and external dynamics, and adding $\gamma_{r}$ to $\gamma_{nr}$ as done in the above method achieves this goal. However, a more rigorous classical approach using the radiation reaction field \cite{Ford84} leads to the same result and therefore justifies combining the two rates in this way to give the effective dipole polarizability of the MNP in Lorentzian form.

\section{Induced dipole field on the QD}
The effective dipole moment of the MNP in the presence of the QD, in the classical picture, is \cite{Dolf10,Ceng07}:
\begin{equation}\label{e10}
P_{MNP} = \epsilon_{0}\epsilon_{b}\alpha(\omega)( E_{inc} + E^{QD}_{ind} ), 
\end{equation}
where $E_{inc} = \frac{E_{0}}{2}(e^{-i\omega t} + e^{i\omega t})$ is the magnitude of the incident field, and 
\begin{equation}\label{e11}
E^{QD}_{ind} = \frac{S_{\alpha} P_{QD}}{4\pi\epsilon_{0}\epsilon_{b}^{\prime} d^{3}}
\end{equation}
is the magnitude of the dipole field induced on the MNP by the QD, where $\epsilon_{b}^{\prime} = (2\epsilon_{b} + \epsilon_{d})/3$ is the effective dielectric constant of the local environment \cite{Ceng07}, and $P_{QD}$ is the dipole moment of the QD.  The magnitude of the dipole field induced on the QD by the MNP is \cite{Dolf10,Waks10}
\begin{equation}\label{e12}
	E_{ind} = \frac{S_{\alpha} P_{MNP}}{4\pi\epsilon_{0}\epsilon_{b}^{\prime} d^{3}},
\end{equation}
where $P_{MNP}$ is the dipole moment of the MNP.
Substituting Eq. \eqref{e9} for $\alpha(\omega)$ in Eq. \eqref{e10}, and replacing $P_{MNP}$ in Eq. \eqref{e12} with Eq. \eqref{e10}, we obtain the positive-frequency component (the term proportional to $e^{i\omega t}$) of the dipole field induced on the QD by the MNP: 
\begin{equation}\label{e13}
E_{ind}^{(+)} = \frac{iS_{\alpha}}{3\epsilon_{b}^{\prime}d^{3}}\frac{\epsilon^{2}_{b}(f+1)^{2}r^{3}
	\eta }
{i\Delta_{pl} + \frac{1}{2}\gamma_{pl} 
} \left(\frac{1}{2} E_{0} +  \frac{S_{\alpha} \mu \langle \hat{\sigma } \rangle}{4\pi\epsilon_{0}\epsilon_{b}^{\prime} d^{3}}\right) ,
\end{equation}
where we have replaced $P_{QD}$ with the quantum formulation, $\mu \langle \hat{\sigma } \rangle$.  Recall that in the laboratory frame, 
$\langle\hat{\sigma }\rangle$ transforms as $\langle\hat{\sigma }\rangle e^{i\omega t}$, see Sec. \ref{sec3}.

\section{The total Hamiltonian}\label{sec3}
The total Hamiltonian, $\hat{\mathcal{H}}$, is comprised of the free Hamiltonians, $\hat{\mathcal{H}}_{pl}$ and $\hat{\mathcal{H}}_{ex}$, of the MNP plasmon and the QD exciton, respectively, the interaction Hamiltonian, $\hat{\mathcal{H}}_{int}$, of the dipole-dipole coupling between the MNP and the QD, and the driving-field Hamiltonians, $\hat{\mathcal{H}}_{pl-dr}$ and $\hat{\mathcal{H}}_{ex-dr}$, of the MNP and the QD, respectively, i.e., 
\begin{equation}\label{e14}
\hat{\mathcal{H}} = \hat{\mathcal{H}}_{pl} + \hat{\mathcal{H}}_{ex} + \hat{\mathcal{H}}_{int} + \hat{\mathcal{H}}_{pl-dr} + \hat{\mathcal{H}}_{ex-dr}.
\end{equation}
In the dipole limit, the free Hamiltonian of the MNP is given by \cite{Gong15}
\begin{equation}\label{e15}
\hat{\mathcal{H}}_{pl}  = \hslash\omega_{pl}\hat{a}^{\dagger}\hat{a},
\end{equation}
where $\hat{a}^{\dagger}$ and $\hat{a}$ obey the following commutation relations:
$[\hat{a},\hat{a}^{\dagger}] = 1, [\hat{a},\hat{a}] = [\hat{a}^{\dagger},\hat{a}^{\dagger}] = 0$.
The QD is modeled as a two-level atom. 
Let $\omega_{0}$ and $\omega_{1}$ denote the ground and excited state energies, respectively, so that the transition frequency of the QD exciton is $\omega_{1}-\omega_{0} = \omega_{ex}$.
The QD Hamiltonian is \cite{Das07}
\begin{subequations}
\begin{align}
\hat{\mathcal{H}}_{ex} & = \hat{\mathcal{H}}_{0}+\hat{\mathcal{H}}_{1}-\hslash\omega_{0}\mathbb{I},\\
		& = \hslash\omega_{0}\hat{\sigma}\hat{\sigma}^{\dagger}+\hslash\omega_{1}\hat{\sigma}^{\dagger}\hat{\sigma}-\hslash\omega_{0}\mathbb{I},\\
		& = \hslash\omega_{ex}\hat{\sigma}^{\dagger}\hat{\sigma},
\end{align}
\end{subequations}
where $\hat{\sigma}^{\dagger}$ and $\hat{\sigma}$ obey the anti-commutation relations:
$\{\hat{\sigma},\hat{\sigma}^{\dagger}\} = \mathbb{I}, \{\hat{\sigma},\hat{\sigma}\} = \{\hat{\sigma}^{\dagger},\hat{\sigma}^{\dagger}\} = 0$, and the commutation relation: $[\hat{\sigma}^{\dagger},\hat{\sigma}] = \hat{\sigma}_{z}$.
The interaction Hamiltonian is given by \cite{Waks10,Gong15}: 
\begin{equation}\label{e17}
\hat{\mathcal{H}}_{int} = -\bm{\hat{\mu}}. \bm{\hat{E}}_{ind},
\end{equation}
where in the Schr\"odinger picture we have that $\bm{\hat{\mu}} = \mu( \hat{\sigma}^{\dagger} + \hat{\sigma} )\hat{\bm{e}_{i}}$ is QD dipole moment operator, $\bm{\hat{E}}_{ind} = E_{ind}( \hat{a}^{\dagger} + \hat{a} )\hat{\bm{e}_{i}}$ is the induced dipole field operator of the MNP (Longitudinal coupling: $i = z$, Transverse coupling: $i = x, y$), and $\mu E_{ind} = \hslash g$.
Expanding Eq. \eqref{e17} leads to
\begin{equation}\label{e18}
\hat{\mathcal{H}}_{int} = -\hslash g (\hat{\sigma}^{\dagger} \hat{a}^{\dagger} +\hat{\sigma} \hat{a} +
\hat{\sigma}^{\dagger}\hat{a}  + \hat{\sigma} \hat{a}^{\dagger} ).
\end{equation}
In our model, $g \sim 4.8$ meV, and $\gamma_{pl} \sim 72$ meV, which shows that we are in the bad-cavity limit \cite{Gong15}, $g \ll \gamma_{pl}$. The approximate time dependencies of the operator product terms in Eq.~\eqref{e18}, when expectation values are taken to solve the system dynamics, are $\sim e^{-i(\omega_{ex}+\omega_{pl}) t}$, $\sim e^{i(\omega_{ex}+\omega_{pl}) t}$, $\sim e^{-i(\omega_{ex}-\omega_{pl})t}$ and $\sim e^{i(\omega_{ex}-\omega_{pl})t}$, respectively. As with the case in previous models \cite{Dolf10,Gong15,Kim14}, we apply the rotating wave approximation (RWA) to Eq. \eqref{e18}, allowing us to drop the first two terms in Eq. \eqref{e18} (the terms oscillating at $\omega_{ex}+\omega_{pl}$), and re-write the Hamiltonian in the Schr\"{o}dinger picture as
\begin{equation}\label{e19}
	\hat{\mathcal{H}}_{int} \approx -\hslash g (
	\hat{\sigma}^{\dagger}\hat{a}  + \hat{\sigma} \hat{a}^{\dagger} ).
\end{equation}
The driving-field Hamiltonian of the MNP is given by \cite{Dolf10,Kim14}: 
\begin{equation}\label{e20}
\hat{\mathcal{H}}_{pl-dr} = - \bm{\hat{\chi}}.\mathbf{E}(t),
\end{equation}
where $\bm{\hat{\chi}} = \chi(\hat{a}^{\dagger} + \hat{a} )\bm{\hat{e}}_{i}$, and 
$\mathbf{E}(t) = E_{0}\cos(\omega t)\bm{\hat{e}}_{i} = \frac{1}{2}E_{o}(e^{-i\omega t} + e^{i\omega t})\bm{\hat{e}}_{i}$ (Longitudinal coupling: $i = z$, Transverse coupling: $i = x, y$).
Expanding Eq. \eqref{e20} and setting 
$\Omega_{pl} = E_{0}\chi/2\hslash$, we obtain
\begin{equation}\label{e21}
\hat{\mathcal{H}}_{pl-dr} = \hslash\Omega_{pl} (\hat{a}e^{-i\omega t}+\hat{a}^{\dagger}e^{i\omega t} + 
\hat{a}e^{i\omega t}+\hat{a}^{\dagger}e^{i\omega t}).
\end{equation}
Taking into account the approximate time dependence of the expectation value of the MNP operator as $\sim e^{i\omega_{pl} t}$, we can ignore the quickly oscillating terms in Eq. \eqref{e21}, i.e., the terms oscillating at $\omega_{pl}+\omega$ (RWA), obtaining
\begin{equation}\label{e22}
	\hat{\mathcal{H}}_{pl-dr} \approx \hslash\Omega_{pl} (\hat{a}e^{-i\omega t} 
	+\hat{a}^{\dagger}e^{i\omega t}).
\end{equation}
The driving-field Hamiltonian of the QD is given by \cite{Dolf10,Kim14}: 
\begin{equation}\label{e23}
	\hat{\mathcal{H}}_{ex-dr} = - \bm{\hat{\mu}}.\mathbf{E}(t).
\end{equation}
Expanding Eq. \eqref{e23} and setting 
$\Omega_{ex} = E_{0}\mu/2\hslash$, we obtain
\begin{equation}\label{e24}
	\hat{\mathcal{H}}_{ex-dr} = \hslash\Omega_{ex} (\hat{\sigma}e^{-i\omega t}+\hat{\sigma}^{\dagger}e^{-i\omega t} + 
\hat{\sigma}e^{i\omega t}+\hat{\sigma}^{\dagger}e^{i\omega t}).
\end{equation}
Once again, we can apply the RWA to Eq. \eqref{e24}, and obtain
\begin{equation}\label{e25}
	\hat{\mathcal{H}}_{ex-dr} \approx \hslash\Omega_{ex} (\hat{\sigma}e^{-i\omega t} 
	+\hat{\sigma}^{\dagger}e^{i\omega t}).
\end{equation}
Therefore, the total Hamiltonian in the Schr\"{o}dinger picture (lab frame) is 
\begin{equation}\label{e26}
\hat{\mathcal{H}} = \hslash\left( 
\omega_{pl}\hat{a}^{\dagger}\hat{a} + \omega_{ex}\hat{\sigma}^{\dagger}\hat{\sigma} -
g (\hat{\sigma}^{\dagger}\hat{a}  + \hat{\sigma} \hat{a}^{\dagger} )
-\Omega_{pl} (\hat{a}e^{-i\omega t} 
+\hat{a}^{\dagger}e^{i\omega t}) - 
\Omega_{ex} (\hat{\sigma}e^{-i\omega t} 
+\hat{\sigma}^{\dagger}e^{i\omega t})
\right).
\end{equation}
The time-dependence in the total Hamiltonian
can be removed by transforming Eq. \eqref{e26} into a rotating frame via the unitary operator, $\mathcal{U}(t) = e^{-i\mathcal{B}t}$, with 
$\mathcal{B} = \hslash\omega \hat{a}^{\dagger}\hat{a} 
+\hslash\omega \hat{\sigma}^{\dagger}\hat{\sigma}$. 
The Hamiltonian in the rotating frame is:
\begin{eqnarray}
	\mathcal{U}^{\dagger}\hat{\mathcal{H}}\mathcal{U} - \mathcal{B}  = 
	\mathcal{U}^{\dagger}\hslash(\omega_{pl}\hat{a}^{\dagger}\hat{a} +\omega_{ex}\hat{\sigma}^{\dagger}\hat{\sigma} ) \mathcal{U} 
	- \hslash(\omega \hat{a}^{\dagger}\hat{a} +\omega \hat{\sigma}^{\dagger}\hat{\sigma}) 
	 -\hslash g \mathcal{U}^{\dagger}(\hat{\sigma} \hat{a}^{\dagger}+\hat{\sigma}^{\dagger}\hat{a})\mathcal{U} \nonumber \\
	-\hslash\Omega_{ex}\mathcal{U}^{\dagger}(\hat{\sigma} e^{-i\omega t} + \hat{\sigma}^{\dagger} e^{i\omega t}) \mathcal{U} \nonumber 
	-\hslash\Omega_{pl}\mathcal{U}^{\dagger}[(\hat{a} e^{-i\omega t} + \hat{a}^{\dagger} e^{i\omega t})]\mathcal{U}  \\
 	= 
	\hslash\left( \Delta_{pl}\hat{a}^{\dagger}\hat{a} + \Delta_{ex}\hat{\sigma}^{\dagger}\hat{\sigma} 
	-g (\hat{\sigma} \hat{a}^{\dagger}+\hat{\sigma}^{\dagger}\hat{a}) 
	-\Omega_{ex}(\hat{\sigma}  + \hat{\sigma}^{\dagger} )  
	-\Omega_{pl}(\hat{a}  + \hat{a}^{\dagger} ) \right),~~~~~~\label{e27}
\end{eqnarray}
i.e., in the rotating frame, the lab frame operators are oscillating with the driving frequency: 
\begin{equation*}
	\hat{a}\rightarrow \hat{a}e^{i\omega t}, \hat{a}^{\dagger}\rightarrow \hat{a}^{\dagger}e^{-i\omega t}, 
	\hat{\sigma}\rightarrow \hat{\sigma} e^{i\omega t}, \hat{\sigma}^{\dagger}\rightarrow \hat{\sigma}^{\dagger}e^{-i\omega t}.
\end{equation*}
We will use this transformed Hamiltonian, Eq. \eqref{e27}, with $\Delta_{ex} = \omega_{ex}-\omega$ and $\Delta_{pl} = \omega_{pl}-\omega$, in the rest of the discussion but we will maintain the symbol, $\hat{\mathcal{H}}$.

\section{Equations of motion}\label{sec4}
We recall from the main text that at optical frequencies the thermal occupation number is negligible, and the phenomenological master equation
for the system, $\hat{ \rho}$, in the rotating frame and in Lindblad form is \cite{Wals08,Carm99}
\begin{equation}\label{e28}
	\dot{\hat{ \rho }  } = \frac{i}{\hslash}[\hat{ \rho },\hat{ \hham }  ] 
	- \frac{\gamma_{pl}}{2}\mathcal{D}_{pl}[\hat{a},\hat{ \rho } ]
	- \frac{\gamma_{ex}}{2}\mathcal{D}_{ex}[\hat{\sigma},\hat{ \rho } ], 
\end{equation}
where 
\begin{equation}\label{e29}
	\mathcal{D}_{j}[\hat{c}_{j},\hat{ \rho } ] =  \hat{ \rho }\hat{c}_{j}^{\dagger}\hat{c}_{j}  + \hat{c}_{j}^{\dagger}\hat{c}_{j}\hat{ \rho } - 2\hat{c}_{j}\hat{ \rho }\hat{c}_{j}^{\dagger},
\end{equation}
with $j = pl, ex$, so that $\hat{c}_{pl} = \hat{a}$, and $\hat{c}_{ex} = \hat{\sigma}$.

Multiplying Eq. \eqref{e28} on the left by $\hat{a}$ and performing a trace over the entire MNP-QD system (using the cyclic property of the trace and $[\hat{a},\hat{\sigma} ] = [\hat{a}^{\dagger},\hat{\sigma}^{\dagger} ] = 0$ for the Schrödinger picture operators), we obtain an equation of motion for $\langle \hat{a} \rangle = \text{tr}(\hat{a}\hat{\rho})$ as:
\begin{equation}\label{e30}
\text{tr}(\hat{a}\dot{\hat{\rho }}) =  i\text{tr}\left(\hat{a}[\hat{ \rho }, \Delta_{pl}\hat{a}^{\dagger}\hat{a} 
-g (\hat{\sigma} \hat{a}^{\dagger}+\hat{\sigma}^{\dagger}\hat{a}) 
-\Omega_{pl}(\hat{a}  + \hat{a}^{\dagger} )  ]\right)
- \frac{\gamma_{pl}}{2}\text{tr}\left(\hat{a}\hat{ \rho }\hat{a}^{\dagger}\hat{a}  + \hat{a}\hat{a}^{\dagger}\hat{a}\hat{ \rho } - 2\hat{a}\hat{a}\hat{ \rho }\hat{a}^{\dagger} \right).
\end{equation}
Using the cyclic property of the trace: $\text{tr}(abc) = \text{tr}(bca) = \text{tr}(cab)$ and the commutation relations of the MNP field operators, we obtain: 
\[ \langle \overset{.}{\hat{a}} \rangle = \text{tr}(\hat{a}\dot{\hat{\rho }}), 
\text{tr}\left(\hat{a}\hat{ \rho }\hat{a}^{\dagger}\hat{a}  + \hat{a}\hat{a}^{\dagger}\hat{a}\hat{ \rho } - 2\hat{a}\hat{a}\hat{ \rho }\hat{a}^{\dagger} \right) = \text{tr}\left(\hat{a}\hat{ \rho }\hat{a}^{\dagger}\hat{a}  + \hat{a}\hat{ \rho }\hat{a}\hat{a}^{\dagger} - 2\hat{a}\hat{ \rho }\hat{a}^{\dagger}\hat{a} \right) = \text{tr}\left( \hat{a}\hat{ \rho }[\hat{a},\hat{a}^{\dagger} ]\right) = \langle \hat{a} \rangle, \]
\[ 
\text{tr}\left(\hat{a}[\hat{ \rho }, \Delta_{pl}\hat{a}^{\dagger}\hat{a}]\right) = \Delta_{pl}\text{tr}\left(  \hat{a}\hat{ \rho }\hat{a}^{\dagger}\hat{a} - \hat{a}\hat{ \rho }
\hat{a}\hat{a}^{\dagger} \right) = \Delta_{pl}\text{tr}\left(\hat{a}\hat{ \rho }[\hat{a}^{\dagger},\hat{a} ]  \right) = -\Delta_{pl}\langle \hat{a} \rangle, 
 \]
\[\text{tr}\left(\hat{a}[\hat{ \rho },
-g (\hat{\sigma} \hat{a}^{\dagger}+\hat{\sigma}^{\dagger}\hat{a})]\right) = -g\text{tr}\left( 
\hat{\sigma}\hat{ \rho }\hat{a}^{\dagger}\hat{a} - 
\hat{\sigma}\hat{ \rho }\hat{a}\hat{a}^{\dagger} 
\right)  = -g\text{tr}\left(\hat{\sigma}\hat{ \rho }[\hat{a}^{\dagger},\hat{a} ]   \right)
= g\langle \hat{\sigma} \rangle,
\]
\[ \text{tr}\left(\hat{a}[\hat{ \rho },
-\Omega_{pl}(\hat{a}  + \hat{a}^{\dagger} )  ]  \right)
=  -\Omega_{pl}\text{tr}\left(\hat{ \rho }\hat{a}^{\dagger}\hat{a} - 
\hat{ \rho }\hat{a}\hat{a}^{\dagger}   \right) = -\Omega_{pl}\text{tr}\left(\hat{ \rho }[\hat{a}^{\dagger},\hat{a}]   \right) = \Omega_{pl},
\]
so that Eq. \eqref{e30} simplifies to: 
\begin{equation}\label{e31y}
\langle \overset{.}{\hat{a}} \rangle  = 
-\Big(i\Delta_{pl}+\frac{1}{2}\gamma_{pl}\Big)\langle \hat{a} \rangle 
+ i\Big(g\langle \hat{\sigma} \rangle + \Omega_{pl}\Big).
\end{equation}
At steady-state, $\langle \overset{.}{\hat{a}} \rangle = 0$, we obtain
\begin{equation}\label{e32}
\langle \hat{a} \rangle  = \frac{i[\Omega_{pl} + g\langle \hat{\sigma} \rangle ]}{i\Delta_{pl} +\frac{1}{2}\gamma_{pl} }.
\end{equation}

Multiplying Eq. \eqref{e28} on the left by $\hat{a}^{2}$ and performing a trace over the entire MNP-QD system (using the cyclic property of the trace and $[\hat{a},\hat{\sigma} ] = [\hat{a}^{\dagger},\hat{\sigma}^{\dagger} ] = 0$ for the Schrödinger picture operators), we obtain an equation of motion for $\langle \hat{a}^{2} \rangle = \text{tr}(\hat{a}^{2}\hat{\rho})$ as:
\begin{equation}\label{e30x}
	\text{tr}(\hat{a}^{2}\dot{\hat{\rho }}) =  i\text{tr}\left(\hat{a}^{2}[\hat{ \rho }, \Delta_{pl}\hat{a}^{\dagger}\hat{a} 
	-g (\hat{\sigma} \hat{a}^{\dagger}+\hat{\sigma}^{\dagger}\hat{a}) 
	-\Omega_{pl}(\hat{a}  + \hat{a}^{\dagger} )  ]\right)
	- \frac{\gamma_{pl}}{2}\text{tr}\left(\hat{a}^{2}\hat{ \rho }\hat{a}^{\dagger}\hat{a}  + \hat{a}^{2}\hat{a}^{\dagger}\hat{a}\hat{ \rho } - 2\hat{a}^{2}\hat{a}\hat{ \rho }\hat{a}^{\dagger} \right).
\end{equation}
Using the cyclic property of the trace: $\text{tr}(abc) = \text{tr}(bca) = \text{tr}(cab)$ and the commutation relations of the MNP field operators, we obtain: 
\[ \partial_{t}\langle \hat{a}^{2} \rangle = \text{tr}(\hat{a}^{2}\dot{\hat{\rho }}),\]
\[ \text{tr}\left(\hat{a}^{2}\hat{ \rho }\hat{a}^{\dagger}\hat{a}  + \hat{a}^{2}\hat{a}^{\dagger}\hat{a}\hat{ \rho } - 2\hat{a}^{2}\hat{a}\hat{ \rho }\hat{a}^{\dagger} \right) = \text{tr}\left(\hat{ \rho }\hat{a}^{\dagger}\hat{a}\hat{a}^{2}  + \hat{ \rho }\hat{a}^{2}\hat{a}^{\dagger}\hat{a} - 2\hat{ \rho }\hat{a}^{\dagger}\hat{a}\hat{a}^{2} \right) = \text{tr}\left( \hat{a}\hat{ \rho }[\hat{a}^{2},\hat{a}^{\dagger} ]\right) = 2\langle \hat{a}^{2} \rangle, \]
\[ 
\text{tr}\left(\hat{a}^{2}[\hat{ \rho }, \Delta_{pl}\hat{a}^{\dagger}\hat{a}]\right) = \Delta_{pl}\text{tr}\left(  \hat{a}^{2}\hat{ \rho }\hat{a}^{\dagger}\hat{a} - \hat{a}^{2}\hat{ \rho }
\hat{a}\hat{a}^{\dagger} \right) = \Delta_{pl}\text{tr}\left(\hat{a}\hat{ \rho }[\hat{a}^{\dagger},\hat{a}^{2} ]  \right) = -2\Delta_{pl}\langle \hat{a}^{2} \rangle, 
\]
\[\text{tr}\left(\hat{a}^{2}[\hat{ \rho },
-g (\hat{\sigma} \hat{a}^{\dagger}+\hat{\sigma}^{\dagger}\hat{a})]\right) = -g\text{tr}\left( 
\hat{\sigma}\hat{ \rho }\hat{a}^{\dagger}\hat{a}^{2} - 
\hat{\sigma}\hat{ \rho }\hat{a}^{2}\hat{a}^{\dagger} 
\right)  = -g\text{tr}\left(\hat{\sigma}\hat{ \rho }[\hat{a}^{\dagger},\hat{a}^{2} ]   \right)
= 2g\langle \hat{\sigma}\hat{a} \rangle,
\]
\[ \text{tr}\left(\hat{a}^{2}[\hat{ \rho },
-\Omega_{pl}(\hat{a}  + \hat{a}^{\dagger} )  ]  \right)
=  -\Omega_{pl}\text{tr}\left(\hat{ \rho }\hat{a}^{\dagger}\hat{a}^{2} - 
\hat{ \rho }\hat{a}^{2}\hat{a}^{\dagger}   \right) = -\Omega_{pl}\text{tr}\left(\hat{ \rho }[\hat{a}^{\dagger},\hat{a}^{2}]   \right) = 2\Omega_{pl}\langle \hat{a} \rangle,
\]
so that Eq. \eqref{e30x} simplifies to: 
\begin{equation}\label{e31}
	\partial_{t}\langle {\hat{a}^{2}} \rangle  = 
	-2\Big(i\Delta_{pl}+\frac{1}{2}\gamma_{pl}\Big)\langle \hat{a}^{2} \rangle 
	+ i2g\langle \hat{\sigma}  \hat{a} \rangle + i2\Omega_{pl}\langle \hat{a} \rangle.
\end{equation}
At steady-state, $\partial_{t}\langle {\hat{a}^{2}} \rangle = 0$, we obtain
\begin{equation}\label{e32x}
	\langle \hat{a}^{2} \rangle  \approx \frac{i[\Omega_{pl} + g\langle \hat{\sigma} \rangle ]\langle \hat{a} \rangle}{i\Delta_{pl} +\frac{1}{2}\gamma_{pl} } = \langle \hat{a} \rangle^{2} = 
	-\frac{[\Omega_{pl}^{2} + 2\Omega_{pl}g\langle \hat{\sigma} \rangle ]}{(i\Delta_{pl} +\frac{1}{2}\gamma_{pl})^{2} },
\end{equation}
where we have used the approximation: $\langle \hat{\sigma}  \hat{a} \rangle \approx \langle \hat{\sigma} \rangle\langle \hat{a} \rangle$, 
which assumes that the joint expectation value of the MNP and QD operators can be factored 
due to the strong plasmon damping that inhibits the persistence of quantum correlations between the MNP and the QD.
Similarly, using the approximations: $\langle \hat{\sigma}  \hat{a}^{2} \rangle \approx \langle \hat{\sigma} \rangle\langle \hat{a}^{2} \rangle$ and 
$\langle \hat{\sigma}  \hat{a}^{3} \rangle \approx \langle \hat{\sigma} \rangle\langle \hat{a}^{3} \rangle$
in the equations of motion for $\partial_{t}\langle {\hat{a}^{3}} \rangle$ and $\partial_{t}\langle {\hat{a}^{4}} \rangle$ 
lead to: 
\[\langle \hat{a}^{3} \rangle \approx \langle \hat{a} \rangle^{3} 
 = 	-i\frac{[\Omega_{pl}^{3} + 3\Omega_{pl}^{2}g\langle \hat{\sigma} \rangle ]}{(i\Delta_{pl} +\frac{1}{2}\gamma_{pl})^{3} }\]
and 
\[\langle \hat{a}^{4} \rangle \approx \langle \hat{a} \rangle^{4}
= 	-\frac{[\Omega_{pl}^{4} + 4\Omega_{pl}^{3}g\langle \hat{\sigma} \rangle ]}{(i\Delta_{pl} +\frac{1}{2}\gamma_{pl})^{4} },\] respectively, at steady-state.

Multiplying Eq. \eqref{e28} on the left by $\hat{a}^{\dagger}\hat{a}$ and performing a trace over the entire MNP-QD system (using the cyclic property of the trace and $[\hat{a},\hat{\sigma} ] = [\hat{a}^{\dagger},\hat{\sigma}^{\dagger} ] = 0$ for the Schrödinger picture operators), we obtain an equation of motion for $\langle \hat{a}^{\dagger}\hat{a} \rangle = \text{tr}(\hat{a}^{\dagger}\hat{a}\hat{\rho})$ as:
\begin{eqnarray}
	\text{tr}(\hat{a}^{\dagger}\hat{a}\dot{\hat{\rho }}) &= &  i\text{tr}\left(\hat{a}^{\dagger}\hat{a}[\hat{ \rho }, \Delta_{pl}\hat{a}^{\dagger}\hat{a} 
	-g (\hat{\sigma} \hat{a}^{\dagger}+\hat{\sigma}^{\dagger}\hat{a}) 
	-\Omega_{pl}(\hat{a}  + \hat{a}^{\dagger} )  ]\right) \label{e30xx} \\
	& & - \frac{\gamma_{pl}}{2}\text{tr}\left(\hat{a}^{\dagger}\hat{a}\left(\hat{ \rho }\hat{a}^{\dagger}\hat{a}  + \hat{a}^{\dagger}\hat{a}\hat{ \rho } - 2\hat{a}\hat{ \rho }\hat{a}^{\dagger} \right)\right). \nonumber
\end{eqnarray}
Using the cyclic property of the trace: $\text{tr}(abc) = \text{tr}(bca) = \text{tr}(cab)$ and the commutation relations of the MNP field operators, we obtain: 
\[ \partial_{t}\langle \hat{a}^{\dagger}\hat{a} \rangle = \text{tr}(\hat{a}^{\dagger}\hat{a} \dot{\hat{\rho }}), \]
\[\text{tr}\left(\hat{a}^{\dagger}\hat{a}\hat{ \rho }\hat{a}^{\dagger}\hat{a}  + \hat{a}^{\dagger}\hat{a}\hat{a}^{\dagger}\hat{a}\hat{ \rho } - 2\hat{a}^{\dagger}\hat{a}\hat{a}\hat{ \rho }\hat{a}^{\dagger} \right) = 2\text{tr}\left(\hat{a}^{\dagger}\left(\hat{ \rho }\hat{a}\hat{a}^{\dagger}  - \hat{ \rho }\hat{a}^{\dagger}\hat{a}\right)\hat{a}\right) = \text{tr}\left( \hat{ \rho }\hat{a}^{\dagger}[\hat{a},\hat{a}^{\dagger} ]\hat{a}\right) = 2\langle \hat{a}^{\dagger}\hat{a} \rangle, \]
\[ 
\text{tr}\left(\hat{a}^{\dagger}\hat{a}[\hat{ \rho }, \Delta_{pl}\hat{a}^{\dagger}\hat{a}]\right) = 
\Delta_{pl}\text{tr}\left(\hat{a}^{\dagger}\hat{a}\left(\hat{ \rho }\hat{a}^{\dagger}\hat{a} - \hat{a}^{\dagger}\hat{a}\hat{ \rho }\right)\right)  = 0, 
\]
\[\text{tr}\left(\hat{a}^{\dagger}\hat{a}[\hat{ \rho },
-g (\hat{\sigma} \hat{a}^{\dagger}+\hat{\sigma}^{\dagger}\hat{a})]\right)  = -g\text{tr}\left(\hat{a}^{\dagger}\hat{ \rho }[\hat{a}^{\dagger},\hat{a} ]\hat{\sigma} + \hat{\sigma}^{\dagger}\hat{ \rho }[\hat{a},\hat{a}^{\dagger} ]\hat{a}   \right)
= g\left(\langle \hat{a}^{\dagger}\hat{\sigma} \rangle - \langle \hat{\sigma}^{\dagger}\hat{a} \rangle\right),
\]
\[ \text{tr}\left(\hat{a}^{\dagger}\hat{a}[\hat{ \rho },
-\Omega_{pl}(\hat{a}  + \hat{a}^{\dagger} )  ]  \right)
 = -\Omega_{pl}\text{tr}\left(\hat{ \rho }[\hat{a},\hat{a}^{\dagger}]\hat{a} +  \hat{ \rho }\hat{a}^{\dagger}[\hat{a}^{\dagger},\hat{a}]  \right) = \Omega_{pl}(\langle \hat{a}^{\dagger} \rangle - \langle \hat{a} \rangle ),
\]
so that Eq. \eqref{e30xx} simplifies to: 
\begin{equation}\label{e31x}
	\partial_{t}\langle \hat{a}^{\dagger}{\hat{a}} \rangle  = 
	-\gamma_{pl}\langle \hat{a}^{\dagger}{\hat{a}} \rangle 
	+ 2g\Im[\langle \hat{\sigma}^{\dagger}  \hat{a} \rangle] + 2\Omega_{pl}\Im[\langle \hat{a} \rangle].
\end{equation}
At steady-state, $\partial_{t}\langle \hat{a}^{\dagger}{\hat{a}} \rangle = 0$, we obtain
\begin{equation}\label{e32xx}
	\langle \hat{a}^{\dagger}{\hat{a}} \rangle  \approx \frac{2}{\gamma_{pl}} \Im[(g\langle \hat{\sigma}^{\dagger} \rangle + \Omega_{pl})\langle \hat{a} \rangle ]  = \langle \hat{a}^{\dagger} \rangle \langle{\hat{a}} \rangle  \approx \frac{\Omega_{pl}^{2} + g^{2}\langle \hat{\sigma}^{\dagger}\hat{\sigma} \rangle + 2\Omega_{pl}g\Re\langle \hat{\sigma} \rangle   }{\Delta_{pl}^{2} + \frac{1}{4}\gamma_{pl}^{2}  },
\end{equation}
where we have made the approximation: $\langle \hat{\sigma}^{\dagger}  \hat{a} \rangle$ $\approx$ $\langle \hat{\sigma}^{\dagger} \rangle\langle \hat{a} \rangle$, as well as the approximation:  $\langle \hat{\sigma}^{\dagger}\rangle \langle \hat{\sigma} \rangle \approx \langle \hat{\sigma}^{\dagger}\hat{\sigma} \rangle$, since the incoherent scattering component ($\langle \hat{\sigma}^{\dagger}\hat{\sigma} \rangle - \langle \hat{\sigma}^{\dagger}\rangle \langle \hat{\sigma} \rangle$) is negligible in the weak-driving field limit. The approximation, $\langle \hat{\sigma}^{\dagger}  \hat{a} \rangle \approx \langle \hat{\sigma}^{\dagger} \rangle\langle \hat{a} \rangle$, is what led to the result: $\langle \hat{a}^{\dagger}\rangle \langle \hat{a} \rangle \approx \langle \hat{a}^{\dagger}\hat{a} \rangle$. When it is applied to higher-order expectation values, we have the following: 
$\langle (\hat{a}^{\dagger})^{2}\rangle \langle \hat{a}^{2} \rangle \approx \langle (\hat{a}^{\dagger})^{2}\hat{a}^{2} \rangle, \langle (\hat{a}^{\dagger})^{3}\rangle \langle \hat{a}^{3} \rangle \approx \langle (\hat{a}^{\dagger})^{3}\hat{a}^{3} \rangle$, and $\langle (\hat{a}^{\dagger})^{4}\rangle \langle \hat{a}^{4} \rangle \approx \langle (\hat{a}^{\dagger})^{4}\hat{a}^{4} \rangle$. The good agreement between our analytical calculations and QuTIP simulations in the weak-driving field limit shows that these approximations are valid. 

The positive-frequency component of dipole field induced on the QD by the MNP in the steady-state is given by \cite{Gong15,Kim14}:
\begin{equation}\label{e33}
E_{ind}^{(+)} = \frac{\hslash g}{\mu}\langle \hat{a} \rangle.
\end{equation}
Substituting Eq. \eqref{e32} into Eq. \eqref{e33} and comparing the resulting equation to Eq. \eqref{e13}, using the definition of $\Omega_{pl}$ in terms of $\chi$ and $E_0$, we obtain
\begin{subequations}
	\begin{align}
\frac{\hslash g^{2}}{\mu}  =  \frac{S_{\alpha}\epsilon^{2}_{b}(f+1)^{2}r^{3}
	\eta }{ 3\epsilon_{b}^{\prime}d^{3}}
\times \frac{S_{\alpha} \mu }{4\pi\epsilon_{0}\epsilon_{b}^{\prime} d^{3}}  
   \implies	&	g  = \frac{1}{3}(f+1)\left(\frac{\mu S_{\alpha} }{d^{3} } \right)\left(\frac{\epsilon_{b} }{\epsilon^{\prime}_{b} } \right) \sqrt{ \frac{3\eta r^{3} }{4\pi\varepsilon_{0}\hslash }  }, \label{e34a} \\
\frac{\hslash g}{\mu} \times \frac{\chi}{\hslash}
= \frac{S_{\alpha}\epsilon^{2}_{b}(f+1)^{2}r^{3}
	\eta }{ 3\epsilon_{b}^{\prime}d^{3}}
	\implies	&	\chi = \frac{1}{3}(f+1)\epsilon_{b}\sqrt{12\pi\varepsilon_{0}\hslash
			\eta r^{3}  }. \label{e34b}
	\end{align}
\end{subequations}
Having derived $\chi$, we can return to Eq. \eqref{e8}, to finally obtain
\begin{equation}\label{e35a}
	\gamma_{r} = \frac{ \chi^{2} \sqrt{\epsilon_{b}} }{ 3\pi\epsilon_{0} \hslash   }\left(\frac{\omega_{pl}}{c}\right)^{3} = 
	 \frac{ (f+1)^{2}\epsilon_{b}^{2.5}\times 12\pi\varepsilon_{0}\hslash
	 	\eta r^{3}  }{ 3\pi\epsilon_{0} \hslash \times 9  }\left(\frac{\omega_{pl}}{c}\right)^{3} 
		= \frac{ 4}{9}(f+1)^{2}\eta n^{2}\Big(k\Big|_{\omega = \omega_{pl}}r\Big)^{3}, 
\end{equation}
with the wavenumber, $k = \omega n/c$, and $\epsilon_{b} = n^{2}$. 

Multiplying Eq. \eqref{e28} on the left by $\hat{\sigma}$ and performing a trace over the entire MNP-QD system (using the cyclic property of the trace and $[\hat{a},\hat{\sigma} ] = [\hat{a}^{\dagger},\hat{\sigma}^{\dagger} ] = 0$ for the Schrödinger picture operators), we obtain an equation of motion for $\langle \hat{\sigma} \rangle = \text{tr}(\hat{\sigma}\hat{\rho})$ as:
\begin{eqnarray}\label{e35}
	\text{tr}(\hat{\sigma}\dot{\hat{\rho }}) &=&  i\text{tr}\left(\hat{\sigma}[\hat{ \rho }, \Delta_{ex}\hat{\sigma}^{\dagger}\hat{\sigma} 
	-g (\hat{\sigma} \hat{a}^{\dagger}+\hat{\sigma}^{\dagger}\hat{a}) 
	-\Omega_{ex}(\hat{\sigma}  + \hat{\sigma}^{\dagger} )  ]\right) \nonumber \\
	& & - \frac{\gamma_{ex}}{2}\text{tr}\left(\hat{\sigma}\hat{ \rho }\hat{\sigma}^{\dagger}\hat{\sigma}  + \hat{\sigma}\hat{\sigma}^{\dagger}\hat{\sigma}\hat{ \rho } - 2\hat{\sigma}\hat{\sigma}\hat{ \rho }\hat{\sigma}^{\dagger} \right).
\end{eqnarray}
Using the cyclic property of the trace and the commutation and anti-commutation relations of the QD lowering and raising operators, in addition to the following properties: $\hat{\sigma}\hat{\sigma} = 
\hat{\sigma}^{\dagger}\hat{\sigma}^{\dagger} = 0, 
\hat{\sigma}\hat{\sigma}^{\dagger}\hat{\sigma} = \hat{\sigma}, 
\hat{\sigma}^{\dagger}\hat{\sigma}\hat{\sigma}^{\dagger} = \hat{\sigma}^{\dagger}$, 
we obtain: 
\[ \langle \overset{.}{\hat{\sigma}} \rangle = \text{tr}(\hat{\sigma}\dot{\hat{\rho }}),\] 
\[\text{tr}\left(\hat{\sigma}\hat{ \rho }\hat{\sigma}^{\dagger}\hat{\sigma}  + \hat{\sigma}\hat{\sigma}^{\dagger}\hat{\sigma}\hat{ \rho } - 2\hat{\sigma}\hat{\sigma}\hat{ \rho }\hat{\sigma}^{\dagger} \right) = \text{tr}\left(\hat{\sigma}\hat{ \rho }\hat{\sigma}^{\dagger}\hat{\sigma}  + \hat{\sigma}\hat{ \rho }\hat{\sigma}\hat{\sigma}^{\dagger} - 0 \right) = \text{tr}\left( \hat{\sigma}\hat{ \rho }\{\hat{\sigma}^{\dagger},\hat{\sigma} \}\right) = \langle \hat{\sigma} \rangle, \]
\[\text{tr}\left(\hat{\sigma}[\hat{ \rho }, \Delta_{ex}\hat{\sigma}^{\dagger}\hat{\sigma}]\right) = \Delta_{ex}\text{tr}\left(\hat{\sigma}^{\dagger}\hat{\sigma}\hat{\sigma}\hat{ \rho } - \hat{\sigma}
\hat{\sigma}^{\dagger}\hat{\sigma}\hat{ \rho } \right) = \Delta_{ex}\text{tr}\left(0 -  \hat{\sigma}\hat{ \rho } \right) = -\Delta_{ex}\langle \hat{\sigma } \rangle, 
\]
\[\text{tr}\left(\hat{\sigma}[\hat{ \rho },
-g (\hat{\sigma} \hat{a}^{\dagger}+\hat{\sigma}^{\dagger}\hat{a})]\right) = -g\text{tr}
\left( 
\hat{\sigma}\hat{ \rho }\hat{\sigma}^{\dagger}\hat{a} - \hat{\sigma}\hat{\sigma}^{\dagger}\hat{a}\hat{ \rho }
\right)\]
\[= -g\text{tr}\left(\hat{\sigma}^{\dagger}\hat{\sigma}\hat{a}\hat{ \rho} -(1-\hat{\sigma}^{\dagger}\hat{\sigma})\hat{a}\hat{ \rho }  \right)
= g\langle \hat{a} \rangle - 2g\langle \hat{a}\hat{\sigma}^{\dagger}\hat{\sigma} \rangle,
\]
\[ \text{tr}\left(\hat{\sigma}[\hat{ \rho },
-\Omega_{ex}(\hat{\sigma}  + \hat{\sigma}^{\dagger} )  ]  \right)
=  -\Omega_{ex}\text{tr}\left(\hat{ \rho }\hat{\sigma}\hat{\sigma} + 
\hat{\sigma}\hat{ \rho }\hat{\sigma}^{\dagger} 
- \hat{\sigma}\hat{\sigma}\hat{ \rho }
- \hat{\sigma}\hat{\sigma}^{\dagger}\hat{ \rho }
\right) = \Omega_{ex}(1 - 2\langle  \hat{\sigma}^{\dagger}\hat{\sigma} \rangle ),
\]
so that Eq. \eqref{e35} simplifies to: 
\begin{equation}\label{e36}
\langle \overset{.}{\hat{\sigma}} \rangle = 
-\Big(i\Delta_{ex}+\frac{1}{2}\gamma_{ex}\Big)\langle \hat{\sigma} \rangle 
-i g\langle \hat{a} \rangle\langle \hat{\sigma}_{z} \rangle + i\Omega_{ex} (1-2\langle \hat{\sigma}^{\dagger}\hat{\sigma} \rangle),
\end{equation}
where we have assumed that the joint expectation value of the MNP and QD operator, $\langle \hat{a}\hat{\sigma}^{\dagger}\hat{\sigma} \rangle$, factors as $\langle \hat{a}\rangle\langle\hat{\sigma}^{\dagger}\hat{\sigma} \rangle$ 
due to the presence of strong damping from the MNP that inhibits the persistence of quantum correlations between the MNP and the QD. Since we are in the adiabatic limit, where $\gamma_{pl}>>\gamma_{ex}$ (i.e. the dynamics of the MNP in the timescale of the QD dynamics is approximately stationary), we can substitute Eq. \eqref{e32} for 
$\langle \hat{a} \rangle$ in Eq. \eqref{e36} to obtain:
\begin{subequations}
\begin{align}
\langle \overset{.}{\hat{\sigma}} \rangle & = 
-\Big(i\Delta_{ex}+\frac{1}{2}\gamma_{ex}\Big)\langle \hat{\sigma} \rangle 
-i g\left(\frac{i\Omega_{pl} + ig\langle \hat{\sigma} \rangle }{i\Delta_{pl} +\frac{1}{2}\gamma_{pl} } \right)  \langle \hat{\sigma}_{z} \rangle + i\Omega_{ex} (1-2\langle \hat{\sigma}^{\dagger}\hat{\sigma} \rangle),\\
& = -\left[ i\Delta_{ex}+\frac{1}{2}\gamma_{ex}
-\frac{ g^{2} \langle \hat{\sigma}_{z} \rangle }{i\Delta_{pl} +\frac{1}{2}\gamma_{pl} } \right] \langle  \hat{\sigma} \rangle + i\left[ \Omega_{ex} + \frac{ig\Omega_{pl}}{i\Delta_{pl} +\frac{1}{2}\gamma_{pl} } \right] (1-2\langle \hat{\sigma}^{\dagger}\hat{\sigma} \rangle), \\
& = -\Big(i[ \Delta_{ex} + \mathcal{F}\Delta_{pl}\langle \hat{\sigma}_{z} \rangle  ] 
+\frac{1}{2}[\gamma_{ex} - \mathcal{F}\gamma_{pl}\langle \hat{\sigma}_{z} \rangle  ] \Big)\langle \hat{\sigma} \rangle +i\Omega\Big(1 -2\langle \hat{\sigma}^{\dagger}\hat{\sigma} \rangle \Big), \label{e40c}
\end{align}
\end{subequations}
with 
\[ 
\mathcal{F} = \frac{g^{2}}{ \Delta_{pl}^{2} + \frac{1}{4}\gamma_{pl}^{2}}, ~~~~
\Omega  = \Omega_{ex}\left[1 + \frac{ig\chi}{\mu\Big(i\Delta_{pl}+\frac{1}{2}\gamma_{pl}\Big)}\right].
\]
Under the weak-field approximation, $\langle \hat{\sigma}_{z} \rangle \approx -1$, and we obtain the steady-state solution of $\langle \hat{\sigma}\rangle$ from Eq. \eqref{e40c} as
\begin{equation}\label{e41}
\langle \hat{\sigma} \rangle  \approx \frac{i\Omega[1 - 2\langle \hat{\sigma}^{\dagger}\hat{\sigma} \rangle ] }{i\Delta + \frac{1}{2}\Gamma },
\end{equation}
with 
\[ 
\Gamma  = \gamma_{ex} + \mathcal{F}\gamma_{pl}, ~~~
\Delta  = \Delta_{ex} - \mathcal{F}\Delta_{pl}. 
\]

Multiplying Eq. \eqref{e28} on the left by $\hat{\sigma}_{z}$ and performing a trace over the entire MNP-QD system (using the cyclic property of the trace and $[\hat{a},\hat{\sigma} ] = [\hat{a}^{\dagger},\hat{\sigma}^{\dagger} ] = 0$ for the Schrödinger picture operators), we obtain an equation of motion for $\langle \hat{\sigma}_{z} \rangle = \text{tr}(\hat{\sigma}_{z}\hat{\rho})$ as:
\begin{eqnarray}\label{e42}
	\text{tr}(\hat{\sigma}_{z}\dot{\hat{\rho }}) &=&  i\text{tr}\left(\hat{\sigma}_{z}[\hat{ \rho }, \Delta_{ex}\hat{\sigma}^{\dagger}\hat{\sigma} 
	-g (\hat{\sigma} \hat{a}^{\dagger}+\hat{\sigma}^{\dagger}\hat{a}) 
	-\Omega_{ex}(\hat{\sigma}  + \hat{\sigma}^{\dagger} )  ]\right)\nonumber \\
&&	- \frac{\gamma_{ex}}{2}\text{tr}\left(\hat{\sigma}_{z}\hat{ \rho }\hat{\sigma}^{\dagger}\hat{\sigma}  + \hat{\sigma}_{z}\hat{\sigma}^{\dagger}\hat{\sigma}\hat{ \rho } - 2\hat{\sigma}_{z}\hat{\sigma}\hat{ \rho }\hat{\sigma}^{\dagger} \right).
\end{eqnarray}
Using $\hat{\sigma}_{z} = 2\hat{\sigma}^{\dagger}\hat{\sigma}-1$ and the afore-mentioned properties of $\hat{\sigma}^{\dagger}$ and $\hat{\sigma}$, we obtain: 
\[ \langle \overset{.}{\hat{\sigma}}_{z} \rangle = \text{tr}(\hat{\sigma}_{z}\dot{\hat{\rho }}), 
\text{tr}\left(\hat{\sigma}_{z}\hat{ \rho }\hat{\sigma}^{\dagger}\hat{\sigma}  + \hat{\sigma}_{z}\hat{\sigma}^{\dagger}\hat{\sigma}\hat{ \rho } - 2\hat{\sigma}_{z}\hat{\sigma}\hat{ \rho }\hat{\sigma}^{\dagger} \right) = 2\text{tr}\left(\hat{\sigma}^{\dagger}\hat{\sigma}\hat{ \rho }\hat{\sigma}^{\dagger}\hat{\sigma}  + \hat{\sigma}^{\dagger}\hat{\sigma}\hat{ \rho } \right) = 2(\langle \hat{\sigma}_{z} \rangle + 1), \]
\[\text{tr}\left(\hat{\sigma}_{z}[\hat{ \rho }, \Delta_{ex}\hat{\sigma}^{\dagger}\hat{\sigma}]\right) = \Delta_{ex}\text{tr}\left(\hat{\sigma}_{z}\hat{ \rho }\hat{\sigma}^{\dagger}\hat{\sigma} - \hat{\sigma}_{z}
\hat{\sigma}^{\dagger}\hat{\sigma}\hat{ \rho } \right) =  \Delta_{ex}\text{tr}\left(2\hat{\sigma}^{\dagger}\hat{\sigma}\hat{ \rho }\hat{\sigma}^{\dagger}\hat{\sigma} - 2\hat{\sigma}^{\dagger}\hat{\sigma}
\hat{\sigma}^{\dagger}\hat{\sigma}\hat{ \rho } \right) = 0, 
\]
\[\text{tr}\left(\hat{\sigma}_{z}[\hat{ \rho },
-g (\hat{\sigma} \hat{a}^{\dagger}+\hat{\sigma}^{\dagger}\hat{a})]\right) = -g\text{tr}
\left( 
\hat{\sigma}_{z}\hat{ \rho }\hat{\sigma}^{\dagger}\hat{a} + \hat{\sigma}_{z}\hat{ \rho }\hat{\sigma}^{\dagger}\hat{a} - 
\hat{\sigma}_{z}\hat{\sigma}^{\dagger}\hat{a}\hat{ \rho }- 
\hat{\sigma}_{z}\hat{\sigma}^{\dagger}\hat{a}\hat{ \rho }
\right)  
= -2g(\langle \hat{\sigma}\hat{a}^{\dagger} \rangle - \langle \hat{\sigma}^{\dagger}\hat{a} \rangle),
\]
\[ \text{tr}\left(\hat{\sigma}_{z}[\hat{ \rho },
-\Omega_{ex}(\hat{\sigma}  + \hat{\sigma}^{\dagger} )  ]  \right)
=  -\Omega_{ex}\text{tr}\left(
\hat{ \rho }\hat{\sigma}\hat{\sigma}_{z} + 
\hat{\sigma}^{\dagger}\hat{\sigma}_{z}\hat{ \rho } 
- \hat{\sigma}_{z}\hat{\sigma}\hat{ \rho }
- \hat{\sigma}_{z}\hat{\sigma}^{\dagger}\hat{ \rho }
\right)
 = -2\Omega_{ex}[\langle  \hat{\sigma} \rangle - \langle  \hat{\sigma}^{\dagger} \rangle ],
\]
so that Eq. \eqref{e42} simplifies to: 
\begin{equation}\label{e43}
\langle \overset{.}{\hat{\sigma}}_{z} \rangle =
-\gamma_{ex}( \langle  \hat{\sigma}_{z} \rangle + 1  ) -2i(\Omega_{ex} + g\langle\hat{a}^{\dagger} \rangle)\langle \hat{\sigma} \rangle
+2i\langle \hat{\sigma}^{\dagger} \rangle(\Omega_{ex} + g\langle\hat{a} \rangle),
\end{equation}
where we have again factored the joint expectation values of the MNP and QD operators. 
Using the adiabatic approximation for $\langle \hat{a}\rangle$, we have the following approximation for terms in Eq.~\eqref{e43}: 
\[(\Omega_{ex} + g\langle\hat{a}^{\dagger} \rangle)\langle \hat{\sigma} \rangle \approx \left[\Omega_{ex} + g\left( \frac{-i[\Omega_{pl} + g\langle \hat{\sigma}^{\dagger} \rangle ]}{-i\Delta_{pl} +\frac{1}{2}\gamma_{pl} }  \right)  \right]\langle \hat{\sigma} \rangle, 
\]
\[\langle \hat{\sigma}^{\dagger} \rangle(\Omega_{ex} + g\langle\hat{a} \rangle) \approx \langle \hat{\sigma}^{\dagger} \rangle\left[\Omega_{ex} + g\left( \frac{i[\Omega_{pl} + g\langle \hat{\sigma}^{\dagger} \rangle ]}{i\Delta_{pl} +\frac{1}{2}\gamma_{pl} }  \right)  \right], 
\]
\[-2i(\Omega_{ex} + g\langle\hat{a}^{\dagger} \rangle)\langle \hat{\sigma} \rangle + 2i\langle \hat{\sigma}^{\dagger} \rangle(\Omega_{ex} + g\langle\hat{a} \rangle)  = 2i(\Omega\langle \hat{\sigma}^{\dagger} \rangle
-\Omega^{*}\langle \hat{\sigma} \rangle) - \mathcal{F}\gamma_{pl}\langle (\hat{\sigma}_{z} \rangle + 1),
\]
allowing us to further simplify Eq. \eqref{e43} to obtain
\begin{equation}\label{e44}
		\langle \overset{.}{\hat{\sigma}_{z}} \rangle  = 
-\Gamma(\langle \hat{\sigma}_{z} \rangle + 1 )
+ 4 \Im[\Omega^{*}\langle\hat{\sigma}\rangle].
\end{equation}
From  Eq. \eqref{e44}, the steady-state solution of $\langle \hat{\sigma}^{\dagger}\hat{\sigma} \rangle$ is 
\begin{equation}
\langle \hat{\sigma}^{\dagger}\hat{\sigma} \rangle = \frac{2 \Im[\Omega^{*}\langle\hat{\sigma}\rangle]  }{\Gamma }. 
\end{equation}
Expanding $2 \Im[\Omega^{*}\langle\hat{\sigma}\rangle]$ using Eq. \eqref{e41}, we obtain
\[2 \Im[\Omega^{*}\langle\hat{\sigma}\rangle] = 
\mathcal{P}\Gamma(1 - 2\langle \hat{\sigma}^{\dagger}\hat{\sigma} \rangle ),
\]
where 
\[\mathcal{P} =  \frac{2|\Omega/\Gamma|^{2} }{1 + 2(\Delta/\Gamma)^{2}  }
\]
is the saturation parameter, so that
\begin{equation}
\langle \hat{\sigma}^{\dagger}\hat{\sigma} \rangle \approx 
\frac{\mathcal{P}}{1 + 2\mathcal{P}}.
\end{equation}
The above equation is compatible with the weak-field approximation ($\Gamma \gg \Omega$) as $\mathcal{P} \ll 1$ and therefore $\langle \hat{\sigma}^{\dagger}\hat{\sigma} \rangle \approx 0$.

\section{Photodetection}
The normalized steady-state correlation functions, $g^{(2)}(0), g^{(3)}(0)$ and $g^{(4)}(0)$, are given by \cite{Miri14}
\begin{subequations}
\begin{align}
g^{(2)}(0) & =  \frac{\langle (\hat{b}^{\dagger})^{2}\hat{b}^{2} \rangle }{ \langle \hat{b}^{\dagger}\hat{b} \rangle^{2}  }, \label{53a} \\
g^{(3)}(0) & =  \frac{ \langle (\hat{b}^{\dagger})^{3}\hat{b}^{3} \rangle }{ \langle \hat{b}^{\dagger}\hat{b} \rangle^{3}  },\\
g^{(4)}(0) & =  \frac{ \langle (\hat{b}^{\dagger})^{4}\hat{b}^{4} \rangle }{ \langle \hat{b}^{\dagger}\hat{b} \rangle^{4}  }, 
\end{align} 
\end{subequations}
where 
\begin{subequations}
\begin{align}
\langle \hat{b}^{\dagger}\hat{b} \rangle & = \langle (\sqrt{\gamma_{r} }\hat{a})^{\dagger}\sqrt{\gamma_{r} }\hat{a} \rangle, \\
                         & = \gamma_{r}\langle\hat{a}^{\dagger}\hat{a}\rangle,\\
                         & \approx \gamma_{r}\left( 	\frac{\Omega_{pl}^{2} + g^{2}\langle \hat{\sigma}^{\dagger}\hat{\sigma} \rangle + 2\Omega_{pl}g\Re\langle \hat{\sigma} \rangle   }{\Delta_{pl}^{2} + \frac{1}{4}\gamma_{pl}^{2}  }    \right)
\end{align}
\end{subequations}
is the mean scattering rate of photons by the MNP \cite{Waks10}. The expectation values: $\langle (\hat{b}^{\dagger})^{2}\hat{b}^{2} \rangle, \langle (\hat{b}^{\dagger})^{3}\hat{b}^{3} \rangle$, and $\langle (\hat{b}^{\dagger})^{4}\hat{b}^{4} \rangle$, proportional to the 2-photon, 3-photon, and 4-photon scattering probabilities, are obtained respectively as: 
\begin{subequations}
	\begin{align}
		\langle (\hat{b}^{\dagger})^{2}\hat{b}^{2} \rangle & = \langle ((\sqrt{\gamma_{r} }\hat{a})^{\dagger})^{2}(\sqrt{\gamma_{r} }\hat{a})^{2} \rangle, \\
		& = \gamma_{r}^{2}\langle (\hat{a}^{\dagger})^{2}\hat{a}^{2} \rangle, \\
		& \approx \gamma_{r}^{2}\left( 	\frac{\Omega_{pl}^{4} + 4\Omega_{pl}^{2}g^{2}\langle \hat{\sigma}^{\dagger}\hat{\sigma} \rangle + 4\Omega_{pl}^{3}g\Re\langle \hat{\sigma} \rangle   }{(\Delta_{pl}^{2} + \frac{1}{4}\gamma_{pl}^{2})^{2}  }    \right),
	\end{align}
\end{subequations}
\begin{subequations}
	\begin{align}
		\langle (\hat{b}^{\dagger})^{3}\hat{b}^{3} \rangle & = \langle ((\sqrt{\gamma_{r} }\hat{a})^{\dagger})^{3}(\sqrt{\gamma_{r} }\hat{a})^{3} \rangle, \\
& = \gamma_{r}^{3}\langle (\hat{a}^{\dagger})^{3}\hat{a}^{3} \rangle, \\
		& \approx \gamma_{r}^{3}\left( \frac{\Omega_{pl}^{6} + 9\Omega_{pl}^{4}g^{2}\langle \hat{\sigma}^{\dagger}\hat{\sigma} \rangle + 6\Omega_{pl}^{5}g\Re\langle \hat{\sigma} \rangle   }{(\Delta_{pl}^{2} + \frac{1}{4}\gamma_{pl}^{2})^{3}  }    \right),
	\end{align}
\end{subequations}
\begin{subequations}
	\begin{align}
		\langle (\hat{b}^{\dagger})^{4}\hat{b}^{4} \rangle & = \langle ((\sqrt{\gamma_{r} }\hat{a})^{\dagger})^{4}(\sqrt{\gamma_{r} }\hat{a})^{4} \rangle, \\
& = \gamma_{r}^{4}\langle (\hat{a}^{\dagger})^{4}\hat{a}^{4} \rangle, \\
		& \approx \gamma_{r}^{4}\left(\frac{\Omega_{pl}^{8} + 16\Omega_{pl}^{6}g^{2}\langle \hat{\sigma}^{\dagger}\hat{\sigma} \rangle + 8\Omega_{pl}^{7}g\Re\langle \hat{\sigma} \rangle   }{(\Delta_{pl}^{2} + \frac{1}{4}\gamma_{pl}^{2})^{4}  }    \right),
	\end{align}
\end{subequations}
leading to:
\begin{subequations}
	\begin{align}
g^{(2)}(0) & = \frac{\Omega_{pl}^{4} + 4\Omega_{pl}^{3}g\Re\langle \hat{\sigma} \rangle + 4\Omega_{pl}^{2}g^{2}\langle \hat{\sigma}^{\dagger}\hat{\sigma} \rangle    }
{\Big( \Omega_{pl}^{2} + 2\Omega_{pl}g\Re\langle \hat{\sigma} \rangle + g^{2}\langle \hat{\sigma}^{\dagger}\hat{\sigma} \rangle     \Big)^{2}  }, \label{58a}\\ 
g^{(3)}(0) &  = \frac{\Omega_{pl}^{6} +  6g\Omega_{pl}^{5}\Re\langle \hat{\sigma} \rangle + 9g^{2}\Omega_{pl}^{4}\langle \hat{\sigma}^{\dagger}\hat{\sigma} \rangle   }
{\Big( \Omega_{pl}^{2} + 2\Omega_{pl}g\Re\langle \hat{\sigma} \rangle + g^{2}\langle \hat{\sigma}^{\dagger}\hat{\sigma} \rangle     \Big)^{3}  }, \\
g^{(4)}(0) &  = \frac{\Omega_{pl}^{8} + 8g\Omega_{pl}^{7}\Re\langle \hat{\sigma} \rangle + 16g^{2}\Omega_{pl}^{6}\langle \hat{\sigma}^{\dagger}\hat{\sigma} \rangle    }
{\Big( \Omega_{pl}^{2} + 2\Omega_{pl}g\Re\langle \hat{\sigma} \rangle + g^{2}\langle \hat{\sigma}^{\dagger}\hat{\sigma} \rangle    \Big)^{4}  }.
	\end{align}
\end{subequations}
We note that the scattering operator, $\hat{b}$, defined in Ref. \cite{Waks10}, and the polarization operator, $\hat{P}^{+}$, defined in Ref. \cite{Dolf10} (which is proportional to the scattered field operator, $\hat{E}^{(+)}$), are related via the Wigner-Weisskopf formula. In Ref. \cite{Dolf10}, 
$\hat{P}^{+}$ is defined as: $\hat{P}^{+} = \chi \hat{a} + \mu \hat{\sigma}$. 
From Eq. \eqref{e8}, $\chi \propto \sqrt{\gamma_{r}}$. A similar version of Eq. \eqref{e8} for the QD leads to $\mu \propto \sqrt{\gamma_{ex}}$, so that $\hat{P}^{+}$ transforms to: $\hat{b} = \sqrt{\gamma_{r}} \hat{a} + \sqrt{\gamma_{ex}} \hat{\sigma}$ up to some fixed proportionality constant, assuming the MNP and QD have a similar resonance frequency, i.e. $\omega_{pl} \approx \omega_{ex}$.

The derivations in the rest of this section, and in most part, in the next section too, follow directly from Section 6.10 of Ref. \cite{Rod00}.
The mean photocount, i.e., the average number of photons arriving at a detector during the integration time, $\mathrm{T}_{int}$, is given by:
\begin{equation}\label{e53}
\langle \hat{m} \rangle = \left\langle
\int_{t}^{t+\mathrm{T}_{int}}\sqrt{\xi}\hat{b}^{\dagger}(t')\sqrt{\xi}\hat{b}(t')dt'\right\rangle.
\end{equation}
In the steady-state, we have
\begin{equation}\label{e54}
	\langle \hat{m} \rangle = \left\langle
		\int_{t}^{t+\mathrm{T}_{int}}\sqrt{\xi}\hat{b}^{\dagger}(t'\rightarrow\infty)\sqrt{\xi}\hat{b}(t'\rightarrow\infty)dt'\right\rangle = 
		\xi\mathrm{T}_{int}\langle\hat{b}^{\dagger}\hat{b}\rangle.
\end{equation}
\begin{figure} [ht!]
	\centering 
	\includegraphics[width =0.77\textwidth]{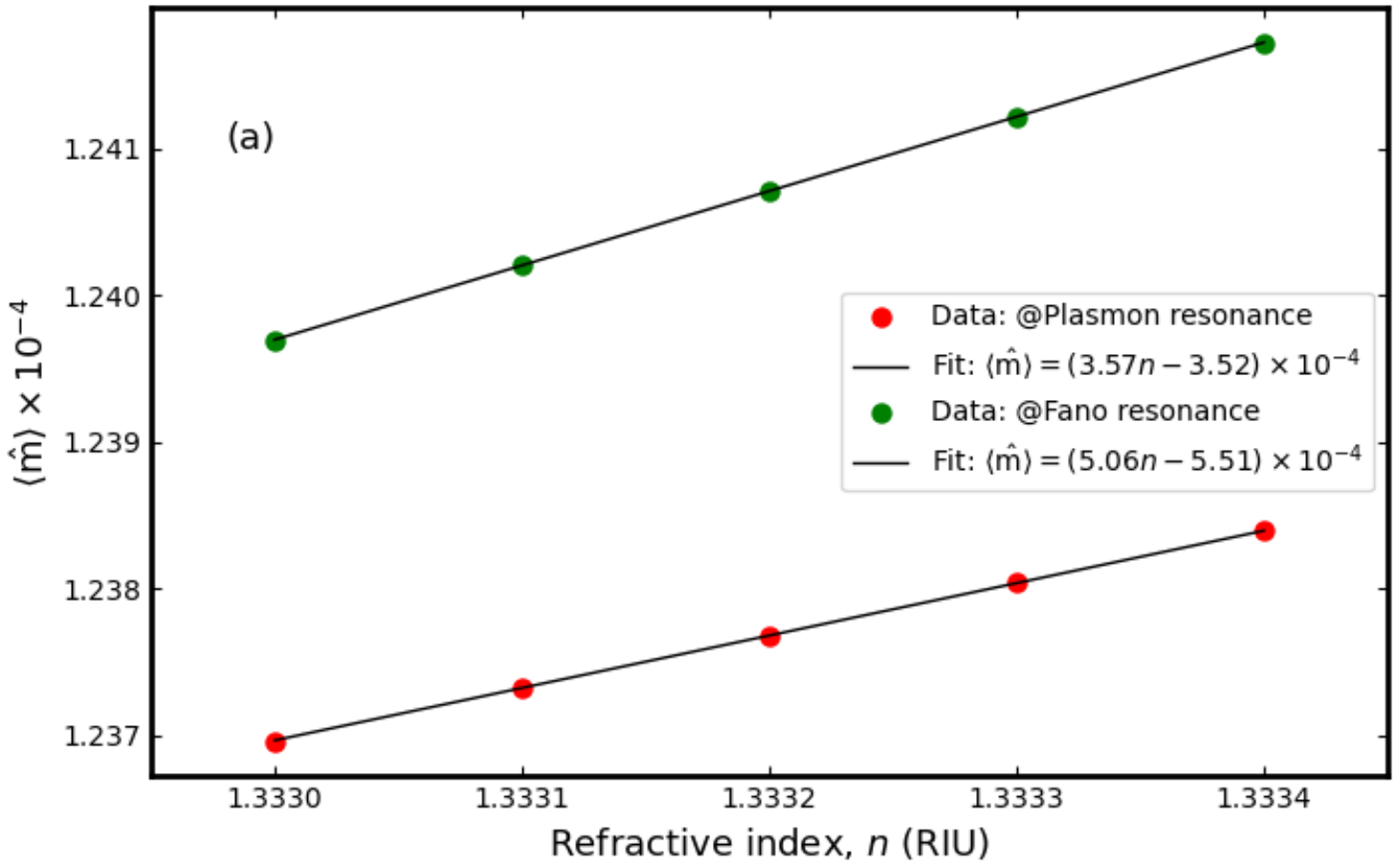}\vspace{0.4cm}\\
	\includegraphics[width = 0.8\textwidth]{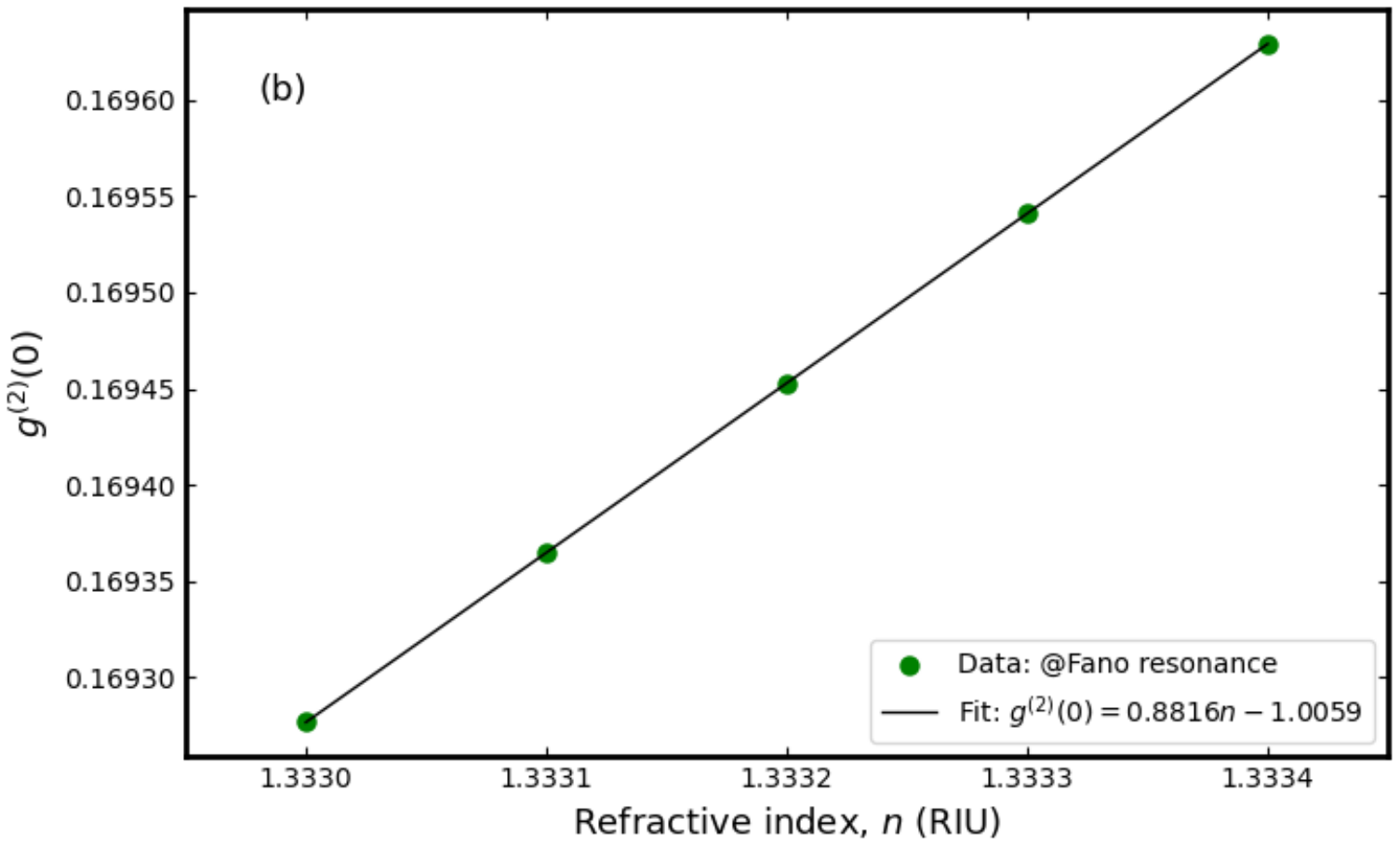}
	\caption{(a) The linear dependence of $\langle \hat{m} \rangle$ on the refractive index, $n$, of the medium, at a driving wavelength near the plasmon resonance (Fano resonance) (as an example). (b) The linear dependence of $g^{(2)}(0)$ on the refractive index, $n$, of the medium, at a driving wavelength near the Fano resonance (as an example). 
	}\label{S1}
\end{figure}
The second factorial moment, $\langle \hat{m}(\hat{m}-1) \rangle$, 
is proportional to the two-photon detection probability, and it is given by \cite{Rod00}:
\begin{equation}
\langle \hat{m}(\hat{m}-1) \rangle = \xi^{2}
\left\langle
\int_{t}^{t+\mathrm{T}_{int}}dt'\int_{t}^{t+\mathrm{T}_{int}}dt''
\hat{b}^{\dagger}(t')\hat{b}^{\dagger}(t'')\hat{b}(t'')\hat{b}(t')\right\rangle.
\end{equation}
Let $t''=t'+\tau$, so that in the steady state we have
\begin{equation}
\langle \hat{m}(\hat{m}-1) \rangle = \xi^{2} \langle \hat{b}^\dag \hat{b}\rangle^2 \int_{t}^{t+\mathrm{T}_{int}}dt'\int_{t}^{t+\mathrm{T}_{int}}dt'' g^{(2)}(\tau).
\end{equation}
This leads to the short time approximation
\begin{equation}
\langle \hat{m}(\hat{m}-1) \rangle_{0} = 
\xi^{2}\mathrm{T}_{int}^{2}
\langle(\hat{b}^{\dagger})^{2}\hat{b}^{2}\rangle,
\end{equation}
where ${\rm T}_{int}$ is chosen small enough such that $g^{(2)}(\tau) \approx g^{(2)}(0) \approx {\rm const.}$ over the integration time. Here, we use a `0' subscript to denote the approximation for the second factorial moment.

The zero-time delayed, second-order correlation function, $g^{(2)}(0)$, is defined in terms of $\langle \hat{m}(\hat{m}-1) \rangle$ and $\langle \hat{m} \rangle$ as \cite{Rod00}: 
\begin{equation}\label{e57}
g^{(2)}(0) = \frac{\langle \hat{m}(\hat{m}-1) \rangle }{\langle \hat{m} \rangle^{2} }.
\end{equation}
When considering all quantities are to be measured over some integration time, ${\rm T}_{int}$, as in an experiment, this becomes
\begin{equation}\label{e57b}
g_D^{(2)}(0) = \frac{\langle \hat{m}(\hat{m}-1) \rangle }{\langle \hat{m} \rangle^{2} },
\end{equation}
where $g_D^{(2)}(0)$ is the detected second-order correlation function. With $\langle \hat{m}(\hat{m}-1) \rangle$ replaced by $\langle \hat{m}(\hat{m}-1) \rangle_0$ and $g_D^{(2)}(0)$ by $g^{(2)}(0)$, for small time ${\rm T}_{int}$, we recover the definition given in Eq.~\eqref{53a}. The linear dependence of $\langle \hat{m} \rangle$ and $g^{(2)}(0)$ on the refractive index $n$ are plotted in Fig. \ref{S1}.

\begin{figure} [ht!]
	\centering 
	\includegraphics[width = 0.6\textwidth]{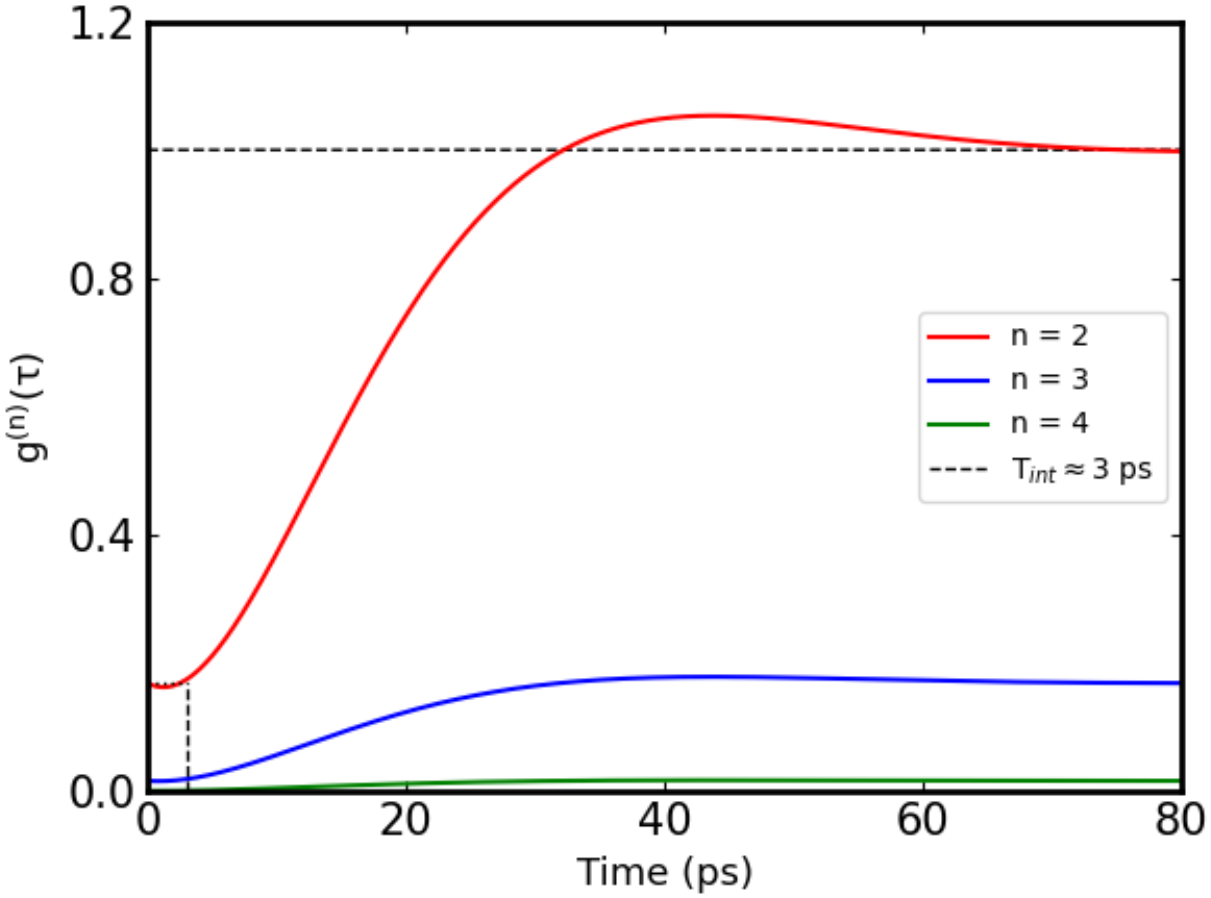}
	\caption{The time-delayed steady-state correlation functions, $g^{(n)}(\tau)$, $n = 2, 3, 4$, at the Fano resonance. 
	}\label{S2}
\end{figure}

The time-delayed steady-state two-photon, $G^{(2)}(\tau)$, three-photon, $G^{(3)}(\tau)$, and four-photon, $G^{(4)}(\tau)$, correlation functions are obtained from QuTiP using the following expressions \cite{Gard04,Nori13}: 
\begin{subequations}
	\begin{align}
		G^{(2)}(\tau) & = \lim_{t\longrightarrow \infty} \langle \hat{b}^{\dagger}(t) \hat{b}^{\dagger}(t+\tau)\hat{b}(t+\tau) \hat{b}(t) \rangle,\\
		& = \text{Tr}[AB\hat{V}(\tau,0)C\hat{\rho} ],\\
		& = \langle A(0) B(\tau) C(0) \rangle, 
	\end{align}
\end{subequations}
with $A = \hat{b}^{\dagger}, B = \hat{b}^{\dagger}\hat{b}, C = \hat{b}$, 
\begin{subequations}
	\begin{align}
		G^{(3)}(\tau) & = \lim_{t\longrightarrow \infty} \langle \hat{b}^{\dagger}(t)\hat{b}^{\dagger}(t+\tau) \hat{b}^{\dagger}(t+\tau)\hat{b}(t+\tau)\hat{b}(t+\tau) \hat{b}(t) \rangle,\\
		& = \text{Tr}[AB\hat{V}(\tau,0)C\hat{\rho} ],\\
		& = \langle A(0) B(\tau) C(0) \rangle, 
	\end{align}
\end{subequations}
with $A = \hat{b}^{\dagger}, B = (\hat{b}^{\dagger})^{2}\hat{b}^{2}, C = \hat{b}$,
\begin{subequations}
	\begin{align}
		G^{(4)}(\tau) & = \lim_{t\longrightarrow \infty} \langle \hat{b}^{\dagger}(t)\hat{b}^{\dagger}(t+\tau)\hat{b}^{\dagger}(t+\tau) \hat{b}^{\dagger}(t+\tau)\hat{b}(t+\tau)\hat{b}(t+\tau)\hat{b}(t+\tau) \hat{b}(t) \rangle,\\
		& = \text{Tr}[AB\hat{V}(\tau,0)C\hat{\rho} ],\\
		& = \langle A(0) B(\tau) C(0) \rangle, 
	\end{align}
\end{subequations}
with $A = \hat{b}^{\dagger}, B = (\hat{b}^{\dagger})^{3}\hat{b}^{3}, C = \hat{b}$, from which we obtain the normalized steady-state correlation functions: 
\begin{equation}
	g^{(n)}(\tau)  = \frac{G^{(n)}(\tau)}{ \langle \hat{b}^{\dagger}\hat{b} \rangle^{n}  }, ~~~n = 2, 3, 4. 
\end{equation} 
Note that for simplicity,  we have considered only a single time delay, $\tau$, for all orders. These correlations functions are plotted in Fig. \ref{S2}.

\section{Standard deviations}
The variance in the mean photon number is \cite{Rod00}
\begin{equation}
(\Delta m )^{2}  = \langle \hat{m}^{2} \rangle -  \langle \hat{m} \rangle^{2},
\end{equation}
where $\hat{m} = \xi\mathrm{T}_{int}\hat{b}^{\dagger}\hat{b}$ is the photocount operator. We can express $\langle \hat{m}^{2} \rangle$ in terms of $g^{(2)}(0)$ and $\langle\hat{m} \rangle$ as follows:
\begin{subequations}
\begin{align}
\langle \hat{m}^{2} \rangle &  = \langle \hat{m}^{2} - \hat{m} + \hat{m} \rangle, \\
& = \langle \hat{m}(\hat{m} - 1)\rangle + \langle\hat{m} \rangle,\\
& = g^{(2)}(0)\langle\hat{m} \rangle^{2} + \langle\hat{m} \rangle,
\end{align}
\end{subequations}
where in the last line we have used Eq.~\eqref{e57}. Here, all quantities should have the same integration time, thus $\langle\hat{m} \rangle$ should be measured over a small enough time, $\mathrm{T}_{int}$, such that $\langle \hat{m}(\hat{m} - 1)\rangle \approx \langle \hat{m}(\hat{m} - 1)\rangle_{0}$ can be used in the calculations, and Eq.~\eqref{58a} together with the higher order expressions give analytical formulas that can be used to obtain the standard deviations in what follows. Once 
$\mathrm{T}_{int}$ grows too much, the non-stationary behaviour 
of $\langle \hat{m}(\hat{m} - 1)\rangle$ begins to form and the 
approximation is no longer valid. As shown in Fig. \ref{S2}, this approximation ($g^{(2)}(\tau) \approx g^{(2)}(0) \approx {\rm const.}$) is valid up to $\mathrm{T}_{int} \approx 3$ ps.
The standard deviation in the mean photocount can then be expressed as 
\begin{equation}
\Delta m = \sqrt{g^{(2)}(0)\langle\hat{m} \rangle^{2} + \langle\hat{m} \rangle - \langle \hat{m} \rangle^{2}  } = \sqrt{\langle \hat{m} \rangle }\sqrt{1 + (g^{(2)}(0)-1)\langle \hat{m} \rangle }.
\end{equation}

Let $\Delta m_{2}$ denote the standard deviation in the second factorial moment. The variance in the second factorial moment is \cite{Rod00}
\begin{equation}\label{e61}
	(\Delta m_{2} )^{2}  = \xi^{4}\left(\langle (\hat{m}(\hat{m}-1)_{L})^{2} \rangle -  \langle \hat{m}(\hat{m}-1)_{L} \rangle\right),
\end{equation}
where $\langle \hat{m}(\hat{m}-1)_{L} \rangle = 
\left\langle
\int_{t}^{t+\mathrm{T}_{int}}dt'\int_{t}^{t+\mathrm{T}_{int}}dt''
\hat{b}^{\dagger}(t')\hat{b}^{\dagger}(t'')\hat{b}(t'')\hat{b}(t')\right\rangle$ is the second factorial moment for a lossless detector.
We can express $(\hat{m}(\hat{m}-1)_{L})^{2}$ in terms of higher-order moments. Let $\langle \hat{m}_{2} \rangle, \langle\hat{m}_{3}\rangle$, and 
$\langle\hat{m}_{4}\rangle$ denote the second, third, and fourth factorial moments. Then we have the higher-order operators:
\begin{subequations}
\begin{align}
\hat{m}_{2} & = \hat{m}(\hat{m}-1) \nonumber\\
           & = \int_{t}^{t+\mathrm{T}_{int}}dt'\int_{t}^{t+\mathrm{T}_{int}}dt''
\hat{b}^{\dagger}(t')\hat{b}^{\dagger}(t'')\hat{b}(t'')\hat{b}(t'),\\ 
\hat{m}_{3} & = \hat{m}(\hat{m}-1)(\hat{m}-2) \nonumber\\
          &  = \int_{t}^{t+\mathrm{T}_{int}}dt'\int_{t}^{t+\mathrm{T}_{int}}dt''\int_{t}^{t+\mathrm{T}_{int}}dt'''
\hat{b}^{\dagger}(t')\hat{b}^{\dagger}(t'')\hat{b}^{\dagger}(t''')\hat{b}(t''')\hat{b}(t'')\hat{b}(t'),\\
\hat{m}_{4} & = \hat{m}(\hat{m}-1)(\hat{m}-2)(\hat{m}-3) \nonumber\\
     &  = \int_{t}^{t+\mathrm{T}_{int}}dt'\int_{t}^{t+\mathrm{T}_{int}}dt''\int_{t}^{t+\mathrm{T}_{int}}dt'''\int_{t}^{t+\mathrm{T}_{int}}dt''''
     \hat{b}^{\dagger}(t')\hat{b}^{\dagger}(t'')\hat{b}^{\dagger}(t''')\hat{b}^{\dagger}(t'''')\hat{b}(t'''')\hat{b}(t''')\hat{b}(t'')\hat{b}(t'),
\end{align} 
\end{subequations}
where
\begin{subequations}
\begin{align}
\hat{m}_{2} & = \hat{m}^{2} - \hat{m}, \\
\hat{m}_{3} & = \hat{m}^{3} - 3\hat{m}^{2} + 2\hat{m},\\
\hat{m}_{4} & = \hat{m}^{4} - 6\hat{m}^{3} + 11\hat{m}^{2} - 6\hat{m},\\
\implies (\hat{m}(\hat{m}-1))^2 &  = \hat{m}^{4} - 2\hat{m}^{3} + \hat{m}^{2} = \hat{m}_{4} + 4\hat{m}_{3} + 2\hat{m}_{2}.
\end{align}
\end{subequations}
The expectation values, $\langle \hat{m}_{2} \rangle, \langle \hat{m}_{3} \rangle$, and $\langle \hat{m}_{4} \rangle$ can be expressed in terms of the steady-state, zero-time delayed correlation functions, 
$g^{(2)}(0), g^{(3)}(0)$, and $g^{(4)}(0)$, for small $\mathrm{T}_{int}$ as:
\[\langle \hat{m}_{2} \rangle = g^{(2)}(0)\langle \hat{m} \rangle^{2},
\langle \hat{m}_{3} \rangle = g^{(3)}(0)\langle \hat{m} \rangle^{3}, 
\langle \hat{m}_{4} \rangle = g^{(4)}(0)\langle \hat{m} \rangle^{4},
\]
so that Eq.~\eqref{e61} can be re-written as:
\begin{subequations}
\begin{align}
(\Delta m_{2})^{2} & = \xi^{4}\left(g^{(4)}(0)\langle \hat{m}_{L} \rangle^{4} +4g^{(3)}(0)\langle \hat{m}_{L} \rangle^{3} +2g^{(2)}(0)\langle \hat{m}_{L} \rangle^{2} -  [g^{(2)}(0)]^{2}\langle \hat{m}_{L} \rangle^{4} \right) , \\
\implies \Delta m_{2} & = \langle \hat{m} \rangle^{2}
\sqrt{g^{(4)}(0) - [g^{(2)}(0)]^{2}   + 4g^{(3)}(0) \left(\frac{\xi}{ \langle \hat{m} \rangle}\right)  + 2g^{(2)}(0)\left(\frac{\xi}{ \langle \hat{m} \rangle} \right) ^{2}  },
\end{align}
\end{subequations}
with $\langle \hat{m} \rangle = \xi\langle \hat{m}_{L} \rangle$.

Given that:
\[g^{(2)}(0) = \frac{\langle \hat{m}_{2} \rangle }{\langle \hat{m} \rangle^{2} },
\]
the standard deviation in $g^{(2)}(0)$ is obtained using the uncertainty propagation rule \cite{Forn08}, as
\begin{subequations}
\begin{align}
\Delta g^{(2)}(0) & = \sqrt{\left(\dfrac{\partial g^{(2)}(0)}{\partial\langle \hat{m}_{2} \rangle} \Delta m_{2}  \right)^{2} + \left(\dfrac{\partial g^{(2)}(0)}{\partial\langle \hat{m} \rangle} \Delta m  \right)^{2}   },\\
& = \sqrt{\left(\frac{\Delta m_{2} }{\langle \hat{m} \rangle^{2}}   \right)^{2} + \left(\frac{-2\langle \hat{m}_{2} \rangle \Delta m }{\langle \hat{m} \rangle^{3}} \right)^{2}   },\\
& = \left(2g^{(2)}(0)\frac{\Delta m }{ \langle \hat{m} \rangle }\right)\sqrt{1 + \left(\frac{\Delta m_{2} }{2g^{(2)}(0) \langle \hat{m} \rangle\Delta m }\right)^{2} }.
\end{align}
\end{subequations}
